\documentclass[a4paper,11pt]{article}
\pdfoutput=1 

\usepackage{jheppub} 
\usepackage[T1]{fontenc}
\usepackage{booktabs}
\usepackage{xspace}
\usepackage{dcolumn}
\usepackage{hyperref}
\usepackage{caption}
\usepackage
[subrefformat=parens,position=top,skip=-15pt,margin=15pt,justification=justified,singlelinecheck=false]
{subcaption}
\usepackage{slashed}
\usepackage{comment}
\usepackage[colorinlistoftodos, shadow]{todonotes}
\usepackage{amsmath}
\usepackage{graphicx}

\makeatletter
\newcommand\myparagraph{\@startsection{paragraph}{4}{\z@}%
  {-10\p@ \@plus -6\p@ \@minus -3\p@}%
  {3\p@}%
  {\normalfont\itshape}%
}
\makeatother

\newcommand{\id}{{\rm 1\kern-.12em
 \rule{0.3pt}{1.5ex}\raisebox{0.0ex}{\rule{0.1em}{0.3pt}}}}

\allowdisplaybreaks

\def\refeq#1{\mbox{(\ref{#1})}}

\def\refta#1{\mbox{Table~\ref{#1}}}

\def\refse#1{\mbox{Section~\ref{#1}}}
\def\refses#1{\mbox{Sections~\ref{#1}}}
\def\refapp#1{\mbox{App.~\ref{#1}}}

\def\refta#1{\mbox{Table~\ref{#1}}}

\def\refse#1{\mbox{Section~\ref{#1}}}
\def\refses#1{\mbox{Sections~\ref{#1}}}
\def\refapp#1{\mbox{App.~\ref{#1}}}

\def\citere#1{\mbox{Ref.~\cite{#1}}}
\def\citeres#1{\mbox{Refs.~\cite{#1}}}

\newcommand{\cmm}[1]{\ensuremath{#1}\ifmmode\else{}\fi}
\newcommand{\nmc}[2]{\newcommand{#1}{\cmm{#2}}}
\nmc{\al}{\alpha}
\nmc{\be}{\beta}
\nmc{\de}{\delta}
\nmc{\De}{\Delta}
\nmc{\si}{\sigma}
\nmc{\eps}{\epsilon}
\newcommand{\bfphi}{\text{\boldmath{$\phi$}}}
\newcommand{\bfrho}{\text{\boldmath{$\rho$}}}
\newcommand{\bftheta}{\text{\boldmath{$\theta$}}}
\newcommand{\bfzeta}{\text{\boldmath{$\zeta$}}}

\newcommand{\ri}{\mathrm i}
\newcommand{\ru}{\mathrm u}

\nmc{\rd}{\mathrm{d}}

\def\beq{\begin{equation}}
\def\eeq{\end{equation}}

\newcommand{\PH}{\ensuremath{\text{H}}\xspace}
\newcommand{\PHone}{\ensuremath{{\text{H}_1}}\xspace}
\newcommand{\PHtwo}{\ensuremath{{\text{H}_2}}\xspace}
\newcommand{\PHpm}{\ensuremath{\text{H}^\pm}\xspace}
\newcommand{\PHa}{\ensuremath{A_0}\xspace}

\newcommand{\Ph}{\ensuremath{\text{h}}\xspace}
\newcommand{\PA}{\ensuremath{\text{A}}\xspace}

\newcommand{\PW}{\ensuremath{\text{W}}\xspace}
\newcommand{\PZ}{\ensuremath{\text{Z}}\xspace}

\newcommand{\MH}{\ensuremath{M_\PH}\xspace}
\newcommand{\Mh}{\ensuremath{M_\Ph}\xspace}
\newcommand{\MA}{\ensuremath{M_\PA}\xspace}

\newcommand{\MW}{\ensuremath{M_\PW}\xspace}

\newcommand{\MZ}{\ensuremath{M_\PZ}\xspace}

\newcommand{\GeV}{\ensuremath{\,\text{GeV}}\xspace}

\newcommand{\ren}{{\mathrm{R}}}
\newcommand{\bare}{{\mathrm{B}}}

\newcommand{\UV}{{\mathrm{UV}}}
\newcommand{\rw}{{\mathrm{w}}}
\nmc{\alem}{\alpha_{\mathrm{em}}}
\nmc{\sw}{s_{\rw}}
\nmc{\cw}{c_{\rw}}

\nmc{\g}{g_2}
\nmc{\gy}{g_1}
\nmc{\Gf}{G_\mathrm{F}}

\renewcommand{\Re}{\mathop{\mathrm{Re}}\nolimits}

\newcommand{\rT}{\ensuremath{\text{T}}\xspace}

\nmc{\rB}{{\rm B}}
\nmc{\rs}{{\rm s}}

\nmc{\Msb}{M_{\rm sb}}

\nmc{\ftone}{t_{s}}
\nmc{\tab}{t_{\alpha\beta}}

\nmc{\tHoneHone}{t_{H_1H_1}}
\nmc{\tHoneHtwo}{t_{H_1H_2}}
\nmc{\tHtwoHtwo}{t_{H_2H_2}}
\nmc{\tHaHa}{t_{\Ha\Ha}}
\nmc{\tHpmHpm}{t_{H^\pm H^\pm}}
\nmc{\tGzGz}{t_{G_0G_0}}
\nmc{\tGpmGpm}{t_{G^\pm G^\pm}}
\nmc{\tGzHa}{t_{G_0\Ha}}
\nmc{\tGpmHpm}{t_{G^\pm H^\pm}}

\newcommand{\Prophecy}{{\sc Prophecy4f}}

\let\lsim\lesssim

\def\draftdate{\relax}
\def\mda{\relax}
\def\mua{\relax}
\def\mla{\relax}
\def\draft{
\def\thtystars{******************************}
\def\sixtystars{\thtystars\thtystars}
\typeout{}
\typeout{\sixtystars**}
\typeout{* Draft mode!
         For final version remove \protect\draft\space in source file *}
\typeout{\sixtystars**}
\typeout{}
\def\draftdate{\today}
\def\mua{\marginpar[\boldmath\hfil$\uparrow$]%
                   {\boldmath$\uparrow$\hfil}%
                    \typeout{marginpar: $\uparrow$}\ignorespaces}
\def\mda{\marginpar[\boldmath\hfil$\downarrow$]%
                   {\boldmath$\downarrow$\hfil}%
                    \typeout{marginpar: $\downarrow$}\ignorespaces}
\def\mla{\marginpar[\boldmath\hfil$\rightarrow$]%
                   {\boldmath$\leftarrow $\hfil}%
                    \typeout{marginpar: $\leftrightarrow$}\ignorespaces}
\def\Mua{\marginpar[\boldmath\hfil$\Uparrow$]%
                   {\boldmath$\Uparrow$\hfil}%
                    \typeout{marginpar: $\uparrow$}\ignorespaces}
\def\Mda{\marginpar[\boldmath\hfil$\Downarrow$]%
                   {\boldmath$\Downarrow$\hfil}%
                    \typeout{marginpar: $\downarrow$}\ignorespaces}
\def\Mla{\marginpar[\boldmath\hfil{$\Rightarrow$}]%
                   {\boldmath{$\Leftarrow $}\hfil}%
                    \typeout{marginpar:$\leftrightarrow$}\ignorespaces}
\def\muanick{\marginpar[\boldmath\hfil$\uparrow$]%
                   {\boldmath$\textcolor{blue}\uparrow$\hfil}%
                    \typeout{marginpar:\textcolor{blue} $\uparrow$}\ignorespaces}
\def\mdanick{\marginpar[\boldmath\hfil$\downarrow$]%
                   {\boldmath$\textcolor{blue}\downarrow$\hfil}%
                    \typeout{marginpar: $\downarrow$}\ignorespaces}
\def\mlanick{\marginpar[\boldmath\hfil$\rightarrow$]%
                   {\boldmath$\textcolor{blue}\leftarrow $\hfil}%
                    \typeout{marginpar: $\leftrightarrow$}\ignorespaces}
\overfullrule 5pt
\oddsidemargin 15mm
\marginparwidth 29mm
}

\nmc{\Hhat}{\hat H}
\nmc{\hhat}{\hat h}
\nmc{\Phihat}{\hat \Phi}
\nmc{\phihat}{\hat \phi}
\nmc{\chihat}{\hat \chi}
\nmc{\etahat}{\hat \eta}
\nmc{\rhohat}{\hat \rho}
\nmc{\thetahat}{\hat \theta}
\nmc{\sihat}{\hat\sigma}
\nmc{\Phione}{\Phi_1}
\nmc{\Phitwo}{\Phi_2}
\nmc{\Zhat}{\hat Z}
\nmc{\Hahat}{\hat A_0}
\nmc{\Gzhat}{\hat G_0}
\nmc{\Xhat}{\hat X}
\nmc{\Yhat}{\hat Y}
\nmc{\Honehat}{\hat H_1}
\nmc{\Htwohat}{\hat H_2}

\nmc{\vone}{v_1}
\nmc{\vtwo}{v_2}
\nmc{\etaone}{\eta_1}
\nmc{\etatwo}{\eta_2}
\nmc{\etaplet}{\boldsymbol{\eta}}
\nmc{\chione}{\chi_1}
\nmc{\chitwo}{\chi_2}
\nmc{\chiplet}{\boldsymbol{\chi}}
\nmc{\Hone}{H_1}
\nmc{\Htwo}{H_2}
\nmc{\Hplet}{\mathbf{H}}
\nmc{\phionepm}{\phi_1^\pm}
\nmc{\phitwopm}{\phi_2^\pm}
\nmc{\Ha}{A_0}

\nmc{\tb}{t_\be}
\nmc{\ca}{c_\al}
\nmc{\catwo}{c^2_\al}
\nmc{\satwo}{s^2_\al}
\nmc{\sa}{s_\al}
\nmc{\stwoa}{s_\al}
\nmc{\cbe}{c_\be}
\nmc{\ctwobe}{c_{2\be}}
\nmc{\cbetwo}{c^2_\be}
\nmc{\sbe}{s_\be}
\nmc{\sbetwo}{s^2_\be}
\nmc{\cab}{c_{\al\be}}

\nmc{\dth}{\delta t_{H}}
\nmc{\dthhat}{\delta t_{\hat{H}}}
\nmc{\Th}{T^{H}}
\nmc{\Thhat}{T^{\Hhat}}
\nmc{\THone}{{T}^{H_1}}
\nmc{\THtwo}{{T}^{H_2}}
\nmc{\dtHone}{{\delta t}_{H_1}}
\nmc{\dtHtwo}{{\delta t}_{H_2}}
\nmc{\dtHhatone}{{\delta t}_{\hat H_1}}
\nmc{\dtHhattwo}{{\delta t}_{\hat H_2}}
\nmc{\MHone}{{M}_\PHone}
\nmc{\MHtwo}{{M}_\PHtwo}
\nmc{\MHa}{{M}_{\PHa}}
\nmc{\MHpm}{{M}_{\text{H}^\pm}}
\nmc{\MHsone}{{M}^2_\PHone}
\nmc{\MHstwo}{{M}^2_\PHtwo}
\nmc{\MHsa}{{M}^2_{\PHa}}
\nmc{\MHspm}{{M}^2_{\PHpm}}

\newcommand{\THDM}{THDM\xspace}

\newcommand{\MSbar}{\ensuremath{\overline{\text{MS}}}\xspace}
\newcommand{\TS}{\mathrm{TS}}
\newcommand{\FJTS}{\mathrm{FJTS}}
\newcommand{\PRTS}{\mathrm{PRTS}}
\newcommand{\GIVS}{\mathrm{GIVS}}

\newcommand{\MSbarFJTS}{{\MSbar}(\text{FJTS})\xspace}

\newcommand{\PhiM}{\mathbf{\Phi}}
\nmc{\PhioneM}{\mathbf{\Phi}_1}
\nmc{\PhitwoM}{\mathbf{\Phi}_2}

\nmc{\vshift}{\bar{v}}

\title{\Large Electroweak renormalization based on gauge-invariant vacuum
expectation values 
of non-linear Higgs representations: 2.~extended Higgs sectors}
\subheader{\today}

\author{Stefan Dittmaier$^1$,}
\affiliation{%
        $^1$Albert-Ludwigs-Universit\"at Freiburg, %
        Physikalisches Institut, %
        79104 Freiburg, %
        Germany%
}

\author{Heidi Rzehak$^2$}
\affiliation{%
	$^2$University of T\"ubingen, %
	Institute for Theoretical Physics, %
	72076 T\"ubingen, %
	Germany%
}

\emailAdd{stefan.dittmaier@physik.uni-freiburg.de,
heidi.rzehak@itp.uni-tuebingen.de}

\abstract{A recently proposed scheme for a gauge-invariant treatment of
tadpole corrections in spontaneously broken gauge 
theories---called {\it Gauge-Invariant Vacuum expectation value 
Scheme (GIVS)}---is
applied to a singlet Higgs extension of the Standard Model and
to the Two-Higgs Doublet Model.
In contrast to previously used tadpole schemes, the GIVS unifies the gauge-invariance
property with perturbative stability. For the Standard Model this was 
demonstrated for the conversion between on-shell and $\MSbar$ renormalized masses,
where the GIVS leads to very moderate, gauge-independent electroweak corrections.
In models with extended scalar sectors, issues with tadpole renormalization
exist if Higgs mixing angles are renormalized with $\MSbar$ conditions, which is the
major subject of this article.
In detail, we first formulate non-linear representations of the extended scalar sectors,
which is an interesting subject in its own right.
Then we formulate the GIVS which employs these non-linear representations 
in the calculation of the tadpole renormalization constants, 
while actual
higher-order calculations in the GIVS proceed in linear representations as usual.
Finally,
for the considered models we discuss the next-to-leading-order 
(electroweak and QCD) corrections
to the decay processes $\Ph/\PH\to\PW\PW/\PZ\PZ\to4\,$fermions of the 
CP-even neutral Higgs bosons h and H
using $\MSbar$-renormalized Higgs mixing angles with the GIVS and previously used
tadpole treatments.}

\begin{document} 

\mbox{}\hfill FR-PHENO-2022-05

\maketitle
\flushbottom

\section{Introduction}
\label{se:Introduction}

Current particle physics can be, very globally, characterized by two facts:
On the one hand, the Standard Model (SM) of particle
physics describes almost all collider data very well;
on the other hand, the SM is not able to explain some
well established phenomena like Dark Matter and the matter--antimatter asymmetry in the
Universe, and it is not yet clear which is the right 
generalization of the SM to accommodate neutrino masses.
Rather independent of theoretical expectations or speculations,
the Higgs sector of the SM either is modified or extended in more comprehensive
models or at least provides a portal to some new sector.
For this reason,
experimental analyses of the Higgs boson and its properties as well as the search
for further Higgs bosons remain a cornerstone in the LHC physics programme and beyond.

On the theory side, this programme implies an immense effort for delivering
predictions of adequate precision both in the Standard Model (SM) of particle
physics and its most prominent extensions
(see, e.g., \citeres{LHCHiggsCrossSectionWorkingGroup:2011wcg,%
Dittmaier:2012vm,LHCHiggsCrossSectionWorkingGroup:2013rie,%
LHCHiggsCrossSectionWorkingGroup:2016ypw,Amoroso:2020lgh,Heinrich:2020ybq}).
In this context, the calculation of QCD and electroweak (EW) radiative corrections
plays a central role, which in turn involves the issue of renormalization and the choice
of appropriate input parameters (see, e.g., the review~\cite{Denner:2019vbn} for details 
and original references).
The choice of appropriate renormalization and input-parameter 
schemes is particularly important and delicate in SM extensions. 
New-physics parameters are often defined via $\MSbar$ renormalization conditions.
On the one hand, this choice is made for simplicity reason; 
on the other hand, it is also a natural choice if the considered model can be treated 
as an effective low-energy model the parameters of which
are determined via the matching to the complete model at a high-energy scale 
and the evolution of the parameters down to the low-energy scale via renormalization 
group equations, typically given in the $\MSbar$ scheme. 
However, choosing $\MSbar$ renormalization conditions leads
to issues in EW corrections to mass parameters and mixing angles.
These issues are related to
the precise higher-order definition of vacuum expectation values (vevs) of
Higgs fields, a subject 
that in turn is connected to
some scheme choice for treating the tadpole diagrams,
which have only one external Higgs-boson leg.
In the SM, this problem 
shows up as large EW corrections or gauge dependences
in the conversion between $\MSbar$-renormalized and
on-shell(OS)-renormalized masses. 
Recall that predictions entirely based on OS renormalization do not depend
on the tadpole treatment.
In SM extensions, similar issues appear in the renormalization of Higgs mixing angles
and severely limit the use of $\MSbar$ conditions,
as discussed for singlet Higgs extensions
and Two-Higgs Doublet Models (THDMs) in the literature in 
detail~\cite{Krause:2016oke,Denner:2016etu,Altenkamp:2017ldc,Altenkamp:2018bcs,Denner:2018opp}.

Perturbative calculations become significantly more transparent if tadpole
contributions are avoided by an appropriate introduction of tadpole counterterms
$\delta {\cal L}_{\delta t}=\delta t\, h$ in the Lagrangian for each Higgs field 
$\phi(x)=v+h(x)$ that might acquire a non-vanishing vev~$v$.
Adjusting $\delta t$ in such a way that the 
vev of $h$ vanishes, $\langle0|h(x)|0\rangle=0$, redistributes the 
{\it explicit} tadpole terms from loop diagrams to counterterms of
other couplings as {\it implicit} tadpole contributions.
In the last decades mostly two different tadpoles schemes for introducing 
$\delta {\cal L}_{\delta t}$ have been used.
One possibility, called {\it Parameter Renormalized Tadpole Scheme (PRTS)} in the following, 
is to include $\delta t$ in the parameter
renormalization transformation which expresses bare parameters in terms of renormalized parameters,
as, e.g., done in \citeres{Bohm:1986rj,Denner:1991kt}. 
The renormalized vev parameter~$v$ is chosen as the value of $\phi$ in the 
minimum of the renormalized (corrected) 
effective Higgs potential, which absorbs potentially large
corrections to the vev into renormalized parameters.
These corrections are induced by tadpole loops which are gauge dependent in the usual
loop calculation machinery.
Unfortunately this in general results in a
gauge-dependent parametrization of predicted 
observables if $\MSbar$ masses or $\MSbar$ Higgs mixing
angles are used (see discussions in \citeres{Krause:2016oke,Denner:2016etu,Denner:2018opp});
for OS-renormalized mass 
or mixing parameters such gauge dependences do not arise.

Alternatively to the PRTS, a tadpole counterterm can be introduced by shifting the field
$h(x)$ according to $h(x)\to h(x)+\Delta v$ in the bare Lagrangian
with a properly adjusted constant $\Delta v$~\cite{Fleischer:1980ub}.
In the following, we refer to this scheme
as {\it Fleischer--Jegerlehner Tadpole Scheme (FJTS)}%
\footnote{This scheme is equivalent to the $\beta_t$ scheme of
\citere{Actis:2006ra}.}.
Note that the FJTS produces the same result as if
including all explicit tadpole diagrams wherever they appear in a calculation,
because the field shift 
implements an unobservable reparametrization of the
path integral over~$h(x)$ employed in the quantization.
The benefit of this procedure is that no gauge dependences between bare parameters
are introduced, resulting in a 
gauge-independent parametrization of observables
in terms of renormalized parameters.
The downside of this approach is the appearance of potentially large
EW corrections in predictions with $\MSbar$ mass parameters, resulting from the fact
that the original Higgs field $\phi(x)$ is not expanded around the ``true''
(corrected) minimum of the 
effective Higgs potential, as e.g.\ discussed for
$\MSbar$ masses in the SM in 
\citeres{Jegerlehner:2012kn,Kniehl:2015nwa,Kataev:2022dua}.
{More aspects of the technical and conceptual differences in the PRTS and FJTS 
treatments are for instance discussed in 
\citeres{Krause:2016oke,Denner:2016etu,Denner:2019vbn,Dudenas:2020ggt,Dittmaier:2022maf}.}

In \citere{Dittmaier:2022maf}, we have proposed a hybrid scheme
called {\it Gauge-Invariant Vacuum expectation value Scheme (GIVS)}
unifying the good features of the PRTS and FJTS.%
\footnote{{After completing this work, we became aware of \citere{Dudenas:2020ggt}
where the impact of gauge-fixing on tadpole renormalization is discussed in detail.
In their remark~3 of Sect.~D the authors of \citere{Dudenas:2020ggt} suggest a tadpole
scheme that seems to be equivalent to the GIVS in the SM at the one-loop level.
As in the GIVS, the gauge-dependent part of the tadpole function $T^h$ is treated as in
the FJTS, and the remaining contribution according to the PRTS prescription.
The gauge-dependent part of the tadpole function is, however, determined by the difference
of $T^h$ in $R_\xi$~gauge to $T^h$ in the Landau gauge $(\xi=0)$, in which $R_\xi$~gauge-fixing
does not break global gauge symmetry.}}
In the GIVS Higgs fields are expanded about
the ``true'' (corrected) minimum of the 
effective Higgs potential, so that no additional
corrections arise from correcting the expansion point. 
Gauge dependences in tadpole contributions are completely avoided by using non-linear
representations of Higgs fields~\cite{Lee:1972yfa,Grosse-Knetter:1992tbp}.
In these non-linear representations, 
tadpole renormalization constants are gauge independent, because
CP-even neutral components of
Higgs multiplets and, thus, Higgs vacuum expectation values are
gauge invariant. Moreover, Higgs potentials are completely free of 
Goldstone-boson fields.
If a higher-order calculation is performed in this non-linear Higgs representation,
in the GIVS no further effort in the tadpole renormalization is needed, i.e.\ there
the GIVS proceeds as the PRTS of a linear representation.
Since, however, actual higher-order calculations for decay or scattering processes
are more conveniently performed using linear Higgs representations,
in the latter a second type of tadpole renormalization constant is introduced
by field shifts
as in the FJTS to guarantee a full compensation of explicit tadpole diagrams,
rendering the GIVS a hybrid scheme.
The two types of tadpole counterterms can be directly read from PRTS and FJTS
Feynman rules, as e.g.\ given in App.~A of \citere{Denner:2019vbn} for the SM.
As a first application of the GIVS, in \citere{Dittmaier:2022maf}
we have evaluated the differences between
OS- and $\MSbar$-renormalized masses of SM particles, demonstrating
the perturbative stability of the mass conversion in the GIVS, similar to the
PRTS, but without gauge dependences.

In this paper,
we formulate the GIVS for a Higgs Singlet Extension of the SM 
(SESM) inspired by \citeres{Schabinger:2005ei,Patt:2006fw,Bowen:2007ia}
and the THDM~\cite{Gunion:2002zf,Branco:2011iw} 
and investigate the perturbative
stability of radiative corrections to the phenomenologically
important decays of CP-even Higgs bosons into four-fermion final states,
$\Ph/\PH\to\PW\PW/\PZ\PZ\to4f$, when the Higgs mixing angles are
renormalized with $\MSbar$ conditions.
Our study completes the discussions of next-to-leading-order (NLO)
predictions for these decays in the SESM and THDM which
started in \citeres{Altenkamp:2018bcs} and \cite{Altenkamp:2017ldc,Altenkamp:2017kxk}, respectively, with 
$\MSbar$ renormalization schemes for the Higgs mixing angles
and were extended to OS and symmetry-inspired renormalization schemes
in \citere{Denner:2018opp}.
To this end, we have extended the Monte Carlo program
\Prophecy~3.0~\cite{Bredenstein:2006rh,Bredenstein:2006ha,Denner:2019fcr} 
by adding $\MSbar$ schemes based on the
GIVS tadpole treatment for the Higgs mixing angles in the SESM and THDM.
As a byproduct of this work, we had to construct an appropriate
non-linear parametrization of the Higgs sector of the THDM, which 
is more involved than in the SM and the SESM and a subject that might
be of interest in its own right.

The article is organized as follows:
In \refse{se:Representations} we describe the non-linear Higgs 
representations of the SESM and THDM in detail.
The formulation of the GIVS for those models, the issue of tadpoles in the
$\MSbar$ renormalization of the Higgs mixing angle, and the 
phenomenological applications to the four-body Higgs decays
are presented in \refse{se:GIVscheme}.
Our conclusions are given in \refse{se:Conclusions},
and the appendix provides some further information on the 
calculational setup for the evaluation of the 
Higgs-boson decays in the SESM and THDM.

\section{Linear and non-linear Higgs representations}
\label{se:Representations}

In this section, we introduce the linear and non-linear Higgs representations for 
the SESM and the THDM. All parameters and fields are considered as ``bare'' in this section,
i.e.\ the renormalization transformation for introducing renormalized quantities and renormalization
constants, including the choice of the tadpole scheme,
will be the next step after this section.
For the definition of 
field-theoretical quantities we consistently follow the
notation and conventions of \citere{Denner:2019vbn}, and
the transition to the non-linear Higgs representation uses 
\citeres{Lee:1972yfa,Grosse-Knetter:1992tbp,Dittmaier:1995cr,Dittmaier:1995ee}
as guideline.

\subsection{Higgs singlet extension of the Standard Model}
\label{se:SESM1}

\subsubsection{Kinetic Higgs Lagrangian}

Higgs Singlet Extensions of the SM have been introduced in the literature in
different variants, see, e.g., \citeres{Schabinger:2005ei,Patt:2006fw,Bowen:2007ia}.
Here, we consider a simple variant with a complex Higgs doublet $\Phi$
and charge conjugate $\Phi^{\mathrm{c}}$
with the same quantum numbers as in the SM, 
\begin{align}\label{eq:PhiSESM}
\Phi = \left( \begin{array}{c}
\phi^{+} \\ \frac{1}{\sqrt2}(v_2 +\eta_2 +\ri\chi)
\end{array} \right),
\qquad
\Phi^{\mathrm{c}} = \left( \begin{array}{c}
\frac{1}{\sqrt2}(v_2 +\eta_2 -\ri\chi) \\ -\phi^{-}
\end{array} \right),
\end{align}
with the complex would-be Goldstone-boson fields $\phi^+$, $\phi^-=(\phi^+)^*$,
the real would-be Goldstone-boson field $\chi$, the
physical Higgs field $\eta_2$, and the constant $v_2$
parametrizing the vev of the Higgs doublet.
Apart from the doublet $\Phi$, we introduce 
a real Higgs singlet field 
\begin{align}
\si =  v_1+\etaone
\end{align}
with vev $v_1$, as also considered in 
\citeres{Kanemura:2015fra,Bojarski:2015kra,Denner:2017vms,Altenkamp:2018bcs,Denner:2018opp}.
Our conventions closely follow \citeres{Denner:2017vms,Altenkamp:2018bcs,Denner:2018opp}.
As in the treatment of the SM in \citere{Dittmaier:2022maf}, we
switch to the ($2\times2$) matrix notation for the Higgs doublet
\begin{align}\label{eq:PhilinSM}
\PhiM \equiv \left( \Phi^{\mathrm{c}},\Phi \right)
= \frac{1}{\sqrt{2}}\bigl[(v_2 + \eta_2) \id + 2\ri \bfphi\bigr], \qquad
\bfphi \equiv \frac{\phi_{j} \sigma_j}{2} = \frac{\vec\phi\cdot\vec\sigma}{2},
\end{align}
where $\id$ is the $2\times2$ unit matrix,
$\sigma_j$ $(j=1,2,3)$ denote the Pauli matrices, and
$\phi_j$ are the three real
would-be Goldstone-boson degrees of freedom in a more generic notation.
As in \citere{Dittmaier:2022maf}, we use the
summation convention over the Goldstone index $j$,
a vector-like notation
$\vec\phi=(\phi_1,\phi_2,\phi_3)^\rT$, $\vec\sigma=(\sigma_1,\sigma_2,\sigma_3)^\rT$, etc., and
boldface characters like $\bfphi$ for matrix structures.
The new field components $\phi_j$ can be identified according to
\begin{align}
\phi^\pm = (\phi_2 \pm \ri \phi_1)/\sqrt{2}, \qquad \chi=-\phi_3.
\end{align}
Since the additional Higgs singlet is invariant under SU(2)$_{\rw}\times\mathrm{U(1)}_Y$ 
transformations by definition, only the parametrization of $\PhiM$ changes in the transition
to the non-linear representation,
\begin{align}\label{eq:PhinonlinSESM}
\si =  v_1+h_1, \qquad
\PhiM = \frac{1}{\sqrt{2}} (v_2 + h_2) U(\bfzeta),
\end{align}
with the physical Higgs fields $h_1\equiv\eta_1$ and $h_2$,
and the same unitary Goldstone-boson matrix $U(\bfzeta)$ as in the SM~\cite{Dittmaier:2022maf},
\begin{align}\label{eq:Uzeta}
U(\bfzeta) = \exp \left\{2\ri \bfzeta/v\right\}, \qquad
\bfzeta \equiv \frac{\zeta_j\sigma_j}{2},
\end{align}
with the real would-be Goldstone-boson components
$\vec\zeta=(\zeta_1,\zeta_2,\zeta_3)^\rT$ and $v=v_2$.
The relation between the fields in the linear and non-linear transformation can be obtained 
in full analogy to the SM~\cite{Grosse-Knetter:1992tbp,Dittmaier:2022maf},
\begin{align}\label{eq:fieldrelationSM}
 \eta_2 ={}& c_\zeta (v + h_2) - v
=  h_2 - \frac{\zeta^2}{2v} \left(1+\frac{h_2}{v} \right) + {\cal O}(\zeta^4),
\nonumber\\
 \vec \phi ={}& \frac{s_\zeta}{\zeta}(v + h_2) \vec\zeta
= \left(1 + \frac{h_2}{v}\right)\vec\zeta + {\cal O}(\zeta^3),
\end{align}
where
\begin{align}
s_\zeta \equiv \sin\left(\frac{\zeta}{v}\right), \qquad
c_\zeta \equiv \cos\left(\frac{\zeta}{v}\right), \qquad
\zeta \equiv |\vec\zeta\,| = \bigl(\vec\zeta^{\,2}\bigr)^{1/2}.
\end{align}
The gauge transformation of $\PhiM$ in the matrix representation is 
given by~\cite{Grosse-Knetter:1992tbp}
\begin{align}\label{eq:Phigaugetrafo}
  \PhiM &\rightarrow S(\bftheta)\,  \PhiM\, S_Y(\theta_Y)
\end{align}
with the unitary transformation matrices
\begin{align}\label{eq:StrafoM}
  S(\bftheta) =\exp\{\ri g_2 \bftheta\}, \qquad
  S_Y(\theta_Y) =\exp\biggl\{\frac{\ri}{2} g_1\theta_Y \sigma_3\biggr\}, \qquad
\bftheta \equiv \frac{\theta_j\sigma_j}{2},
\end{align}
depending on the group parameters $\bftheta=\vec\theta\cdot\vec\sigma/2$ and $\theta_Y$ for the 
SU(2)$_{\rw}$ and the  U(1)$_Y$ transformations, respectively.
Taking into account that $\sigma$ is gauge invariant, the Higgs and would-be
Goldstone fields of the non-linear representation transform according to
\begin{align}
h_1 \rightarrow h_1, \qquad
h_2 \rightarrow h_2, \qquad
U(\bfzeta) \rightarrow S(\bftheta)\,U(\bfzeta) \,S_Y(\theta_Y),
\label{eq:hUtrafo}
\end{align}
i.e.\ Higgs fields $h_1$ and $h_2$ are gauge invariant, 
while the would-be Goldstone fields $\zeta_j$ change
under gauge transformations.

The kinetic terms of the Higgs Lagrangian just adds a term for the field~$h_1$ to the SM 
contribution,
\begin{align}
{\cal L}_{\text{H,kin}} &{}=
\frac{1}{2}(\partial \sigma)^2+
\frac{1}{2}\text{tr}\left[(D_\mu \PhiM)^\dagger(D^\mu \PhiM)\right],
\end{align}
where 
\begin{align}\label{eq:Dmu}
 D^\mu \PhiM = \partial^\mu \PhiM
-  \ri g_2  {\bf W}^\mu  \PhiM - \ri g_1 \PhiM B^\mu \frac{\sigma_3}{2},\qquad
{\bf W}^\mu \equiv \frac{W_j^\mu\sigma_j}{2},
\end{align}
is the covariant derivative in matrix notation, 
with the matrix-valued SU(2)$_{\rw}$ gauge field ${\bf W}^\mu$ and
the U(1)$_Y$ gauge field $B^\mu$ and respective gauge couplings $g_2$ and $g_1$.
Inserting the field parametrizations \refeq{eq:PhinonlinSESM}, this leads to
\begin{align}
{\cal L}_{\text{H,kin}} &{}=
\frac{1}{2}(\partial h_1)^2+
\frac{1}{2}(\partial h_2)^2 + \frac{g_2^2}{8} (v_2+h_2)^2 \,
{\vec C}^{(\ru)}_\mu \cdot{\vec C}^{(\ru),\mu},
\end{align}
with ${\vec C}^{(\ru)}_\mu$ as given for the SM,
\begin{align}
{\bf C}^{(\ru)}_\mu ={}&
{\bf W}^{(\ru)}_\mu + \frac{g_1}{g_2} B_\mu \frac{\sigma_3}{2}
= \frac{{\vec C}^{(\ru)}_\mu\cdot\vec \sigma}{2},
\label{eq:Cu}
\\
{\bf W}^{(\ru)}_\mu ={}& U(\bfzeta)^\dagger \,{\bf W}_\mu \, U(\bfzeta)
+\frac{\ri}{g_2} U(\bfzeta)^\dagger \, \partial_\mu U(\bfzeta).
\end{align}
More explicit expressions for ${\bf C}^{(\ru)}_\mu$
and ${\vec C}^{(\ru)}_\mu \cdot{\vec C}^{(\ru),\mu}$
can be found in Sect.~2.1 of \citere{Dittmaier:2022maf}.
The gauge-fixing part of the Lagrangian in the non-linear representation, which is the same as 
in the SM, can be found there as well.

\subsubsection{Higgs potential and tadpoles}

In the considered variant of the SESM,
the interactions among the Higgs fields are ruled by the following
Higgs potential,
\begin{align}
  V ={}& -\mu_{2}^2 \Phi^\dagger \Phi  - \mu_{1}^2 \si^2
  +\frac{\lambda_{2}}{4} \big(\Phi^\dagger \Phi\big)^2
  +\lambda_{1} \si^4
  +\lambda_{12} \Phi^\dagger \Phi\,\si^2
  \label{eq:VSESM}
\end{align}
with real coupling constants $\mu_1^2$, $\mu_2^2$, 
$\lambda_1$, $\lambda_2$, $\lambda_{12}$. 
The potential $V$ has the most general form that supports
renormalizability and a $\mathbb{Z}_2$ symmetry under $\si\to-\si$.
As in the SM, $V$ involves would-be Goldstone-boson fields in the
linear Higgs representation, but not in the non-linear representation,
where it is given by
\begin{align}
V ={}& -\frac{\mu_{2}^2}{2} \mathrm{tr}\bigl[\PhiM^\dagger \PhiM\bigr]  - \mu_{1}^2 \si^2
  +\frac{\lambda_{2}}{16} \big(\mathrm{tr}\bigl[\PhiM^\dagger \PhiM\bigr]\big)^2
  +\lambda_{1} \si^4
  +\frac{\lambda_{12}}{2} \mathrm{tr}\bigl[\PhiM^\dagger \PhiM\bigr]\,\si^2
\nonumber\\
={}& -\frac{\mu_{2}^2}{2} (v_2+h_2)^2  - \mu_{1}^2 (v_1+h_1)^2
  +\frac{\lambda_{2}}{16} (v_2+h_2)^4
  +\lambda_{1} (v_1+h_1)^4
\nonumber\\
{}& {}
  +\frac{\lambda_{12}}{2} (v_1+h_1)^2 \,(v_2+h_2)^2.
\end{align}
Both in the linear and non-linear representations, the original 
Higgs-boson fields $(\eta_1,\eta_2)$ and $(h_1,h_2)$, respectively, 
can be rotated
into a field basis of $h$ and $H$, which correspond to mass eigenstates,
\begin{align}\label{eq:RalphaSESM}
\begin{pmatrix} \eta_{1} \\ \eta_{2} \end{pmatrix}
= R(\alpha) \begin{pmatrix} H \\ h \end{pmatrix},
\qquad
\begin{pmatrix} h_{1} \\ h_{2} \end{pmatrix}
= R(\alpha) \begin{pmatrix} H \\ h \end{pmatrix}_{\!\mathrm{nl}},
\qquad
R(\alpha) = \begin{pmatrix} \ca & -\sa\\ \sa & ~\ca \end{pmatrix},
\end{align}
where $\alpha$ is a real-valued mixing angle 
which is determined by the parameters of the Higgs potential.%
\footnote{To avoid confusion between the angle $\alpha$ and the electromagnetic coupling,
we denote the fine-structure constant $\alpha_{\mathrm{em}}$ in the following.}
In the following, we will often use the shorthands 
$s_\alpha\equiv\sin\alpha$, $c_\alpha\equiv\cos\alpha$, 
$t_\alpha\equiv\tan\alpha$ 
and analogously for other angles.
The fact that the field bases $(h,H)$ in the two representations are not the
same is indicated by the subscript ``nl'' in Eq.~\refeq{eq:RalphaSESM}
indicating the non-linear representation.
By convention, $H$ corresponds to the heavier and $h$
to the lighter Higgs state.%
\footnote{In \citere{Denner:2018opp}, the two physical Higgs bosons
were called $\PH_{1/2}$ which corresponds to
$H\equiv H_1$ and $h\equiv H_2$ compared to our conventions here.}
Since $\eta_n=h_n+{\cal O}(\zeta^2)$ ($n=1,2$) holds 
up to terms of higher powers in the would-be Goldstone fields $\zeta_j$,
the mixing angle $\alpha$ is the same in the two representations
(at least in lowest perturbative order).

The tadpole constants for the non-linear and linear Higgs representations
are easily calculated to
\begin{align}
   \Gamma^{h_n}_{\mathrm{nl}} = T^{h_n}_{\mathrm{nl}}
  ={}& \frac{1}{16 \pi^2 v_2}\biggl\{ \lambda_n^h v_2^2 A_0(\Mh^2) + \lambda_n^H v_2^2 A_0(\MH^2) 
- 4 \delta_{n2} \sum_f N^{\mathrm{c}}_f m_f^2 A_0(m_f^2) 
\nonumber \\& 
\quad {} + \delta_{n2} \MZ^2\left[ 3 A_0(\MZ^2) -2\MZ^2\right] 
+ 2\delta_{n2} \MW^2\left[ 3 A_0(\MW^2) -2\MW^2\right] 
\biggr\}
\label{eq:TnlSESM}
\end{align}
and
\begin{align}
\Gamma^{\eta_n} = T^{\eta_n} = T^{h_n}_{\mathrm{nl}} + \lambda_n^{G} v_2^2 \Delta v_\xi,
\label{eq:TSESM}
\end{align}
respectively,
where the gauge-dependent constant $\Delta v_\xi$ is defined as in the SM~\cite{Dittmaier:2022maf},
\begin{align}
\Delta v_\xi = \frac{1}{16 \pi^2 v}\left\{
\frac{1}{2} A_0(\xi_Z \MZ^2) +  A_0(\xi_W \MW^2)\right\},
\label{eq:Deltavxi}
\end{align}
with $\xi_Z$ and $\xi_W$ representing the arbitrary $R_\xi$ gauge parameters
of the Z- and W-boson fields.
As mentioned before, in the SESM we have $v=v_2$.
In Eq.~\refeq{eq:TnlSESM} $\delta_{n2}$ represents the usual Kronecker $\delta$, and 
the scalar coupling factors $\lambda_n^h$, $\lambda_n^H$, 
and $\lambda_n^{G}$ with $n=1,2$ are
\begin{align}
  \lambda_1^h &= 
\frac{12s_\alpha^2 v_1\lambda_1}{v_2}
+\left( \frac{c_\alpha^2 v_1}{v_2} - 2 c_\alpha  s_\alpha\right)\lambda_{12}, &
  \lambda_2^h &= 
 \frac{3 c_\alpha^2 \lambda_2}{4}
-\left(\frac{2 c_\alpha  s_\alpha v_1}{v_2}  - s_\alpha^2\right)\lambda_{12} ,
\nonumber\\
  \lambda_1^H &= 
 \frac{12 c_\alpha^2 v_1\lambda_1}{v_2} 
+ \left(\frac{ s_\alpha^2 v_1}{v_2}
+ 2 c_\alpha s_\alpha \right)\lambda_{12}, &
  \lambda_2^H &= 
 \frac{3s_\alpha^2 \lambda_2}{4} 
+\left( \frac{2 c_\alpha s_\alpha v_1}{v_2} 
+ c_\alpha^2\right) \lambda_{12} , \nonumber\\
\lambda_1^{G} &= \frac{2v_1\lambda_{12}}{v_2}, &
\lambda_2^{G} &= \frac{\lambda_2}{2}.
\label{eq:tadcoefsSESM}
\end{align}
In Eqs.~\refeq{eq:TnlSESM} and \refeq{eq:Deltavxi}, we made use of the scalar
one-point one-loop integral in $D=4-2\eps$ dimensions,
\begin{align}\label{eq:A0}
A_0(m^2) =
\displaystyle\frac{(2\pi\mu)^{4-D}}{\ri\pi^{2}}\int \rd^{D}q\, \frac{1}{q^2-m^{2}+\ri0}
= m^2\left[ \Delta + \ln\left(\frac{\mu^2}{m^2}\right) + 1 \right] + {\cal O}(\eps)
\end{align}
with the arbitrary reference mass $\mu $ and the standard UV divergence
\beq\label{eq:Delta}
\Delta = \frac{2}{4-D} + \ln 4\pi - \gamma_{\mathrm{E}},
\eeq
in which $\gamma_{\mathrm{E}}$ denotes the Euler--Mascheroni constant.

Obviously, the tadpole constants $T^{h_n}_{\mathrm{nl}}$ of the non-linear representation
are gauge independent, while the $T^{\eta_n}$ of the linear representation involve
gauge-dependent contributions proportional to $\Delta v_\xi$. 
The gauge independence of $T^{h_n}_{\mathrm{nl}}$ holds to all perturbative orders
by virtue of the Nielsen identities~\cite{Kluberg-Stern:1974iel,Nielsen:1975fs,Gambino:1999ai} 
as discussed in Sect.~2.2 of \citere{Dittmaier:2022maf}
for the SM.

\subsection{Two-Higgs-Doublet Model}
\label{se:THDM1}

\subsubsection{Kinetic Higgs Lagrangian}

Two-Higgs Doublet Models (THDMs)~\cite{Gunion:2002zf,Branco:2011iw} 
contain two complex Higgs doublets $\Phi_n$ $(n=1,2)$, both
with weak hypercharge $Y_{\rw,\Phi_n}=+1$. 
As also done in previous literature on the renormalization of the 
THDM~\cite{Kanemura:2004mg,Lopez-Val:2009xtx,Kanemura:2014dja,%
Krause:2016oke,Denner:2016etu,Denner:2017vms,Altenkamp:2017ldc,Denner:2018opp,%
Krause:2018wmo,Krause:2019qwe}, 
we only consider CP-conserving Higgs potentials with additional
discrete symmetries that allow only for one type of Yukawa coupling per Higgs
doublet and fermion type to avoid flavour-changing neutral currents at tree level.
In the two-component notation, the doublets $\Phi_n$ and their charge conjugates
$\Phi_n^{\mathrm{c}}=i\sigma_2\Phi^*_n$
are parametrized according to
\begin{align}
\Phi_n = \begin{pmatrix}
\phi_n^+ \\  \frac{1}{\sqrt{2}}(v_n+\eta_n+\ri\chi_n)
\end{pmatrix}, \qquad
\Phi^{\mathrm{c}}_n = \left( \begin{array}{c}
\frac{1}{\sqrt2}(v_n +\eta_n -\ri\chi_n) \\ -\phi_n^{-}
\end{array} \right),
\label{eq:THDMPhi}
\end{align}
where $\eta_n$ are the CP-even neutral Higgs fields that mix to the two
physical CP-even neutral Higgs bosons h and H and from which the corresponding
vevs $v_n$ have been split off.
The remaining component fields $\phi_n^+$, $\phi_n^-=(\phi_n^+)^*$, and $\chi_n$
are superpositions of would-be Goldstone-bosons fields and 
the fields of the physical Higgs bosons,
as described further below.
Analogously to Eq.~\eqref{eq:PhilinSM} in the SESM,
in the linear parametrization  
the two Higgs doublets of the THDM can be written as $2 \times 2$ matrices,
\begin{align}
\label{eq:Philin}
\PhiM_n \equiv \left( \Phi^{\mathrm{c}}_n,\Phi_n \right)
= \frac{1}{\sqrt{2}}\bigl[(v_n + \eta_n) \id + 2\ri \bfphi_{n}\bigr], \qquad
\bfphi_n \equiv \frac{\phi_{nj} \sigma_j}{2},
\end{align}
with the same fields $\eta_n$ as in Eq.~\refeq{eq:THDMPhi}, but
the components $\phi_n^\pm$ and $\chi_n$ are absorbed into the two
field matrices $\bfphi_{n}$ with the 
real component fields $\phi_{nj}$ ($n = 1, 2$; $j = 1,2,3$).
In analogy to the SM, the two sets of fields are related according to
\begin{align}
\phi^\pm_n = (\phi_{n2} \pm \ri \phi_{n1})/\sqrt{2}, \qquad \chi_n=-\phi_{n3}.
\end{align}

In the non-linear representation, we have to take care of the fact
that after splitting off the
Goldstone-boson matrix $U(\bfzeta)$ from $\PhiM_n$
with  $U(\bfzeta)$ given in Eq.~\eqref{eq:Uzeta},
apart from the physical CP-even neutral fields $h_n$ 
some dependence on three further real physical Higgs fields $\rho_j$ $(j=1,2,3)$ remains,
which correspond to one CP-odd neutral 
and two charged Higgs bosons.
We, thus, can make the following ansatz for the non-linearly parametrized
Higgs matrix fields,
\begin{align}\label{eq:Phinonlin}
\PhiM_n = U(\bfzeta) \, \PhiM_n^{(\ru)}, \qquad 
\PhiM_n^{(\ru)} = \frac{1}{\sqrt{2}} \, 
\bigl[(v_n + h_n) \id +\ri c_{{nj}}\sigma_j \rho_j \bigr],
\end{align}
where the coefficients $c_{nj}$ ($n = 1, 2$; $j = 1,2,3$) are real constants, 
which can be determined
upon requiring canonical field normalization. 
To this end, it is sufficient to take the terms in $\partial_\mu\PhiM_n$
that are linear in all scalar fields and to evaluate the (bilinear) 
terms in the kinetic Lagrangian of the Higgs fields for free propagation.
These kinetic terms are canonically normalized and do not mix the free scalar fields
$\zeta_j$, $h_n$, and $\rho_j$ if we set
\begin{align}\label{eq:Coeffs}
  c_{1j} = -\frac{v_2}{v} \equiv -\sin\beta, \qquad
  c_{2j} =  \frac{v_1}{v} \equiv  \cos\beta, \qquad
  v \equiv  \sqrt{v_1^2 + v_2^2}.
\end{align}
Since the constants $c_{{nj}}$ do not depend on $j$, we can write
\begin{align}\label{eq:Phinonlin2}
\PhiM_n^{(\ru)} = \frac{1}{\sqrt{2}} \, 
  \bigl[(v_n + h_n) \id
  +  2\ri c_{n} \bfrho \bigr], \qquad \bfrho \equiv \frac{\rho_j \sigma_j}{2}, \qquad
c_{n} \equiv c_{{nj}}.
\end{align}
The gauge-invariant trace of $\PhiM_m^\dagger\PhiM_n$ reads
\begin{align}
\mathrm{tr}\big[ \PhiM_m^\dagger\PhiM_n \big] =
\mathrm{tr}\big[ (\PhiM_m^{(\ru)})^\dagger\PhiM_n^{(\ru)} \big] =
(v_m+h_m)(v_n+h_n) +  c_{m} c_{n} \vec\rho^{\,2}.
\end{align}
The relations between the fields of the two representations are given by
\begin{align}
  \eta_n &= c_\zeta (v_n + h_n) 
- v_n -  s_\zeta c_n \,\frac{\vec\rho\cdot\vec\zeta}{\zeta},
\qquad
  \vec\phi_{n} = c_\zeta c_n \vec\rho 
   + s_\zeta \left[\left(v_n + h_n\right)\frac{\vec\zeta}{\zeta} 
   + c_n \frac{\vec\rho\times\vec\zeta}{\zeta}\,\right],
  \label{eq:etanphin}
\end{align}
where $\times$ denotes the usual cross product of 3-dimensional vectors.

Analogously to Eq.~\eqref{eq:Phigaugetrafo}, the Higgs doublets transform as
\begin{align}
  \PhiM_n &\rightarrow S(\bftheta) \, \PhiM_n \,S_Y(\theta_Y)
\end{align}
under SU(2)$_{\rw}\times\mathrm{U(1)}_Y$ transformations with $S$ and $S_Y$ given in 
Eq.~\eqref{eq:StrafoM}. Applying these transformations to Eq.~\eqref{eq:Phinonlin}, 
we can determine the gauge transformation of $\PhiM_n^{(\ru)}$ by isolating
the transformation of $U(\bfzeta)$ as described in \refeq{eq:hUtrafo},
\begin{align}
  \PhiM_n &\rightarrow \Bigl[S(\bftheta) U(\bfzeta) S_Y(\theta_Y)\Bigr]
\,\Bigl[S_Y(-\theta_Y) \PhiM_n^{(\ru)} \,S_Y(\theta_Y)\Bigr].
\end{align}
The resulting 
gauge transformation rules of the fields $h_n$, $\rho_j$, and $\zeta_j$
are given by
\begin{align}
  h_n &\rightarrow h_n, 
\\\nonumber
  \sigma_j \rho_j &\rightarrow S_Y(-\theta_Y) \sigma_j \rho_j S_Y(\theta_Y) 
= \biggl(\sum_{j=1,2}\sigma_j \rho_j\biggr) S_Y(2\theta_Y) +\sigma_3 \rho_3
\\
\Rightarrow\;
  \rho^\pm &{}=\frac{1}{\sqrt{2}}(\rho_2\pm\ri\rho_1) \rightarrow \exp\{\mp\ri g_1\theta_Y\}\rho^\pm, \quad
  \rho_3 \rightarrow \rho_3.
\end{align}
That means that the fields $h_n$ and $\rho_3$ are completely gauge invariant 
while $\rho_1$ and $\rho_2$ are only invariant under the SU(2)$_\rw$ gauge transformations. 
The fields $\rho_1,$ $\rho_2$ mix into each other 
under the U(1)$_Y$ hypercharges transformations, but $\rho^\pm$ correspond to eigenstates
of weak hypercharge.
As SU(2)$_\rw$ singlets, the fields $\rho^\pm$, thus, have electric charge
$Q_{\rho^\pm} = Y_{\rw,\rho^\pm}/2=\pm1$ according to the Gell-Mann--Nishijima relation.
The 
vev parameters $v_n$ again are associated with the
gauge-invariant fields $h_n$.
The Goldstone-boson fields $\zeta_j$ transform under  SU(2)$_{\rw}\times\mathrm{U(1)}_Y$ 
exactly as in the SM.

Using the matrix representation \eqref{eq:Philin} for $\PhiM_n$, the Higgs kinetic terms
are given by
\begin{align}
  {\cal L}_{\text{H,kin}} ={}&
\frac{1}{2}\text{tr}\left[(D_\mu \PhiM_n)^\dagger(D^\mu \PhiM_n)\right], 
\nonumber\\
={}& 
\frac{1}{2}(\partial h_n)^2 + \frac{1}{2}(\partial\vec\rho\,)^2 
+ \frac{g_2^2}{8} \left[ (v_n+h_n)^2 + \vec\rho^{\,2} \,\right]
{\vec C}^{(\ru)}_\mu \cdot{\vec C}^{(\ru),\mu}
\nonumber\\
& {} +\frac{g_2}{2} \, {\vec C}^{(\ru)}_\mu \cdot \bigl[
c_n(\partial^\mu h_n) \vec\rho  - c_n h_n (\partial^\mu\vec\rho\,)
+(\vec\rho\times\partial^\mu\vec\rho\,) \bigr]
\nonumber\\
& {}+ \frac{g_1^2}{2} \,B^2\left(\rho_1^2+\rho_2^2\right)
 + g_1 B^\mu \bigl[ (\partial_\mu\rho_1)\rho_2 - (\partial_\mu\rho_2)\rho_1 \bigr]
\nonumber\\
& {} + \frac{g_1 g_2}{2} \, B_\mu \Bigl[
c_n h_n \left( \rho_1 C^{(\ru),\mu}_2 -\rho_2 C^{(\ru),\mu}_1 \right)
+\rho_3 \, \vec \rho \cdot {\vec C}^{(\ru),\mu}
- \vec \rho^{\;2}\, C^{(\ru),\mu}_3  \Bigr]
\end{align}
where a summation over $n=1,2$ is implicitly understood
and ${\vec C}^{(\ru)}_\mu$ is again as defined in Eq.~\refeq{eq:Cu}.
Up to quadratic order in $\zeta_j$, the Higgs kinetic Lagrangian reads
\begin{align}
%
%
  {\cal L}_{\text{H,kin}} ={}& 
\frac{1}{2}(\partial h_n)^2 
+ \frac{1}{2}(\partial\vec\rho\,)^2 
+ \frac{1}{2v^2} \left[ (v_n+h_n)^2 + \vec\rho^{\,2} \,\right]
\biggl\{
(\partial_\mu \vec \zeta\,)\cdot(\partial^\mu \vec \zeta \,)
+ \frac{g_2^2 v^2}{4}\, \vec C_\mu \cdot \vec C^\mu
\nonumber \\
& \hspace*{2em}
+g_1 g_2 B_\mu \left[ - W^\mu_3 \zeta^2
+  (\vec W^\mu \cdot \vec \zeta\,) \zeta_3 \right]
-g_2^2 v \,\vec C_\mu\cdot(\vec W^\mu \times \vec \zeta\,) 
\nonumber \\
& \hspace*{2em}
- g_2 v \,\vec C_\mu\cdot \partial^\mu \vec \zeta 
- g_2 (\vec C_\mu-2\vec W_\mu)\cdot (\vec \zeta \times \partial^\mu \vec \zeta\,)
\biggr\}
\nonumber\\
&  {} + \frac{g_2}{2}
\biggl\{\vec C_\mu - \frac{2}{g_2 v} \, \partial_\mu\vec\zeta
   -\frac{2}{v}\,\vec W_\mu\times\vec\zeta
- \frac{2}{g_2 v^2} \,\bigl( \vec\zeta\times\partial_\mu\vec\zeta\;\bigr)
- \frac{2}{v^2} \Bigl[\zeta^2\,\vec W_\mu
	- \bigl(\vec W_\mu\cdot\vec\zeta\;\bigr)\,\vec\zeta \, \Bigr]
\biggr\}
\nonumber\\
&  {} \qquad
\cdot
\bigl[c_n(\partial^\mu h_n) \vec\rho  -c_n h_n (\partial^\mu\vec\rho\,)
+(\vec\rho\times\partial^\mu\vec\rho\,) 
+g_1 B^\mu \rho_3 \vec\rho\,
\bigr]
\nonumber\\
  & {} + \frac{g_1^2}{2} \,B^2\left(\rho_1^2+\rho_2^2\right)
+ g_1 B^\mu \bigl[ (\partial_\mu\rho_1)\rho_2 - (\partial_\mu\rho_2)\rho_1 \bigr]
\nonumber\\
&  {}
+\frac{g_1 g_2}{2} \, B_\mu \biggl\{
  c_n h_n \biggl[ \left( \rho_1 C^\mu_2 - \rho_2 C^\mu_1\right) - \frac{2}{g_2 v} \,\left(\rho_1 \partial^\mu \zeta_2 -\rho_2 \partial^\mu \zeta_1\right)
\nonumber\\
&  {} \qquad\quad
    -\frac{2}{v}\,\bigl[(\vec \rho \cdot \vec \zeta\,) W^\mu_3 -
     (\vec \rho \cdot \vec W^\mu\,) \zeta_3\,\bigr]
 -\frac{2}{g_2 v^2}\,\bigl[(\vec \rho \cdot \partial^\mu \vec \zeta\,) \zeta_3 -
   (\vec \rho \cdot \vec \zeta \,) \partial^\mu \zeta_3\bigr]
\nonumber\\
&  {} \qquad \quad  
 +\frac{2}{v^2} \bigl[\bigl(\vec W^\mu \cdot \vec \zeta\;\bigr)\bigl(\rho_1 \zeta_2 -\rho_2 \zeta_1\,\bigr)
   - \zeta^2 \bigl(\rho_1  W^\mu_2 -\rho_2  W^\mu_1\,\bigr)
   \bigr]\biggr]
  \nonumber\\
& \hspace*{2em}
  -\vec \rho^{\;2} \biggl[
    C^\mu_3 - \frac{2}{g_2 v} \, \partial^\mu\zeta_3
   -\frac{2}{v}\,\left( W^\mu_1 \zeta_2 - W^\mu_2 \zeta_1\right)
- \frac{2}{g_2 v^2} \,\left( \zeta_1 \partial^\mu \zeta_2 -  \zeta_2 \partial^\mu \zeta_1\,\right)
\nonumber\\
&  {}\qquad\quad
- \frac{2}{v^2} \bigl[\zeta^2\,W^\mu_3
	- \bigl(\vec W^\mu \cdot \vec \zeta \;\bigr)\zeta_3 \bigr]
\biggr]\biggr\}
+ {\cal O}(\zeta^3).
\end{align}
The gauge-fixing part of the Lagrangian in the non-linear representation is the same as 
in the SM, see Sect.~2.1 of \citere{Dittmaier:2022maf}.

\subsubsection{Higgs potential and tadpoles}

In order to rule out flavour-changing neutral currents at tree level,
we assume the $\mathbb{Z}_2$ symmetry $\Phi_1\to-\Phi_1$ and
$\Phi_2\to\Phi_2$ that is only softly broken by the $m_{12}^2$ term
in the Higgs potential, which in terms of bare parameters and fields reads
\begin{align}
V={}&
m_{11}^2\Phi_{1}^{\dagger}\Phi_{1}+m_{22}^2\Phi_{2}^{\dagger}\Phi_{2}
-m_{12}^2\big(\Phi_{1}^{\dagger}\Phi_{2}+\Phi_{2}^{\dagger}\Phi_{1}\big)
\notag\\
&{}+\frac{\lambda_{1}}{2}\big(\Phi_{1}^{\dagger}\Phi_{1}\big)^2
+\frac{\lambda_{2}}{2}\big(\Phi_{2}^{\dagger}\Phi_{2}\big)^2
+\lambda_{3}\big(\Phi_{1}^{\dagger}\Phi_{1}\big)\big(\Phi_{2}^{\dagger}\Phi_{2}\big)
\notag\\
&{}
+\lambda_{4}\big(\Phi_{1}^{\dagger}\Phi_{2}\big)\big(\Phi_{2}^{\dagger}\Phi_{1}\big)
+\frac{\lambda_{5}}{2}\left[\big(\Phi_{1}^{\dagger}\Phi_{2}\big)^2
+\big(\Phi_{2}^{\dagger}\Phi_{1}\big)^2\right].
\label{eq:thdmpot}
\end{align}
Moreover, we assume all couplings in $V$ to be real in order to conserve CP.
In the non-linear Higgs representation this translates to
\begin{align}
V={}&
\frac{m_{11}^2}{2}\mathrm{tr}\bigl[\PhiM_{1}^{\dagger}\PhiM_{1}\bigr]
+\frac{m_{22}^2}{2}\mathrm{tr}\bigl[\PhiM_{2}^{\dagger}\PhiM_{2}\bigr]
-{m_{12}^2}\mathrm{tr}\bigl[\PhiM_{1}^{\dagger}\PhiM_{2}\bigr]
\notag\\
&{}+\frac{\lambda_{1}}{8}\big(\mathrm{tr}\bigl[\PhiM_{1}^{\dagger}\PhiM_{1}\bigr]\big)^2
+\frac{\lambda_{2}}{8}\big(\mathrm{tr}\bigl[\PhiM_{2}^{\dagger}\PhiM_{2}\bigr]\big)^2
+\frac{\lambda_{3}}{4}\mathrm{tr}\bigl[\PhiM_{1}^{\dagger}\PhiM_{1}\bigr]\mathrm{tr}\bigl[\PhiM_{2}^{\dagger}\PhiM_{2}\bigr]
\notag\\
&{}
+\lambda_{4}\mathrm{tr}\bigl[\PhiM_{1}^{\dagger}\PhiM_{2}\Omega_+\bigr]\mathrm{tr}\bigl[\PhiM_{1}^{\dagger}\PhiM_{2}\Omega_-\bigr]
+\frac{\lambda_{5}}{2}\left[\big(\mathrm{tr}\bigl[\PhiM_{1}^{\dagger}\PhiM_{2} \Omega_+ \bigr]\big)^2
+\big(\mathrm{tr}\bigl[\PhiM_{1}^{\dagger}\PhiM_{2}\Omega_-\bigr]\big)^2\right]
\end{align}
with the two-dimensional projection operators $\Omega_\pm = \frac{1}{2}(1\pm \sigma_3)$,
which select the original Higgs doublet $\Phi$ or its charge conjugate
from the matrix field $\PhiM$.
Obviously, the unitary Goldstone-boson matrix $U(\bfzeta)$ again drops out in $V$ in this
representation.

In the linear Higgs representation,
the CP-even Higgs fields $\eta_n$ of the doublets $\Phi_n$
are related to the fields $H$, $h$
corresponding to mass eigenstates of the CP-even neutral Higgs bosons
by a rotation with an angle $\alpha$
as given in Eq.~\refeq{eq:RalphaSESM}.
Similarly, there is an analogous rotation about an angle $\beta$ between
the CP-odd neutral fields $\chi_n$ to the CP-odd
neutral Higgs field $A_0$ and the neutral would-be Goldstone boson field $G_0$.
Similarly, in the charged sector the charged fields $\phi_n^\pm$
and the charged Higgs fields $H^\pm$ are rotated into the charged 
would-be Goldstone boson fields $G^\pm$,
\begin{align}
\begin{pmatrix} -\phi_{13} \\ -\phi_{23} \end{pmatrix}
= \begin{pmatrix} \chi_{1} \\ \chi_{2} \end{pmatrix}
= R(\beta) \begin{pmatrix} G_{0} \\ A_{0} \end{pmatrix},
\qquad
\begin{pmatrix} \phi^\pm_{1} \\ \phi^\pm_{2} \end{pmatrix}
= R(\beta) \begin{pmatrix} G^\pm \\ H^\pm \end{pmatrix}.
\end{align}
The matrix $R(\beta)$ is defined as $R(\alpha)$ in Eq.~\refeq{eq:RalphaSESM}, 
with $\alpha$ replaced by $\beta$.
In lowest order, the angle $\beta$ is given by the ratio of the vev
parameters $v_2$ and $v_1$, as already indicated in Eq.~\refeq{eq:Coeffs},
i.e.\ we have
\begin{align}
t_\beta\equiv\tan\beta = \frac{v_2}{v_1}.
\end{align}
The parameters $\beta$, which is usually taken as independent input parameter,
and $v = \sqrt{v_1^2 + v_2^2}$,
which is directly related to the W-boson mass by $\MW=g_2 v/2$,
then fix the vevs $v_1$, $v_2$ in terms of $\MW$, $g_2$, and $\beta$.

Owing to the relations $\eta_n=h_n+\dots$ 
up to higher powers in the fields, the fields $h_n$ 
of the non-linear Higgs representation
are transformed into the fields $h$ and $H$ of the CP-even Higgs bosons 
by a rotation with the same angle $\alpha$ as for the $\eta_n$
in the linear representation (at least in lowest perturbative order),
i.e.\ Eq.~\refeq{eq:RalphaSESM} holds in the THDM as well.
The rotation of the CP-odd field components $\phi_{nj}$ to fields
corresponding to mass eigenstates and would-be Goldstone-boson fields
is already contained in the factorization of the Goldstone matrix
$U(\bfzeta)$ from the matrix fields $\PhiM_n$ in the non-linear
representation. The relations
$\phi_{nj}=c_n\rho_j+v_n\zeta_j/v+\dots$, which are valid up to 
higher powers in the fields, just provide this rotation,
\begin{align}
\begin{pmatrix} \phi_{1j} \\ \phi_{2j} \end{pmatrix}
= R(\beta) \begin{pmatrix} \zeta_j \\ \rho_j \end{pmatrix} + \dots.
\end{align}
We can, thus, identify
\begin{align}
\begin{pmatrix} -\zeta_3 \\ -\rho_3 \end{pmatrix}
= \begin{pmatrix} G_{0} \\ A_{0} \end{pmatrix}_{\!\mathrm{nl}},
\qquad
\begin{pmatrix} \zeta^\pm \\ \rho^\pm \end{pmatrix}
= \begin{pmatrix} G^\pm \\ H^\pm \end{pmatrix}_{\!\mathrm{nl}}
\end{align}
with the same angle $\beta$ as in the linear representation 
(at least in lowest order).

Finally, we again calculate and compare the tadpole contributions. 
They are given by
\begin{align}
   \Gamma^{h_n}_{\mathrm{nl}} = T^{h_n}_{\mathrm{nl}}
  =\frac{1}{16 \pi^2 v}\biggl\{& \lambda_n^h v^2 A_0(\Mh^2) + \lambda_n^H v^2 A_0(\MH^2) 
+ \lambda_n^{A} v^2 A_0(\MA^2) + \lambda_n^{H^\pm} v^2 A_0(\MHpm^2) 
\nonumber \\& 
- 4 \sum_f \xi_{n}^f N^{\mathrm{c}}_f m_f^2 A_0(m_f^2) 
+ \frac{v_n \MZ^2}{v}  \bigl[3 A_0(\MZ^2) - 2 \MZ^2 \bigr] 
\nonumber \\&  
+ \frac{2v_n \MW^2}{v} \bigl[3 A_0(\MW^2) -2 \MW^2 \bigr]\biggr\} 
\end{align}
and 
\begin{align}
\Gamma^{\eta_n} = T^{\eta_n} 
= T^{h_n}_{\mathrm{nl}} 
+  \lambda_n^G v^2 \Delta v_\xi
\end{align}
for the non-linear and the linear representation, respectively, 
where $\Delta v_\xi$ given in Eq.~\eqref{eq:Deltavxi}
and $v^2 = v_1^2+ v_2^2$ in the THDM.
The parameters $\xi_{n}^f$ are the coupling strengths to the fermions relative to the SM value of $m_f/v$. 
The factors $\xi_{n}^f$ depend on the type of the THDM model and are related to the 
parameters $\xi_{H}^f$ and $\xi_{h}^f$ given 
in Tab.~2
of \citere{Altenkamp:2017ldc} by
\begin{align}
\begin{pmatrix} \xi_{1}^f \\ \xi_{2}^f \end{pmatrix}
= R(\alpha)
\begin{pmatrix} \xi_{H}^f \\ \xi_{h}^f \end{pmatrix}.
\end{align}
The trilinear coupling factors $\lambda_n^h$, $\lambda_n^H$, $\lambda_n^{A}$, 
and $ \lambda_n^{\pm}$ with $n=1,2$ are
%
\begin{align}
  \lambda_1^h &= \frac{3}{2} s_\alpha^2 c_\beta  \lambda_1 +
  \frac{c_\alpha}{2} \left[c_\beta c_\alpha  - 2 s_\beta s_\alpha\right]\lambda_{345}, &
  \lambda_2^h &=  \frac{3}{2} c_\alpha^2 s_\beta \lambda_2 
   + \frac{s_\alpha}{2} \left[s_\beta s_\alpha  - 2 c_\beta c_\alpha\right] \lambda_{345},\nonumber\\
  \lambda_1^H &= \frac{3}{2} c_\alpha^2 c_\beta  \lambda_1  
  + \frac{s_\alpha}{2} \left[c_\beta s_\alpha  + 2 s_\beta c_\alpha\right] \lambda_{345}, &
  \lambda_2^H &= 
   \frac{3}{2} s_\alpha^2 s_\beta \lambda_2 
   + \frac{c_\alpha}{2} \left[s_\beta c_\alpha  + 2 c_\beta s_\alpha\right] \lambda_{345},\nonumber\\
\lambda_1^{A} &= \frac{c_\beta}{2} \left[s_\beta^2 \lambda_1 + c_\beta^2 \lambda_{345} -2 \lambda_5 \right],&
\lambda_2^{A} &=  \frac{s_\beta}{2} \left[c_\beta^2\lambda_2 +s_\beta^2\lambda_{345} -2 \lambda_5 \right],\nonumber\\  
\lambda_1^{H^\pm} &=  c_\beta \left[s_\beta^2 (\lambda_1 -\lambda_{345}) + \lambda_3 \right],&
\lambda_2^{H^\pm} &=  s_\beta \left[c_\beta^2 (\lambda_2 -\lambda_{345}) + \lambda_3 \right],\nonumber\\
\lambda_1^{G} &= c_\beta \left[c_\beta^2 \lambda_1 + s_\beta^2 \lambda_{345} \right],&
\lambda_2^{G} &= s_\beta \left[s_\beta^2\lambda_2 + c_\beta^2\lambda_{345} \right], 
\end{align}
where the shorthand $\lambda_{345} = \lambda_3 + \lambda_4 + \lambda_5$ was used with
$\lambda_i$, $i = 1,\dots,5$, denoting the quartic couplings from the Higgs potential.
The tadpole constants $T^{h_n}_{\mathrm{nl}}$ are gauge independent,
a fact that even holds to all perturbative orders
by virtue of the Nielsen identities~\cite{Kluberg-Stern:1974iel,Nielsen:1975fs,Gambino:1999ai} 
(see discussion in Sect.~2.2 of \citere{Dittmaier:2022maf} for the SM).

\section{Gauge-invariant vacuum expectation value renormalization}
\label{se:GIVscheme}

\subsection{Higgs singlet extension of the Standard Model}
\label{se:SESMren}

\subsubsection{Schemes for tadpole and vacuum expectation value renormalization}

\begin{sloppypar}
To prepare the renormalization of the scalar sector of the SESM,
we split the bare scalar fields $\si_\bare$, $\Phi_\bare$ into bare field excitations 
$\eta_{1,\bare}$, $\eta_{2,\bare}$, etc.,
and bare constants $v_{1,0}$ and $v_{2,0}$,
\end{sloppypar}
\begin{align}
\si_\bare = v_{1,0}+\eta_{1,\bare},
\qquad
\Phi_\bare=\left(\begin{array}{c} \phi_\bare^+ \\
\frac{1}{\sqrt{2}}(v_{2,0}+\eta_{2,\bare}+\ri\chi_\bare)
\end{array}\right).
\label{eq:SESMPhiBsigma0}
\end{align}
The bases of bare Higgs-boson fields $(\eta_{1,\bare},\eta_{2,\bare})$
and ($h_\bare,H_\bare$) are related to each other
by rotations about the bare mixing angle $\alpha_0$, as defined in Eq.~\refeq{eq:RalphaSESM}.
As in the SM case described 
in \citere{Dittmaier:2022maf}, we need not consider renormalization in the non-linear
Higgs representation, because only the tadpole constants
$T^{h}_{\mathrm{nl}}$ and $T^{H}_{\mathrm{nl}}$  will be required from this representation.

Since the gauge structure of the SESM is the same as in the SM,
and since only $\Phi$, but not $\sigma$, interacts with the EW
gauge bosons, the W-boson and Z-boson masses $\MW$ and $\MZ$ are
determined by $v_2$ and the gauge couplings.

Different renormalization schemes for the SESM were proposed in
\citeres{Kanemura:2015fra,Bojarski:2015kra,Denner:2017vms,%
Altenkamp:2018bcs,Denner:2018opp}, where the most delicate 
parameter to be renormalized is the mixing angle $\alpha$.
\citere{Denner:2018opp} contains a comprehensive overview
of different ways to renormalize $\alpha$
and discusses the renormalization scheme dependence of some
Higgs-boson production and decay processes obtained
with $\alpha$ renormalized with OS,
$\MSbar$, or symmetry-inspired schemes.

Before briefly recapitulating the
application of the FJTS and PRTS to the SESM and formulating
the GIVS for the SESM, we first summarize the common
features of these schemes.
Inserting the field decomposition \refeq{eq:SESMPhiBsigma0} into the bare
Lagrangian ${\cal L}$, produces a term 
$t_{\eta_1,0}\eta_1+t_{\eta_2,0}\eta_2$ in the Lagrangian ${\cal L}$ with
\begin{align}
t_{\eta_1,0} & {} = 
v_{1,0} \left(2\mu_{1,0}^2 - 4\lambda_{1,0} v_{1,0}^2
- \lambda_{12,0} v_{2,0}^2 \right),
\nonumber
\\
t_{\eta_2,0} & {} = 
v_{2,0} \left(\mu_{2,0}^2 - \frac{1}{4}\lambda_{2,0} v_{2,0}^2 
- \lambda_{12,0} v_{1,0}^2\right)
\end{align}
at the one-loop level.
Since the fields $\eta_1$ and $\eta_2$ correspond to components of Higgs
fields that develop vevs, the corresponding one-point vertex functions
$\Gamma^{\eta_1}$ and $\Gamma^{\eta_2}$ only vanish in higher orders
after imposing appropriate renormalization conditions.
To this end, a contribution 
$\delta t_{\eta_1}\eta_1+\delta t_{\eta_2}\eta_2$ in the 
counterterm Lagrangian $\delta {\cal L}$ is required with appropriate
tadpole renormalization constants $\de t_{\eta_n}$, 
and we demand for the renormalized one-point functions $\Gamma_{\ren}^{\eta_n}$
\begin{align}
\Gamma_{\ren}^{\eta_n} = T^{\eta_n} + \delta t_{\eta_n} = 0
\quad\Rightarrow\quad
\de t_{\eta_n} = - T^{\eta_n}, \qquad
n=1,2.
\label{eq:SESMtadCT1}
\end{align}
The tadpole renormalization constants $\de t_{\eta_n}$ are generated 
in different ways by different choices of
the bare tadpole constants $t_{\eta_n,0}$, possibly accompanied by
further field redefinitions of the bare Higgs fields $\eta_{n,\bare}$.
These choices define different tadpole schemes.
It is convenient to rotate the two vertex functions
and the corresponding
tadpole renormalization constants $\delta t_{\eta_k}$
into the field basis $H$, $h$ of mass eigenstates,
\begin{align}
\begin{pmatrix} T^{\eta_1} \\ T^{\eta_2} \end{pmatrix}
= R(\alpha)
\begin{pmatrix} T^{H} \\ T^{h} \end{pmatrix},
\qquad
\begin{pmatrix} \de t_{\eta_1} \\ \de t_{\eta_2} \end{pmatrix}
= R(\alpha)
\begin{pmatrix} \de t_{H} \\ \de t_{h} \end{pmatrix}.
\label{eq:SESMtadCT2}
\end{align}
In this field basis, the tadpole renormalization condition reads
\begin{alignat}{3}
\Gamma_{\ren}^{H} ={}& T^{H} + \delta t_{H} = 0
& \quad\Rightarrow\quad &&
\de t_{H} =& - T^{H},
\nonumber\\
\Gamma_{\ren}^{h} ={}& T^{h} + \delta t_{h} = 0
& \quad\Rightarrow\quad &&
\de t_{h} =& - T^{h}.
\label{eq:SESMtadCT3}
\end{alignat}

\myparagraph{FJTS:}

In complete analogy to the SM case, in the FJTS the bare tadpole constants
are set to zero by definition, 
$t_{\eta_n,0}=0$, determining the vev
parameters according to
\begin{align}
v_{1,0} = \sqrt{ \frac{\lambda_{2,0}\mu_{1,0}^2-2\lambda_{12,0}\mu_{2,0}^2}%
	         {2(\lambda_{1,0}\lambda_{2,0} - \lambda_{12,0}^2)} }, 
\qquad
v_{2,0} = \sqrt{ \frac{2(2\lambda_{1,0}\mu_{2,0}^2-\lambda_{12,0}\mu_{1,0}^2)}%
                 {\lambda_{1,0}\lambda_{2,0} - \lambda_{12,0}^2} }.
\end{align}
The tadpole counterterms $\delta t_{\eta_1}\eta_1+\delta t_{\eta_2}\eta_2$ in the 
counterterm Lagrangian $\delta {\cal L}$ are generated from field shifts
\begin{align}
\eta_{n,\bare}&{} \;\to\;\eta_{n,\bare} + \De v_n^\FJTS, \qquad n=1,2,
\label{eq:etanshift}
\end{align}
in the bare Lagrangian, or equivalently
\begin{align}
h_{\bare}&{} \;\to\; h_{\bare} + \De v_h^\FJTS, \qquad 
H_{\bare}{} \;\to\; H_{\bare} + \De v_H^\FJTS
\label{eq:hHshift}
\end{align}
in the field basis corresponding to mass eigenstates.
The constants $\Delta v_{1,2}^\FJTS$ and $\Delta v_{h/H}^\FJTS$ are related to
each other like the corresponding fields,
\begin{align}
\begin{pmatrix} \Delta v_1^\FJTS \\ \Delta v_2^\FJTS \end{pmatrix}
= R(\alpha)
\begin{pmatrix} \Delta v_H^\FJTS \\ \Delta v_h^\FJTS \end{pmatrix}.
\label{eq:Dv12FJTS}
\end{align}
The field shifts lead to the terms 
$-\Delta v_{h}\Mh^2 h-\Delta v_{H}\MH^2 H$ linear in the fields
$h$, $H$ in the Lagrangian, so that the identification with the
tadpole counterterms fixes the constants $\Delta v_{h/H}^\FJTS$ to
\begin{align}
\Delta v_{H}^\FJTS {} = -\frac{\delta t_H}{\MH^2} = \frac{T^H}{\MH^2}, 
\qquad
\Delta v_{h}^\FJTS {} = -\frac{\delta t_h}{\Mh^2} = \frac{T^h}{\Mh^2}. 
\label{eq:DvH1H2FJTS}
\end{align}
We finally note that $\Delta v_1^\FJTS$ is gauge independent
because of the gauge invariance of $\langle\sigma_\bare\rangle$,
but $\Delta v_2^\FJTS$ is gauge dependent.

In the FJTS, the parameters $v_{n,0}$ 
correspond to the location of the minimum
of the tree-level Higgs potential. To define the fields
$\eta_{n,\bare}$, $h_\bare$, $H_\bare$ as excitations about the minimum of the
effective Higgs potential, the shifts 
$\Delta v_{1,2}^\FJTS$ and $\Delta v_{h/H}^\FJTS$ had to be introduced,
which are a source of potentially large corrections in the FJTS if
$\MSbar$ parameters are used.

\myparagraph{PRTS:}

In the PRTS, the identification of fields as excitations about
the minimum of the effective Higgs potential is achieved without further
field redefinition, i.e.\ with an appropriate definition of the 
bare parameters $t_{\eta_n,0}$ and $v_{n,0}$.
The condition that the renormalized vev 
minimizes the effective Higgs potential implies that the sum of bare
and loop-induced tadpole contributions vanish,
\begin{align}
0 = t_{\eta_n,0} + T^{\eta_n}.
\end{align}
Splitting the bare tadpole parameters 
$t_{\eta_n,0} = t_{\eta_n} + \delta t_{\eta_n}^\PRTS$
into a renormalized parts $t_{\eta_n}$ and a tadpole renormalization 
constants $\delta t_{\eta_n}^\PRTS$ and using Eq.~\refeq{eq:SESMtadCT1},
this leads to
\begin{align}
0 = t_{\eta_n} + \delta t_{\eta_n}^\PRTS+ T^{\eta_n} = t_{\eta_n},
\end{align}
i.e.\ the renormalized parts $t_{\eta_n}$ vanish.
In turn, this means that the bare tadpole parameters directly provide the
tadpole renormalization constants, $t_{\eta_n,0} = \delta t_{\eta_n}^\PRTS$,
which are given by
\begin{align}
\delta t_{\eta_1}^\PRTS & {} = 
v_{1,0} \left(2\mu_{1,0}^2 - 4\lambda_{1,0} v_{1,0}^2- \lambda_{12,0} v^2_{2,0} \right),
\nonumber\\
\delta t_{\eta_2}^\PRTS & {} = 
v_{2,0} \left(\mu_{2,0}^2 - \frac{1}{4}\lambda_{2,0} v_{2,0}^2 - \lambda_{12,0} v_{1,0}^2\right).
\label{eq:dtSESMPRTS}
\end{align}

The bare vev parameters $v_{n,0}=v_n+\delta v_n$
are split into renormalized parts $v_n$ and renormalized constants $\delta v_n$ accordingly,
where the renormalized parameters $v_n$ are directly related to measured quantities. 
As in the SM, $v_2$ is directly related to the W-boson mass and the gauge coupling $g_2$,
while $v_1$ is related to 
the masses $\Mh$, $\MH$ of the light and heavy Higgs bosons~h and H,
the renormalized mixing angle $\alpha$, and to one of the scalar couplings $\lambda_k$,
for which we choose $\lambda_{12}$ as in \citere{Altenkamp:2018bcs}.
These relations are demanded both for bare and renormalized quantities,
\begin{align}
v_{1,0} ={}& \frac{s_{2\alpha_0}(M_{\PH,0}^2-M_{\Ph,0}^2)}{4v_{2,0}\lambda_{12,0}},&
v_{2,0} ={}& \frac{2M_{\PW,0}}{g_{2,0}} = \frac{2M_{\PW,0}s_{\rw,0}}{e_0}, 
\label{eq:vSESMPRTS}
\\
v_1 ={}& \frac{s_{2\alpha}(\MH^2-\Mh^2)}{4v_2\lambda_{12}},&
v_2 ={}& \frac{2\MW}{g_2} = \frac{2\MW\sw}{e}, 
\end{align}
where $e$ is the elementary charge and $\sw$ the sinus of the weak mixing angle 
$\theta_\rw$, which is fixed by
the ratio of the W-/Z-boson masses, $\cw=\cos\theta_\rw=\MW/\MZ$.
The relation of $v_n$ to measured quantities which serve as input for the model 
then fixes the renormalization constants $\delta v_n$ via the renormalization
conditions for the input quantities.

Having fixed the renormalization of the vev parameters $v_n$ and the tadpoles,
the renormalization of the Higgs sector can be completed by splitting
the bare versions of the original parameters of the Higgs potential,
$\mu_{1,0}^2$, $\mu_{2,0}^2$, $\lambda_{1,0}$, $\lambda_{2,0}$, $\lambda_{12,0}$,
into renormalized parameters and renormalization constants as usual.
As done in \citere{Altenkamp:2018bcs}, we take the parameters
$\Mh$, $\MH$, $\alpha$, $\lambda_{12}$, and $v_2=2\MW\sw/e$ as input to 
parametrize the Higgs sector, so that the corresponding renormalization constants 
are directly fixed by renormalization conditions.
The corresponding bare parameters are tied to the bare original Higgs parameters
via the diagonalization of the bare Higgs mass matrix. 
In total, the five relations providing the link between the two sets of bare parameters,
$\mu_{1,0}^2$, $\mu_{2,0}^2$, $\lambda_{1,0}$, $\lambda_{2,0}$, $\lambda_{12,0}$
and $M_{\Ph,0}$, $M_{\PH,0}$, $\alpha_0$, $\lambda_{12,0}$, $v_{2,0}$,
are given by Eq.~\refeq{eq:dtSESMPRTS}, the 
first relation in Eq.~\refeq{eq:vSESMPRTS},
and the following two relations,
\begin{align}
\lambda_{1,0} ={} & 
\frac{1}{8v_{1,0}^2} \left( c_{\alpha,0}^2 M_{\PH,0}^2 
	+ M_{\Ph,0}^2 s_{\alpha,0}^2 + \frac{\delta t^\PRTS_{\eta_1}}{v_{1,0}} \right),
\nonumber\\
\lambda_{2,0} ={} & \frac{2}{v_{2,0}^2} \left( c_{\alpha,0}^2 M_{\Ph,0}^2 
	+ M_{\PH,0}^2 s_{\alpha,0}^2 + \frac{ \delta t^\PRTS_{\eta_2}}{v_{2,0}} \right).
\end{align}
The corresponding relations between renormalized parameters are obtained by replacing
the bare parameters by renormalized ones and by setting the tadpole constants
$\delta t^\PRTS_{\eta_n}$ to zero; these relations are given in 
Eq.~(2.15) of \citere{Altenkamp:2018bcs}.
The bare and renormalized relations, finally fix the renormalization constants
$\delta\mu_{1}^2$, $\delta\mu_{2}^2$, $\delta\lambda_{1}$, $\delta\lambda_{2}$, 
$\delta\lambda_{12}$ in terms of
$\delta\Mh^2$, $\delta\MH^2$, $\delta\alpha$, $\delta\lambda_{12}$, $\delta v_2$
as given in Eq.~(3.7) of \citere{Altenkamp:2018bcs}.
From the results given there, we can deduce that the tadpole contributions to all
counterterms can be obtained from the bare Lagrangian without tadpole terms
by the substitutions
\begin{align}
\mu_{1,0}^2 &{} \;\to\; \mu_{1,0}^2 + \frac{3 \de t_{\eta_1}^\PRTS}{4 v_1},
& 
\mu_{2,0}^2 &{} \;\to\; \mu_{2,0}^2 + \frac{3 \de t_{\eta_2}^\PRTS}{2 v_2},
\nonumber\\
\lambda_{1,0} &{} \;\to\; \lambda_{1,0} + \frac{\de t_{\eta_1}^\PRTS}{8 v_1^3},
& 
\lambda_{2,0} &{} \;\to\; \lambda_{2,0} + \frac{2 \de t_{\eta_2}^\PRTS}{v_2^3},
&
\lambda_{12,0} &{} \;\to\; \lambda_{12,0}.
\label{eq:SESMdtPTRSgeneration}
\end{align}

In complete analogy to the SM, the PRTS induces gauge-dependent
relations between the bare input parameters and the bare original parameters
of the Higgs sector,
leading to a gauge-dependent relation between predictions for
observables and input parameters if $\MSbar$-renormalized masses or
an $\MSbar$-renormalized Higgs mixing angle $\alpha$ are used.

\myparagraph{GIVS:}

As in the SM, the GIVS is fully equivalent to the formulation of the
PRTS in the non-linear Higgs representation. 
In some slight abuse of notation, we keep the letters $H$, $h$ for the
Higgs fields corresponding to mass eigenstates also in the non-linear Higgs
representation.
This should not cause confusion, since the fields $H$, $h$ of the non-linear 
representation are only appearing as labels for vertex functions
$\Gamma^{\dots}_{\mathrm{nl}}$ and tadpoles $T^{\dots}_{\mathrm{nl}}$,
where the non-linear representation is explicitly indicated.
The PRTS tadpole renormalization constants of the non-linear representation
are determined by the unrenormalized 
Higgs one-point functions,
\begin{align}\label{eq:dtPRTSnlSESM}
\delta t^\PRTS_{H,\mathrm{nl}} = -T^{H}_{\mathrm{nl}}, \qquad 
\delta t^\PRTS_{h,\mathrm{nl}} = -T^{h}_{\mathrm{nl}}. 
\end{align}
Transferred to the
linear representation, we keep the PRTS part of the tadpole
renormalization constants, defining 
\begin{align}\label{eq:dtGIVS1SESM}
\delta t^\GIVS_{H,1} = \delta t^\PRTS_{H,\mathrm{nl}} = -T^{H}_{\mathrm{nl}}, \qquad 
\delta t^\GIVS_{h,1} = \delta t^\PRTS_{h,\mathrm{nl}} = -T^{h}_{\mathrm{nl}}, 
\end{align}
and supplement them with FJTS parts,
\begin{align}\label{eq:dtGIVS2SESM}
\de t^\GIVS_{H,2} ={}& -\MH^2\Delta v^\GIVS_H = T^{H}_{\mathrm{nl}}-T^{H}
= -s_\alpha \MH^2 \Delta v_\xi, 
\nonumber\\
\de t^\GIVS_{h,2} ={}& -\Mh^2\Delta v^\GIVS_h = T^{h}_{\mathrm{nl}}-T^{h}
= -c_\alpha \Mh^2 \Delta v_\xi,
\end{align}
where we have used Eq.~\refeq{eq:TSESM} with Eq.~\refeq{eq:tadcoefsSESM} 
for the differences $T^{H}_{\mathrm{nl}}-T^{H}$ and
$T^{H}_{\mathrm{nl}}-T^{H}$.
Specifically, we get 
\begin{align}
\Delta v^\GIVS_H ={} s_\alpha \Delta v_\xi,
\qquad
\Delta v^\GIVS_h ={} c_\alpha \Delta v_\xi.
\end{align}
In total, we obtain the tadpole renormalization constants
from the sum of the two respective parts
in order to cancel all explicit tadpole diagrams,
\begin{align}\label{eq:dtGIVSSESM}
\de t^\GIVS_{H} = \de t^\GIVS_{H,1} + \de t^\GIVS_{H,2} = -T^{H}, \qquad
\de t^\GIVS_{h} = \de t^\GIVS_{h,1} + \de t^\GIVS_{h,2} = -T^{h}.
\end{align}
In analogy to the SM, the tadpole contributions
$\delta t^\GIVS_{H,1}$ and $\delta t^\GIVS_{h,1}$ enter relations between the bare
parameters of the Higgs sector, causing no problems with gauge dependences, 
since these terms are gauge independent.
On the other hand, the gauge-dependent parts $\delta t^\GIVS_{H,2}$ and $\delta t^\GIVS_{h,2}$
have no effects on predictions for observables,
because they only enter the calculation via field shifts.

The generation of the two types of tadpole counterterms can be deduced from the 
generation of the PRTS and FJTS tadpole counterterms exactly as in the SM~\cite{Dittmaier:2022maf},
resulting in the following rules to be applied to the bare Lagrangian,
\begin{align}
\hspace*{2em}
\mu_{1,0}^2 &{} \;\to\; \mu_{1,0}^2 + \frac{3 \de t_{\eta_1,1}^\GIVS}{4 v_1},
& 
\mu_{2,0}^2 &{} \;\to\; \mu_{2,0}^2 + \frac{3 \de t_{\eta_2,1}^\GIVS}{2 v_2},
\nonumber\\
\lambda_{1,0} &{} \;\to\; \lambda_{1,0} + \frac{\de t_{\eta_1,1}^\GIVS}{8 v_1^3},
& 
\lambda_{2,0} &{} \;\to\; \lambda_{2,0} + \frac{2 \de t_{\eta_2,1}^\GIVS}{v_2^3},
& 
\lambda_{12,0} &{} \;\to\; \lambda_{12,0},
\hspace*{2em}
\nonumber\\
\eta_{1,\bare}&{} \;\to\eta_{1,\bare} + \De v_1^\GIVS, 
& \hspace*{-4em}
\eta_{2,\bare}&{} \;\to\eta_{2,\bare} + \De v_2^\GIVS,
\end{align}
where the 
renormalization constants of the field basis $\eta_1$, $\eta_2$ are related to
the ones of the $H$, $h$ basis as follows,
\begin{align}\label{eq:dtGIVSTHDM}
\begin{pmatrix} \delta t_{\eta_1,1}^\GIVS \\ \delta t_{\eta_2,1}^\GIVS \end{pmatrix}
&{}= R(\alpha)
\begin{pmatrix} \delta t_{H,1}^\GIVS \\ \delta t_{h,1}^\GIVS \end{pmatrix},
\nonumber\\
\begin{pmatrix} \Delta v_{1}^\GIVS \\ \Delta v_{2}^\GIVS \end{pmatrix}
&{}= R(\alpha)
\begin{pmatrix} \Delta v_{H}^\GIVS \\ \Delta v_{h}^\GIVS \end{pmatrix}
= - R(\alpha)
\begin{pmatrix} \delta t_{H,2}^\GIVS/\MH^2 \\ 
\delta t_{h,2}^\GIVS/\Mh^2 \end{pmatrix}.
\end{align}

\subsubsection{Mixing-angle renormalization}
\label{se:SESMmixingangle}

Finally, we address the $\MSbar$ renormalization variants for the
mixing angle $\alpha$ with the different tadpole treatments
if $\alpha$ is used as one of the basic input parameters of the SESM.
For the FJTS and PRTS we translate the results given in
\citeres{Denner:2017vms,Altenkamp:2018bcs,Denner:2018opp}
into the conventions and notation used in this paper.
In detail, we trade the original parameters $\mu_1^2$, $\mu_2^2$, 
$\lambda_1$, $\lambda_2$, and $\lambda_{12}$ of the Higgs potential~$V$
given in Eq.~\refeq{eq:VSESM} for the parameters
$v_2$ (fixed by the unit charge $e$ and the masses $\MW$ and $\MZ$),
$\Mh$, $\MH$, $s_\alpha$, and 
one of the quartic scalar couplings, which is either taken as
$\lambda_{12}$ as in \citere{Altenkamp:2018bcs} 
or as $\lambda_{1}$ as in \citere{Denner:2018opp}.
Since the $\MSbar$ renormalization of $\lambda_{12}$ or $\lambda_{1}$
is independent of the tadpole scheme, we do not discuss the
renormalization of those parameters below and refer to \citeres{Altenkamp:2018bcs,Denner:2018opp}
for details.

In the $\MSbar$ scheme, the renormalization constant
$\de\alpha$ for the angle $\alpha$ can be determined according to
\begin{align}
\de\al_{\MSbar}={}& 
\frac{1}{4}\left( \delta Z_{Hh}-\delta Z_{hH} \right) \Big|_{\UV},
\label{eq:alphaMSren}
\end{align}
where $\delta Z_{ij}$ are the 
OS field renormalization constants
for the $H$, $h$ fields defined by
\begin{align}
\begin{pmatrix} H_{\bare} \\ h_{\bare} \end{pmatrix} = 
\begin{pmatrix} 
1 + \frac{1}{2}\de Z_{HH} &   \frac{1}{2}\de Z_{Hh}\\
         \frac{1}{2}\de Z_{hH} &  1+\frac{1}{2}\de Z_{hh}
\end{pmatrix}
\begin{pmatrix} H \\ h \end{pmatrix},
\end{align}
and the subscript ``UV'' indicates that in dimensional regularization
only the UV-divergent parts proportional to $\Delta$ as given in
\refeq{eq:Delta} are taken into account.
Recall that the reference mass scale $\mu$ takes over the role
of the renormalization scale in $\MSbar$ schemes.
The constants $\delta Z_{ij}$ are determined by the
Higgs-boson self-energies $\Sigma^{ij}(p^2)$
with momentum transfer~$p$
evaluated at the on-shell points $p^2=M_{\PH}^2$ or $p^2=M_{\Ph}^2$.
In the following, we decompose 
$\Sigma^{ij}(p^2) = \Sigma^{ij}_{\mathrm{1PI}}(p^2)+\Sigma^{ij}_{\mathrm{tad}}$
into the contribution $\Sigma^{ij}_{\mathrm{1PI}}(p^2)$
induced by one-particle-irreducible (1PI) diagrams
and the contribution $\Sigma^{ij}_{\mathrm{tad}}$ of all
(momentum-independent) explicit tadpole diagrams and tadpole counterterms.
In detail, the diagonal quantities $\delta Z_{ii}$ 
involve only derivatives w.r.t.\ $p^2$ and thus do not 
receive tadpole contributions, but the mixing constants
\begin{align}
\delta Z_{Hh} = \frac{2\Re\{\Sigma^{Hh}(M_{\Ph}^2)\}}{M_{\PH}^2-M_{\Ph}^2}, \qquad
\delta Z_{hH} = \frac{2\Re\{\Sigma^{Hh}(M_{\PH}^2)\}}{M_{\Ph}^2-M_{\PH}^2}
\end{align}
involve tadpole contributions,%
\footnote{Following the conventions of \citeres{Denner:2018opp,Denner:2019vbn,Dittmaier:2022maf},
our unrenormalized one-loop self-energy functions $\Sigma^{ij}$
include one-particle-irreducible
diagrams as well as explicit tadpole loops and tadpole counterterm
contributions, as pictorially indicated in Eq.~(141) of \citere{Denner:2019vbn}.}
which are the source for the
different $\MSbar$ results for $\de\alpha$ in the
various tadpole schemes.
The tadpole contributions to $\de\al_{\MSbar}$ are given by
\begin{align}
\de\al_{\MSbar,\mathrm{tad}}={}& 
\frac{\Sigma^{Hh}_{\mathrm{tad}}}{M_{\PH}^2-M_{\Ph}^2}
\bigg|_{\UV},
\label{eq:datadSESM}
\end{align}
where we have used that $\Sigma^{Hh}_{\mathrm{tad}}$ is a real $p^2$-independent quantity.

Inserting all tadpole counterterms contributing to
the $hH$~mixing self-energy in the various tadpole schemes into 
Eq.~\refeq{eq:datadSESM}, we find
\begin{align}
%
\de\al^\FJTS_{\MSbar,\mathrm{tad}}={}& 
\frac{e(C_{hhH}\,\Delta v_h^\FJTS + C_{hHH}\,\Delta v_H^\FJTS)}{\MH^2-\Mh^2} \bigg|_{\UV},
\label{eq:dalphatadFJTSSESM}
\\
\de\al^\PRTS_{\MSbar,\mathrm{tad}}={}& 0,
\label{eq:dalphatadPRTSSESM}
\\
%
\de\al^\GIVS_{\MSbar,\mathrm{tad}}={}& 
\frac{e\left(c_\alpha C_{hhH} + s_\alpha C_{hHH}\right)\Delta v_\xi}{\MH^2-\Mh^2} \bigg|_{\UV},
\label{eq:dalphatadGIVSSESM}
\end{align}
with the shorthands
\begin{align}
C_{hhH} ={}& \frac{s_\alpha}{e} \left( 2 \Mh^2 + \MH^2 \right) 
\left( \frac{2v_2 \lambda_{12}}{\Mh^2 - \MH^2} - \frac{c_\alpha^2}{v_2} \right),
\nonumber\\
C_{hHH} ={}& \frac{c_\alpha}{e} \left( \Mh^2 + 2 \MH^2 \right) 
\left( \frac{2v_2 \lambda_{12}}{\MH^2 - \Mh^2} - \frac{s_\alpha^2}{v_2} \right)
\end{align}
for the scalar self-couplings $e(C_{hhH}h^2 H+C_{hHH}h H^2)/2$ in the Lagrangian.

As already explained in 
\citeres{Denner:2017vms,Altenkamp:2018bcs,Denner:2018opp},
changing the tadpole treatment in the $\MSbar$ renormalization of the mixing angle~$\alpha$ 
changes the physical meaning of $\alpha$, similar to a change in its
renormalization condition. This change leads to a finite shift in the numerical
value of the renormalized parameter~$\alpha^{\TS}_{\MSbar}$ in the 
tadpole scheme TS when TS is changed.
This shift, which is related to the finite parts of tadpole contributions,
was calculated in \citere{Denner:2017vms} by considering NLO
corrections to specific vertices. In \citere{Altenkamp:2018bcs}
arguments for a universal relation between 
$\alpha^\PRTS_{\MSbar}$ and $\alpha^\FJTS_{\MSbar}$ were given, based on the observation that
the FJTS is equivalent to a procedure in which no tadpole renormalization
is performed at all. 
In the following we derive the general relation between the input parameters
of the SESM when changing the tadpole scheme, again basing the arguments on the
equivalence of the FJTS and performing no tadpole renormalization.

While the bare original Higgs parameters
$\mu_{1,0}^2$, $\mu_{2,0}^2$, $\lambda_{1,0}$, $\lambda_{2,0}$, $\lambda_{12,0}$
are the same in the FJTS and PRTS, this is in general not the case between the derived 
bare parameters 
$M_{\Ph,0}$, $M_{\PH,0}$, $\alpha_0$, $\lambda_{12,0}$ (or $\lambda_{1,0}$), 
and $v_{2,0}$ corresponding to
the input parameters.
This is due to the fact that tadpole terms $\delta t^\PRTS_{\eta_n}$ appear 
in the relations linking the two sets of bare parameters in the PRTS, which do not
appear in the FJTS.
As pointed out in the previous section, these PRTS tadpole terms can be introduced 
into the relations without tadpole terms, which are valid in the FJTS,
via the substitutions \refeq{eq:SESMdtPTRSgeneration}.
Applying this consideration to the bare mixing angle $\alpha_0$,
we get the one-loop relation
\begin{align}
\alpha_{0}^\FJTS &{} \;=\; 
\alpha_{0}\Big|_{\footnotesize\refeq{eq:SESMdtPTRSgeneration},\, p_0\to p_0^\PRTS},
\end{align}
which means that the substitution \refeq{eq:SESMdtPTRSgeneration} is applied
to $\alpha_0$ expressed in terms of the bare parameters 
$\{p_0\}$ = $\{\mu_{1,0}^2$, $\mu_{2,0}^2$, $\lambda_{1,0}$, $\lambda_{2,0}$, 
$\lambda_{12,0}\}$ in the absence of tadpole terms.
Analogous substitutions hold for the other bare parameters such as the Higgs masses
$M_{\Ph,0}$ and $M_{\PH,0}$. 
For the mixing angle $\alpha_0$ the explicit result of the substitution is
\begin{align}
\alpha_{0}^\PRTS &{} \;=\; \alpha_{0}^\FJTS
+ \frac{e}{\MH^2-\Mh^2} \left( C_{hhH}\,\frac{\delta t_h^\FJTS}{\Mh^2} 
+ C_{hHH}\,\frac{\delta t_H^\FJTS}{\MH^2} \right),
\label{eq:SESMalphaconvPRTSFJTS}
\end{align}
the derivation of which is somewhat cumbersome but straightforward.
The relation between the two renormalized parameters
$\alpha^\PRTS$ and $\alpha^\FJTS$ simply follows from the renormalization transformation
\begin{align}
\alpha_{0}^\TS = \alpha^\TS + \delta\alpha^\TS
\end{align}
with TS being the PRTS or FJTS.
The shift between the $\MSbar$-renormalized parameters in the PRTS and FJTS, 
thus, reads
\begin{align}
\alpha^\PRTS_{\MSbar} - \alpha^\FJTS_{\MSbar} ={}&
\left(\alpha_0^\PRTS - \alpha_0^\FJTS \right)
-\left(\delta\alpha^\PRTS_{\MSbar} - \delta\alpha^\FJTS_{\MSbar} \right)
\label{eq:SESMalphaconvPRTSFJTS2}
\end{align}
and can be explicitly calculated using Eq.~\refeq{eq:SESMalphaconvPRTSFJTS}
and the difference between the corresponding renormalization constants
$\delta\alpha^\TS_{\MSbar}$.
As explained above,
the constants $\delta\alpha^\TS_{\MSbar}$ receive contributions from the 
1PI self-energy part
$\Sigma^{Hh}_{1\mathrm{PI}}$ and from tadpole loops and tadpole renormalization constants.
Since the self-energy contributions do not depend on the tadpole scheme,
the difference $\delta\alpha^\PRTS_{\MSbar} - \delta\alpha^\FJTS_{\MSbar}$ 
in Eq.~\refeq{eq:SESMalphaconvPRTSFJTS2} can be calculated from the 
tadpole contributions given in Eqs.~\refeq{eq:dalphatadPRTSSESM} and \refeq{eq:dalphatadFJTSSESM}
alone. Combining all those parts, we finally get
\begin{align}
%
\alpha^\PRTS_{\MSbar} - \alpha^\FJTS_{\MSbar} ={}&
- \frac{e}{\MH^2-\Mh^2} \left( C_{hhH}\,\frac{T^h}{\Mh^2} 
+ C_{hHH}\,\frac{T^H}{\MH^2} \right) \bigg|_{\mathrm{finite}},
\label{eq:SESMalphaconvPRTSFJTS3}
\end{align}
where the subscript ``finite'' indicates that the UV-divergent part proportional to 
the parameter $\Delta$ defined in Eq.~\refeq{eq:Delta} is omitted.

\begin{sloppypar}
The conversion between the FJTS and the GIVS proceeds along the same lines.
In Eqs.~\refeq{eq:SESMalphaconvPRTSFJTS}--\refeq{eq:SESMalphaconvPRTSFJTS2}
we merely have to change the label PRTS to GIVS on all parameters
and to replace the tadpole renormalization constants $\delta t^\PRTS_X$
($X=\eta_1,\eta_2,H,h$) by $\delta t^\GIVS_{X,1}$, because only part~1 of 
$\delta t^\GIVS_X$ enters the relations between bare parameters in the GIVS.
This finally results in
\end{sloppypar}
\begin{align}
\alpha^\GIVS_{\MSbar} - \alpha^\FJTS_{\MSbar} ={}&
- \frac{e}{\MH^2-\Mh^2} \left( C_{hhH}\,\frac{T^h_{\mathrm{nl}}}{\Mh^2} 
+ C_{hHH}\,\frac{T^H_{\mathrm{nl}}}{\MH^2} \right) \bigg|_{\mathrm{finite}},
\end{align}
which is identical to the conversion \refeq{eq:SESMalphaconvPRTSFJTS3} between
the PRTS and the FJTS except that the tadpole contributions are calculated in the
non-linear Higgs representation, rendering  the relation between
$\alpha^\GIVS_{\MSbar}$ and $\alpha^\FJTS_{\MSbar}$ gauge independent.

\subsubsection[NLO decay widths for $\Ph/\PH\to4f$ in the SESM]%
{\boldmath{NLO decay widths for $\Ph/\PH\to4f$ in the SESM}}
\label{se:SESMnumerics}

In \citere{Altenkamp:2018bcs}, the renormalization of the SESM is described in detail
for an $\MSbar$-renormalized Higgs mixing angle $\alpha$, both with
the PRTS and the FJTS for treating tadpoles. 
In \citere{Denner:2018opp} other renormalization schemes based on OS conditions or 
prescriptions inspired by symmetries are formulated in addition.
In those articles, explicit NLO predictions for the important class of Higgs decay processes
$\Ph/\PH\to\PW\PW/\PZ\PZ\to4f$ are discussed, in particular the renormalization scale
and renormalization scheme dependences of these decay widths for some selected
SESM scenarios. 
The underlying calculations are the basis for the implementation of these
predictions in the public Monte Carlo program 
\Prophecy~3.0~\cite{Bredenstein:2006rh,Bredenstein:2006ha,Denner:2019fcr}.

Here we continue the discussion of \citeres{Altenkamp:2018bcs,Denner:2018opp}
by adding results for the scheme with an
$\MSbar$-renormalized mixing angle $\alpha$ with GIVS tadpole treatment.
To this end, we have implemented this new scheme into \Prophecy\ as new option.%
\footnote{An updated public version of \Prophecy\ will appear on https://prophecy4f.hepforge.org/ soon;
meanwhile a non-public version of the program may be obtained from the
authors on request.}
The numerical input for the scenarios called BHM200$^\pm$, BHM400, BHM600
as well as the details of the calculational setup
can be found in \citeres{Altenkamp:2018bcs,Denner:2018opp}.
We just recall that the mass of the lighter Higgs boson is
set to $\Mh=125.1\GeV$, the mass of the heavier is indicated in the name of the
scenario (i.e.\ $\MH=200\GeV$ for BHM200$^\pm$, etc.), and 
the mixing angle $\alpha$ 
{is chosen as $\pm0.29$, $0.26$, and $0.22$ in BHM200$^\pm$, BHM400, and
BHM600, respectively, so that $\alpha$ is of}
the order of magnitude that is maximally allowed by LHC Higgs analyses.
Results of the scheme conversion of the input values for the $\MSbar$ parameters $\alpha$ 
and $\lambda_{12}$ are given in \refapp{app:SESM}.
As in \citeres{Altenkamp:2018bcs,Denner:2018opp}, the PRTS is evaluated in 
't~Hooft--Feynman gauge ($\xi_a=1$).

Tables~\ref{tab:SESM-H24f} and \ref{tab:SESM-H14f} summarize the NLO predictions for
the partial decay widths of $\Ph\to4f$ and $\PH\to4f$, respectively,
based on the various $\MSbar$ schemes using the OS scheme
as reference scheme in which the input parameters are defined.
\begin{table}
\centerline{\renewcommand{\arraystretch}{1.25} \small \tabcolsep 3pt
\begin{tabular}{|c|c||l|l|l|l|}
\hline
&& \multicolumn{2}{c|}{BHM200$^+$} & \multicolumn{2}{c|}{BHM200$^-$}
\\
Ren.\ scheme & tadpoles & \multicolumn{1}{c|}{LO} & \multicolumn{1}{c|}{NLO}  & \multicolumn{1}{c|}{LO} & \multicolumn{1}{c|}{NLO}
\\
\hline
\hline
OS & & $0.84034(3)$ & $0.90553(6)_{-0.0\%}^{+0.0\%}$ &
      $0.84034(3)$ & $0.90552(6)_{-0.0\%}^{+0.0\%}$
\\
\hline
\MSbar& FJTS & $0.82292(3)_{+2.7\%}^{-3.5\%}$ & $0.90550(7)_{+0.0\%}^{+0.7\%}$ &
        $0.82614(3)_{+0.7\%}^{-0.6\%}$ & $0.90558(7)_{-0.1\%}^{+0.0\%}$
\\
& & {\footnotesize\quad $\Delta_{\mathrm{OS}}=-2.07\%$} & {\footnotesize\quad $\Delta_{\mathrm{OS}}=-0.00\%$}
  & {\footnotesize\quad $\Delta_{\mathrm{OS}}=-1.69\%$} & {\footnotesize\quad $\Delta_{\mathrm{OS}}=+0.01\%$}
\\
\hline
\MSbar& PRTS & $0.83361(3)_{-4.4\%}^{+3.2\%}$ & $0.90539(6)_{+0.5\%}^{+0.5\%}$ &
        $0.83261(3)_{-4.5\%}^{+3.3\%}$ & $0.90546(7)_{+0.6\%}^{+0.5\%}$
\\
& & {\footnotesize\quad $\Delta_{\mathrm{OS}}=-0.80\%$} & {\footnotesize\quad $\Delta_{\mathrm{OS}}=-0.02\%$}
  & {\footnotesize\quad $\Delta_{\mathrm{OS}}=-0.92\%$} & {\footnotesize\quad $\Delta_{\mathrm{OS}}=-0.01\%$}
\\
\hline
\MSbar& GIVS & $ 0.83440(3)_{-4.6\%}^{+3.3\%}$ & $0.90540(6)_{+0.6\%}^{+0.5\%}$ &
        $0.83363(3)_{-4.7\%}^{+3.4\%}$ & $0.90545(6)_{+0.6\%}^{+0.6\%}$
\\
& & {\footnotesize\quad $\Delta_{\mathrm{OS}}=-0.71\%$} & {\footnotesize\quad $\Delta_{\mathrm{OS}}=-0.01\%$}
  & {\footnotesize\quad $\Delta_{\mathrm{OS}}=-0.80\%$} & {\footnotesize\quad $\Delta_{\mathrm{OS}}=-0.01\%$}
\\
\hline
\multicolumn{5}{c}{}\\
\hline
&& \multicolumn{2}{c|}{BHM400} & \multicolumn{2}{c|}{BHM600}
\\
Ren.\ scheme & tadpoles & \multicolumn{1}{c|}{LO} & \multicolumn{1}{c|}{NLO}  & \multicolumn{1}{c|}{LO} & \multicolumn{1}{c|}{NLO}
\\
\hline
\hline
OS & &
$0.85548(3)$ & $0.92178(6)_{-0.0\%}^{+0.0\%}$ &
$0.87309(3)$ & $0.94078(7)_{-0.0\%}^{+0.0\%}$
\\
\hline
\MSbar& FJTS &
$0.85349(3)_{+1.6\%}^{-2.1\%}$ & $0.92166(7)_{+0.3\%}^{+0.1\%}$ &
$0.87608(3)_{+1.2\%}^{-1.5\%}$ & $0.94106(7)_{+0.3\%}^{-0.0\%}$
\\
& & {\footnotesize\quad $\Delta_{\mathrm{OS}}=-0.23\%$} & {\footnotesize\quad $\Delta_{\mathrm{OS}}=-0.01\%$}
  & {\footnotesize\quad $\Delta_{\mathrm{OS}}=+0.34\%$} & {\footnotesize\quad $\Delta_{\mathrm{OS}}=+0.03\%$}
\\
\hline
\MSbar& PRTS &
$0.85209(3)_{-0.5\%}^{+0.5\%}$ & $0.92159(7)_{-0.0\%}^{+0.0\%}$ &
$0.87067(3)_{-0.1\%}^{+0.1\%}$ & $0.94060(7)_{-0.0\%}^{+0.0\%}$
\\
& & {\footnotesize\quad $\Delta_{\mathrm{OS}}=-0.40\%$} & {\footnotesize\quad $\Delta_{\mathrm{OS}}=-0.02\%$}
  & {\footnotesize\quad $\Delta_{\mathrm{OS}}=-0.28\%$} & {\footnotesize\quad $\Delta_{\mathrm{OS}}=-0.02\%$}
\\
\hline
\MSbar& GIVS &
$0.85239(3)_{-0.5\%}^{+0.5\%}$ & $0.92160(7)_{-0.0\%}^{+0.0\%}$ &
$0.87087(3)_{-0.1\%}^{+0.1\%}$ & $0.94061(7)_{-0.0\%}^{+0.0\%}$
\\
& & {\footnotesize\quad $\Delta_{\mathrm{OS}}=-0.36\%$} & {\footnotesize\quad $\Delta_{\mathrm{OS}}=-0.02\%$}
  & {\footnotesize\quad $\Delta_{\mathrm{OS}}={\color{blue} -0.25}\%$} & {\footnotesize\quad $\Delta_{\mathrm{OS}}=-0.02\%$}
\\
\hline
\end{tabular}
}
\caption{LO and NLO decay widths $\Gamma^{\Ph\to4f}$[MeV]
of the light SESM Higgs boson $\Ph$ for various
SESM scenarios in different renormalization schemes,
with the OS scheme as input scheme (and full conversion
of the input parameters into the other schemes).
{The scale variation (given in percent as sub- and superscripts) corresponds to
the scales $\mu=\mu_0/2$ and $\mu=2\mu_0$ with central
scale $\mu_0=M_{\Ph}$.
The quantity $\Delta_{\mathrm{OS}}=\Gamma_{\MSbar}/\Gamma_{\mathrm{OS}}-1$ shows the
relative
difference between the $\MSbar$ predictions and the OS value; its spread 
illustrates the scheme dependence of the prediction at LO and NLO.}
}
\label{tab:SESM-H24f}
\end{table}
%
\begin{table}
\centerline{\renewcommand{\arraystretch}{1.25} \small \tabcolsep 3pt
\begin{tabular}{|c|c||l|l|l|l|}
\hline
&& \multicolumn{2}{c|}{BHM200$^+$} & \multicolumn{2}{c|}{BHM200$^-$}
\\
Ren.\ scheme & tadpoles & \multicolumn{1}{c|}{LO} & \multicolumn{1}{c|}{NLO}  & \multicolumn{1}{c|}{LO} & \multicolumn{1}{c|}{NLO}
\\
\hline
\hline
OS & &
$109.430(4)$ & $119.847(8)_{+0.0\%}^{-0.0\%}$ &
$109.430(4)$ & $119.812(8)_{+0.0\%}^{-0.0\%}$
\\
\hline
\MSbar& FJTS &
$134.126(4)_{-23.2\%}^{+30.1\%}$ & $120.403(9)_{-1.0\%}^{-7.1\%}$ &
$129.570(4)_{-6.0\%}^{+5.6\%}$   & $120.132(8)_{+0.6\%}^{-0.1\%}$
\\
& & {\footnotesize\quad $\Delta_{\mathrm{OS}}=+22.6\%$} & {\footnotesize\quad $\Delta_{\mathrm{OS}}=+0.5\%$}
  & {\footnotesize\quad $\Delta_{\mathrm{OS}}=+18.4\%$} & {\footnotesize\quad $\Delta_{\mathrm{OS}}=+0.3\%$}
\\
\hline
\MSbar& PRTS &
$118.976(4)_{+43.6\%}^{-31.9\%}$ & $120.240(7)_{-4.7\%}^{-5.9\%}$ &
$120.397(4)_{+44.6\%}^{-32.4\%}$ & $120.114(7)_{-5.6\%}^{-5.7\%}$
\\
& & {\footnotesize\quad $\Delta_{\mathrm{OS}}=+8.7\%$} & {\footnotesize\quad $\Delta_{\mathrm{OS}}={\color{blue}+0.3}\%$}
  & {\footnotesize\quad $\Delta_{\mathrm{OS}}=+10.0\%$} & {\footnotesize\quad $\Delta_{\mathrm{OS}}=+0.3\%$}
\\
\hline
\MSbar& GIVS &
$117.847(4)_{+44.7\%}^{-32.5\%}$ & $120.207(7)_{-4.6\%}^{-6.3\%}$ &
$118.947(4)_{+46.0\%}^{-33.1\%}$ & $120.093(7)_{-5.4\%}^{-6.3\%}$
\\
& & {\footnotesize\quad $\Delta_{\mathrm{OS}}=+7.7\%$} & {\footnotesize\quad $\Delta_{\mathrm{OS}}={\color{blue} +0.3}\%$}
  & {\footnotesize\quad $\Delta_{\mathrm{OS}}=+8.7\%$} & {\footnotesize\quad $\Delta_{\mathrm{OS}}=+0.2\%$}
\\
\hline
\multicolumn{5}{c}{}\\
\hline
&& \multicolumn{2}{c|}{BHM400} & \multicolumn{2}{c|}{BHM600}
\\
Ren.\ scheme & tadpoles & \multicolumn{1}{c|}{LO} & \multicolumn{1}{c|}{NLO}  & \multicolumn{1}{c|}{LO} & \multicolumn{1}{c|}{NLO}
\\
\hline
\hline
OS & &
$1533.42(4)$ & $1643.86(8)_{+0.0\%}^{-0.0\%}$ &
$4295.9(1)$ & $4532.4(2)_{+0.0\%}^{-0.0\%}$
\\
\hline
\MSbar& FJTS &
$1582.44(4)_{-21.7\%}^{+27.6\%}$ & $1646.83(8)_{-3.6\%}^{-1.5\%}$ &
$4007.1(1)_{-24.8\%}^{+32.5\%}$ & $4509.4(3)_{-6.0\%}^{-0.3\%}$
\\
& & {\footnotesize\quad $\Delta_{\mathrm{OS}}=+3.2\%$} & {\footnotesize\quad $\Delta_{\mathrm{OS}}=+0.2\%$}
  & {\footnotesize\quad $\Delta_{\mathrm{OS}}=-6.7\%$} & {\footnotesize\quad $\Delta_{\mathrm{OS}}=-0.5\%$}
\\
\hline
\MSbar& PRTS &
$1617.26(4)_{+6.3\%}^{-6.2\%}$ & $1648.62(8)_{+0.6\%}^{-0.6\%}$ &
$4530.1(1)_{+2.1\%}^{-2.3\%}$ & $4546.0(2)_{+0.6\%}^{-0.4\%}$
\\
& & {\footnotesize\quad $\Delta_{\mathrm{OS}}=+5.5\%$} & {\footnotesize\quad $\Delta_{\mathrm{OS}}=+0.3\%$}
  & {\footnotesize\quad $\Delta_{\mathrm{OS}}=+5.5\%$} & {\footnotesize\quad $\Delta_{\mathrm{OS}}=+0.3\%$}
\\
\hline
\MSbar& GIVS &
$1609.86(4)_{+6.7\%}^{-6.6\%}$ & $1648.26(8)_{+0.6\%}^{-0.7\%}$ &
$4511.4(1)_{+2.4\%}^{-2.6\%}$ & $4545.1(2)_{+0.6\%}^{-0.4\%}$ 
\\
& & {\footnotesize\quad $\Delta_{\mathrm{OS}}=+5.0\%$} & {\footnotesize\quad $\Delta_{\mathrm{OS}}=+0.3\%$}
  & {\footnotesize\quad $\Delta_{\mathrm{OS}}=+5.0\%$} & {\footnotesize\quad $\Delta_{\mathrm{OS}}=+0.3\%$}
\\
\hline
\end{tabular}
}
\caption{As in \refta{tab:SESM-H24f}, but for the decay width
$\Gamma^{\PH\to4f}$[MeV] of the heavy SESM Higgs boson $\PH$.}
\label{tab:SESM-H14f}
\end{table}%
The values for the widths in the OS and $\MSbar$ schemes with PRTS or FJTS tadpole treatment
are taken from Tabs.~4 and 5 of \citere{Denner:2018opp}, i.e.\ the $\MSbar$ results
with GIVS tadpole treatment extend the results of \citere{Denner:2018opp}.
For completeness, we mention that those results are based on the set of independent parameters
containing a running $\lambda_1$, not $\lambda_{12}$, but the scenarios are still defined
via the values for $\lambda_{12}$.
Interchanging the roles of $\lambda_1$ and $\lambda_{12}$ as running parameters
has only a marginal effect on the presented numbers.

Since the input values of the mixing angle $\alpha$ and the parameter $\lambda_{12}$
are properly converted from the OS scheme to the other $\MSbar$ schemes,
ideal predictions would not show any residual dependence on the renormalization 
scale $\mu$ and on the renormalization scheme.%
\footnote{In the OS scheme, there is a tiny scale dependence in the NLO
predictions resulting from the $\MSbar$ definition of the scalar coupling
$\lambda_{1}$, which formally enters beyond NLO.}
{An estimate for the scheme dependence is given by the spread of the
quantity $\Delta_{\mathrm{OS}}=\Gamma_{\MSbar}/\Gamma_{\mathrm{OS}}-1$, which shows the
relative difference between the $\MSbar$ predictions and the OS value for the integrated widths.}
The results show a significant reduction of the scheme dependence in the transition
from LO to NLO, with a residual scheme dependence mostly of the naively expected size
{of $<0.1\%$ for $\Ph\to4f$ and of $\lsim0.5\%$ for $\PH\to4f$.
Similar comments apply to the residual renormalization scale dependence of the
predictions in the $\MSbar$ schemes, which is, however, generally larger than the uncertainties
due to scheme dependence.}
The seemingly higher stability of the FJTS w.r.t.\ scale variations
in the BHM200$^-$ scenario is accidental, as can be seen from the more detailed
discussion of scale variations in \citere{Altenkamp:2018bcs}.
Systematic differences between the $\MSbar$ predictions based 
on the various tadpole
schemes are only visible for the decay width of the heavy H~boson with high
masses $\MH=400\GeV$ and $600\GeV$. The larger scale dependence of the FJTS numbers
reflect the onset of the potentially larger corrections in this scheme,
which were already mentioned in the introduction.
Such enhanced corrections in the FJTS 
are also expected from the different
decoupling properties of a heavy H~boson, depending on the tadpole scheme,
as worked out in \citere{Dittmaier:2021fls};%
\footnote{To be precise, the heavy field $H$ was integrated out in
\citere{Dittmaier:2021fls} in the non-linear representation of the Higgs
doublet, so that the PRTS of \citere{Dittmaier:2021fls} is identical to
the GIVS proposed in this paper. However, in that publication
only potential non-decoupling effects were considered, i.e.\ terms in the
effective Lagrangian that do not vanish for $\MH\to\infty$ 
(with $s_\alpha$ scaling like $1/\MH$), and in this order of the
$1/\MH$ expansion there is no difference between the PRTS and the GIVS
in the SESM anyhow.}
while decoupling is predicted in the $\MSbar$ scheme with PRTS tadpole treatment
and in the OS scheme, in the FJTS scheme the NLO corrections do not
go over into the SM prediction, but receive contributions enhanced
by terms $\propto\ln(\MH^2/\mu^2)$.
The overall picture of the
shown results, however, demonstrates perturbative stability of all schemes for the
considered scenarios,
with the residual scale dependence indicating a realistic estimate of missing
higher-order corrections. 
We, finally, emphasize that the new GIVS results are very close to
the earlier PRTS results from \citeres{Altenkamp:2018bcs,Denner:2018opp},
as already expected from the analytical results for the tadpole corrections.

\subsection{Two-Higgs-Doublet Model}

\subsubsection{Schemes for tadpole and vacuum expectation value renormalization}
\label{se:GIVschemeTHDM}

Several different renormalization procedures for the THDM
have been proposed in the 
literature~\cite{Kanemura:2004mg,Lopez-Val:2009xtx,Kanemura:2014dja,%
Krause:2016oke,Denner:2016etu,Denner:2017vms,Altenkamp:2017ldc,%
Denner:2018opp,Krause:2018wmo}, where
the most important differences concern the renormalization conditions
imposed on the two mixing angles $\alpha$ and $\beta$.
As in the SESM considered above,
\citere{Denner:2018opp} compares the strengths and weaknesses of the
different types of renormalizations, such as renormalization schemes for 
$\alpha$ and $\beta$ based on OS,
$\MSbar$, or symmetry-inspired conditions,
including in particular the $\MSbar$ renormalization variants with tadpoles
treated in the FJTS or PRTS (see also \citeres{Denner:2017vms,Altenkamp:2017ldc}).

The bare vev parameters $v_{n,0}$, which quantify the vevs
of the bare doublets 
\begin{align}\label{eq:THDMPhiB}
\Phi_{n,\bare} = \begin{pmatrix}
\phi_{n,\bare}^+ \\  \frac{1}{\sqrt{2}}(v_{n,0}+\eta_{n,\bare}+\ri\chi_{n,\bare})
\end{pmatrix}, \qquad n=1,2,
\end{align}
will again be specified in the context of tadpole renormalization. 
Inserting this field decomposition into the bare
Lagrangian ${\cal L}$, produces the tadpole contribution
$t_{\eta_{1,0}}\eta_1+t_{\eta_{2,0}}\eta_2$ 
in the Lagrangian ${\cal L}$ with
\begin{align}
t_{\eta_1,0}={}& - \frac{v_{1,0}}{2}\left(2m^2_{11,0} + \lambda_{1,0}v_{1,0}^2\right)
+ \frac{v_{2,0}}{2} (2m^2_{12,0}-\lambda_{345,0}v_{1,0} v_{2,0}),
\nonumber\\
t_{\eta_2,0}={}& - \frac{v_{2,0}}{2}\left(2m^2_{22,0} + \lambda_{2,0}v_{2,0}^2 \right)
+ \frac{v_{1,0}}{2} (2m^2_{12,0}-\lambda_{345,0} v_{1,0} v_{2,0}).
\label{eq:tn0THDM}
\end{align}
In all versions for tadpole renormalization considered here,
tadpole counterterms $\delta t_{\eta_1}\eta_1+\delta t_{\eta_2}\eta_2$
will be generated in the counterterm Lagrangian $\delta {\cal L}$
with tadpole renormalization constants $\de t_{\eta_n}$ which
cancel the explicit tadpole loop contributions
$T^{\eta_n}$ as expressed in Eq.~\refeq{eq:SESMtadCT1},
or in its equivalent form \refeq{eq:SESMtadCT3} in the
field basis $H$, $h$ with the transition given in Eq.~\refeq{eq:SESMtadCT2}.
As described for the SESM in \refse{se:SESMren} in detail, the PRTS, FJTS,
and GIVS generate the tadpole renormalization constants $\de t_{\eta_n}$ in different
ways out of the bare tadpole constants $t_{\eta_n,0}$ and possible field
redefinitions. The procedure in the THDM follows the same pattern as in the SESM,
so that we spell out 
only the most salient steps in the following.

\myparagraph{FJTS:}

In the FJTS, the bare tadpole constants $t_{\eta_n,0}$ are set to zero 
in Eq.~\refeq{eq:tn0THDM}.
The tadpole counterterms 
$\delta t_{\eta_1}\eta_1+\delta t_{\eta_2}\eta_2$
are generated by the field shifts given in Eq.~\refeq{eq:etanshift},
or equivalently by the shifts \refeq{eq:hHshift} in the field basis of $h$, $H$.
The constants $\Delta v_{1,2}^\FJTS$ and $\Delta v_{h/H}^\FJTS$ are again related to
each other as given in Eq.~\refeq{eq:Dv12FJTS} and related to the
tadpole renormalization constants $\delta t_{h/H}$ and the loop-induced tadpole 
constants $T^{h/H}$ as given in Eq.~\refeq{eq:DvH1H2FJTS}.

\myparagraph{PRTS:}

In the PRTS, the bare tadpole constants $t_{\eta_n,0}=t_{\eta_n}+\delta t_{\eta_n}^\PRTS$
are again split into the renormalized tadpole constants $t_{\eta_n}$,
which are demanded to vanish, and the tadpole renormalization constants
$\delta t_{\eta_n}$, which by virtue of Eq.~\refeq{eq:tn0THDM}
are given by
\begin{align}
\de t_{\eta_1}^{\PRTS}={}& - \frac{v_{1,0}}{2}\left(2m^2_{11,0} + \lambda_{1,0}v_{1,0}^2\right)
+ \frac{v_{2,0}}{2} (2m^2_{12,0}-\lambda_{345,0}v_{1,0} v_{2,0}) = -T^{\eta_1},
\nonumber\\
\de t_{\eta_2}^{\PRTS}={}& - \frac{v_{2,0}}{2}\left(2m^2_{22,0} + \lambda_{2,0}v_{2,0}^2\right)
+ \frac{v_{1,0}}{2} (2m^2_{12,0}-\lambda_{345,0} v_{1,0} v_{2,0}) = -T^{\eta_2}.
\label{eq:dtnTHDMPRTS}
\end{align}
Again, all tadpole terms translate into the $h$, $H$ field basis 
by the rotation with the angle $\alpha$ as specified in Eq.~\refeq{eq:SESMtadCT2}.

The bare vev parameters $v_{n,0}=v_n+\delta v_n$ are decomposed into
renormalized parameters $v_n$ and renormalization constants as usual.
The vev parameters $v_n$ are translated into the parameters
$v$ and $\beta$ which are closely related to input parameters,
\begin{align}
v_{1,0}^2 + v_{2,0}^2 ={}& v_0^2 = \frac{4M_{\PW,0}^2}{g_{2,0}^2} 
= \frac{4M_{\PW,0}^2s_{\rw,0}^2}{e_0^2},  &
\tan\beta_0 ={}& t_{\beta,0} = \frac{v_{2,0}}{v_{1,0}},
\label{eq:v0THDM}
\\
v_{1}^2 + v_{2}^2 ={}& v^2 = \frac{4\MW^2}{g_{2}^2} = \frac{4\MW^2\sw^2}{e^2},  &
\tan\beta ={}& t_{\beta} = \frac{v_{2}}{v_{1}}.
\end{align}
The renormalization constants $\delta v_n$ are, thus, tied to the renormalization constants
$\delta\MW^2$, $\delta\sw$, $\delta Z_e$, and $\delta\beta$, which are directly fixed by
renormalization conditions of the respective parameters.

Equations \refeq{eq:dtnTHDMPRTS} and \refeq{eq:v0THDM} represent four independent
relations between the eight bare original parameters of the Higgs potential 
$m_{11,0}^2$, $m_{22,0}^2$, $m_{12,0}^2$,
$\lambda_{1,0}$, \dots, $\lambda_{5,0}$,
and the parameters that serve as phenomenological input of the THDM.
For the latter, we follow \citeres{Altenkamp:2017ldc,Altenkamp:2017kxk,Denner:2018opp}
and take the parameters
$v_0=2M_{\PW,0}/g_{2,0}$, $\alpha_0$, $\beta_0$,
$M_{\Ph,0}$, $M_{\PH,0}$, $M_{\PA,0}$, $M_{\PHpm,0}$, and
$\lambda_{5,0}$.
The four missing relations follow from the diagonalization of the mass matrices
in the Higgs sector and can, e.g., be written as
\begin{align}
M_{\Ph,0}^2={}& \frac{2c_{\alpha - \beta,0}^2}{s_{2 \beta,0}} m^2_{12,0}
+ \frac{v_0^2}{2} \left( 2\lambda_{1,0} c_{\beta,0}^2 s_{\alpha,0}^2 
+2\lambda_{2,0} c_{\alpha,0}^2 s_{\beta,0}^2-s_{2 \alpha,0} s_{2\beta,0} \lambda_{345,0}\right)
\nonumber\\
& {}
+t_{\eta_1,0}\frac{s_{\alpha,0}^2}{v_0 c_{\beta,0}}
+t_{\eta_2,0}\frac{c_{\alpha,0}^2}{v_0 s_{\beta,0}},
\nonumber\\
M_{\PH,0}^2={}& \frac{2s_{\alpha-\beta,0}^2}{s_{2\beta,0}}  m^2_{12,0}
+ \frac{v_0^2}{2} \left( 2\lambda_{1,0} c_{\beta,0}^2 c_{\alpha,0}^2 
+2\lambda_{2,0} s_{\alpha,0}^2 s_{\beta,0}^2 +s_{2 \alpha,0} s_{2\beta,0} \lambda_{345,0}\right)
\nonumber\\
& {}
+t_{\eta_1,0}\frac{c_{\alpha,0}^2}{v_0 c_{\beta,0}}
+t_{\eta_2,0}\frac{s_{\alpha,0}^2}{v_0 s_{\beta,0}},
\nonumber\\
M_{\PA,0}^2  ={} & \frac{2m^2_{12,0}}{s_{2\beta,0}}- \lambda_{5,0} v_0^2
+ t_{\eta_1,0} \frac{s_{\beta,0}^2}{v_0 c_{\beta,0}}
+ t_{\eta_2,0} \frac{c_{\beta,0}^2}{v_0 s_{\beta,0}},
\nonumber\\
M_{\PHpm,0}^2 ={} & \frac{2m^2_{12,0}}{s_{2\beta,0}}-\frac{v_0^2}{2}(\lambda_{4,0}+\lambda_{5,0})
+ t_{\eta_1,0} \frac{s_{\beta,0}^2}{v_0 c_{\beta,0}}
+ t_{\eta_2,0} \frac{c_{\beta,0}^2}{v_0 s_{\beta,0}},
\end{align}
which can be directly read from Eqs.~(2.16a/b), (2.17a), and (2.18a) of
\citere{Altenkamp:2017ldc}.
These eight bare relations carry over to eight renormalized relations upon replacing the
bare quantities by the respective renormalized ones and setting all tadpole terms to zero.
The bare and renormalized relations then determine the renormalization constants
$\delta m_{11}^2$, $\delta m_{22}^2$, $\delta m_{12}^2$,
$\delta \lambda_{1}$, \dots, $\delta \lambda_{5}$ of the original parameters
in terms of the renormalization constants
$\delta v$, $\delta \alpha$, $\delta \beta$,
$\delta \Mh$, $\delta \MH$, $\delta \MA$, $\delta \MHpm$, $\delta \lambda_{5}$.
From the tadpole renormalization constants contained in these relations,
we can, for instance, read off the substitutions
\begin{align}
m_{11,0}^2 &{} \;\to\; m_{11,0}^2 
- \frac{(9 + 4 c_{2 \beta} - c_{4 \beta})\de t^\PRTS_{\eta_1}}{8c_{\beta} v} 
+ \frac{ c_{\beta}^2 s_{\beta}\de t^\PRTS_{\eta_2}}{v},
\nonumber\\
m_{22,0}^2 &{} \;\to\; m_{22,0}^2 
+ \frac{c_{\beta} s_{\beta}^2\de t^\PRTS_{\eta_1}}{v} 
- \frac{ (9 - 4 c_{2 \beta} - c_{4 \beta}) \de t^\PRTS_{\eta_2}}{8s_{\beta} v} ,
\nonumber\\
m_{12,0}^2 &{} \;\to\; m_{12,0}^2 
+ \frac{s_{\beta}^3 \de t^\PRTS_{\eta_1}}{v} 
+ \frac{ c_{\beta}^3 \de t^\PRTS_{\eta_2}}{v},
\nonumber\\
\lambda_{1,0} &{} \;\to\; \lambda_{1,0} 
+ \frac{ (3 - c_{2 \beta}) \de t^\PRTS_{\eta_1}}{2c_{\beta} v^3} 
- \frac{s_{\beta}\de t^\PRTS_{\eta_2}}{v^3 } ,
\nonumber\\
\lambda_{2,0} &{} \;\to\; \lambda_{2,0} 
- \frac{c_{\beta}\de t^\PRTS_{\eta_1}}{v^3 } 
+ \frac{  (3 + c_{2 \beta}) \de t^\PRTS_{\eta_2}}{2s_{\beta} v^3},
\nonumber\\
\lambda_{3,0} &{} \;\to\; \lambda_{3,0} 
+ \frac{s_{\beta}^2 \de t^\PRTS_{\eta_1}}{c_{\beta}v^3} 
+ \frac{ c_{\beta}^2 \de t^\PRTS_{\eta_2}}{s_{\beta}v^3 },
\qquad
\lambda_{4,0} {} \;\to\; \lambda_{4,0}, \qquad
\lambda_{5,0} {} \;\to\; \lambda_{5,0},
\label{eq:THDMdtPTRSgeneration}
\end{align}
which generate the PRTS
tadpole counterterms out of the bare Lagrangian with vanishing
tadpole terms.

Again, owing to the gauge dependences in the
relations among bare input parameters and the bare original parameters
of the Higgs sector, the PRTS
leads to a gauge-dependent relation between predictions for
observables and input parameters if $\MSbar$-renormalized masses or
$\MSbar$-renormalized Higgs mixing angles $\alpha$, $\beta$ are used.

\myparagraph{GIVS:}

Formally, the construction of the GIVS tadpole renormalization constants for the
fields $H$, $h$ follows exactly the same pattern as in the SESM, described in the
previous section, i.e.\
Eqs.~\refeq{eq:dtPRTSnlSESM}--\refeq{eq:dtGIVS2SESM} and \refeq{eq:dtGIVSSESM} literally apply in the THDM
as well.
Again, the gauge-independent constants
$\delta t^\GIVS_{H,1}$ and $\delta t^\GIVS_{h,1}$ enter relations between the bare
parameters of the Higgs potential, 
the gauge-dependent parts $\delta t^\GIVS_{H,2}$ and $\delta t^\GIVS_{h,2}$
have no effects on predictions for observables,
and the full tadpole constants $\delta t^\GIVS_{H}$ and $\delta t^\GIVS_{h}$
exactly cancel all explicit tadpole diagrams by construction.
Since the rotation \refeq{eq:RalphaSESM}
between the $\eta_1$, $\eta_2$ and $H$, $h$ field bases is
formally the same in the SESM and THDM, the translation
\refeq{eq:dtGIVSTHDM} of the tadpole constants into the $\eta_1$, $\eta_2$
basis holds in the THDM 
without change.
In the determination of the renormalization constants of the mixing angles
$\alpha$ and $\beta$ below, the explicit results 
for $\delta t^\GIVS_{H,2}$ and $\delta t^\GIVS_{h,2}$ will be very useful,
\begin{align}
\Delta v_H^\GIVS ={}& -\frac{\delta t^\GIVS_{H,2}}{\MH^2} 
= \frac{T^H-T^H_{\mathrm{nl}}}{\MH^2} = c_{\beta-\alpha} \Delta v_\xi,
\nonumber\\
\Delta v_h^\GIVS ={}& -\frac{\delta t^\GIVS_{h,2}}{\Mh^2}
= \frac{T^h-T^h_{\mathrm{nl}}}{\Mh^2} = s_{\beta-\alpha} \Delta v_\xi.
\label{eq:DvhHGIVS}
\end{align}
To derive these relations, the tree-level relations between the Higgs masses
$\MH$, $\Mh$ and the original parameters of the THDM Higgs potential
have been used.

What of course changes in the THDM w.r.t.\ SESM, is the generation of tadpole
counterterms from the bare Lagrangian because of the completely different form
of the Higgs potential. 
The generation of the counterterms connected with $\de t^\GIVS_{\eta_n,1}$
follow from the ones for $\de t^\PRTS_{\eta_n}$, as given in
Eq.~\refeq{eq:THDMdtPTRSgeneration}, while the $\de t^\GIVS_{\eta_n,2}$ again
originate from field shifts,
\begin{align}
m_{11,0}^2 &{} \;\to\; m_{11,0}^2 
- \frac{(9 + 4 c_{2 \beta} - c_{4 \beta})\de t^\GIVS_{\eta_1,1}}{8c_{\beta} v} 
+ \frac{ c_{\beta}^2 s_{\beta}\de t^\GIVS_{\eta_2,1}}{v},
\nonumber\\
m_{22,0}^2 &{} \;\to\; m_{22,0}^2 
+ \frac{c_{\beta} s_{\beta}^2\de t^\GIVS_{\eta_1,1}}{v} 
- \frac{ (9 - 4 c_{2 \beta} - c_{4 \beta}) \de t^\GIVS_{\eta_2,1}}{8s_{\beta} v} ,
\nonumber\\
m_{12,0}^2 &{} \;\to\; m_{12,0}^2 
+ \frac{s_{\beta}^3 \de t^\GIVS_{\eta_1,1}}{v} 
+ \frac{ c_{\beta}^3 \de t^\GIVS_{\eta_2,1}}{v},
\nonumber\\
\lambda_{1,0} &{} \;\to\; \lambda_{1,0} 
+ \frac{ (3 - c_{2 \beta}) \de t^\GIVS_{\eta_1,1}}{2c_{\beta} v^3} 
- \frac{s_{\beta}\de t^\GIVS_{\eta_2,1}}{v^3 } ,
\nonumber\\
\lambda_{2,0} &{} \;\to\; \lambda_{2,0} 
- \frac{c_{\beta}\de t^\GIVS_{\eta_1,1}}{v^3 } 
+ \frac{  (3 + c_{2 \beta}) \de t^\GIVS_{\eta_2,1}}{2s_{\beta} v^3},
\nonumber\\
\lambda_{3,0} &{} \;\to\; \lambda_{3,0} 
+ \frac{s_{\beta}^2 \de t^\GIVS_{\eta_1,1}}{c_{\beta}v^3} 
+ \frac{ c_{\beta}^2 \de t^\GIVS_{\eta_2,1}}{s_{\beta}v^3 },
\qquad
\lambda_{4,0} {} \;\to\; \lambda_{4,0}, \qquad
\lambda_{5,0} {} \;\to\; \lambda_{5,0},
\nonumber\\
\eta_{1,\bare}&{} \;\to\eta_{1,\bare} + \De v_1^\GIVS, 
\qquad
\eta_{2,\bare} \;\to\eta_{2,\bare} + \De v_2^\GIVS.
\label{eq:THDMdtGIVSgeneration}
\end{align}

\subsubsection{Mixing-angle renormalization}

The $\MSbar$ renormalization constant $\de\al_{\MSbar}=\alpha_0-\alpha$ of $\alpha$
can be determined as in Eq.~\refeq{eq:alphaMSren} from the
OS field renormalization constants $\de Z_{ij}$ of the $h/H$ system exactly
as in the SESM. As a result, Eq.~\refeq{eq:datadSESM} holds in the THDM
as well. Inserting all tadpole loop and counterterm contributions into
the $hH$~self-energy in Eq.~\refeq{eq:datadSESM}, we get
%
\begin{align}
\de\al^\FJTS_{\MSbar,\mathrm{tad}}={}& 
\frac{e(C_{hhH}\Delta v_h^\FJTS+C_{hHH}\Delta v_H^\FJTS)}{\MH^2-\Mh^2} \bigg|_{\UV},
\label{eq:dalphatadFJTSTHDM}
\\
\de\al^\PRTS_{\MSbar,\mathrm{tad}}={}& 0,
\label{eq:dalphatadPRTSTHDM}
\\
\de\al^\GIVS_{\MSbar,\mathrm{tad}}={}& 
\frac{e(s_{\beta-\alpha}C_{hhH}
+c_{\beta-\alpha}C_{hHH})\Delta v_\xi}{\MH^2-\Mh^2} \bigg|_{\UV},
\label{eq:dalphatadGIVSTHDM}
\end{align}
with the shorthands
\begin{align}
C_{hhH} ={}& 
\frac{c_{\beta-\alpha}}{e s_{2\beta}v}
\left[ (3s_{2\alpha}-s_{2\beta})\left( \MA^2+v^2\lambda_5\right)
        -s_{2\alpha}(2\Mh^2+\MH^2) \right],
\nonumber\\
C_{hHH} ={}& 
\frac{s_{\beta-\alpha}}{e s_{2\beta}v}
\left[ -(3s_{2\alpha}+s_{2\beta})\left( \MA^2+v^2\lambda_5\right)
        +s_{2\alpha}(\Mh^2+2\MH^2) \right]
\end{align}
for the scalar self-couplings $e(C_{hhH}h^2 H+C_{hHH}h H^2)/2$ in the Lagrangian.

The $\MSbar$ renormalization constant 
$\de\be_{\MSbar}=\beta_0-\beta$ of $\beta$
can be obtained according to
\begin{align}
\de\be_{\MSbar}={}& 
\frac{1}{4}\left( \delta Z_{G_0A_0}-\delta Z_{A_0G_0} \right) \Big|_{\UV},
\end{align}
from the field renormalization constants $\de Z_{ij}$ of the 
$G_0/A_0$ system defined by
\begin{align}
\begin{pmatrix} G_{0,\bare} \\ A_{0,\bare} \end{pmatrix} = 
\begin{pmatrix} 
1 + \frac{1}{2}\de Z_{G_0 G_0} &   \frac{1}{2}\de Z_{G_0 A_0}\\
         \frac{1}{2}\de Z_{A_0 G_0} &  1+\frac{1}{2}\de Z_{A_0 A_0}
\end{pmatrix}
\begin{pmatrix} G_0 \\ A_0 \end{pmatrix}.
\end{align}
For field renormalization constants $\de Z_{ij}$ relevant for 
external physical A$_0$ bosons, OS renormalization conditions for momentum
transfer $p^2=\MHa^2$ can be chosen as usual; 
for the Goldstone field 
$G_0$ OS conditions may be taken
for any virtuality $p^2=M^2$ in order to absorb UV divergences
(without changing the final result).
Taking $M=0$, we, thus, can define
\begin{align}
\delta Z_{A_0 G_0} = \frac{2\Sigma^{A_0 G_0}(0)}{M_{\PHa}^2}, \qquad
\delta Z_{G_0 A_0} = -\frac{2\Re\{\Sigma^{A_0 G_0}(M_{\PHa}^2)\}}{M_{\PHa}^2}.
\end{align}
The tadpole contributions to $\de\be_{\MSbar}$ are then given by
\begin{align}
\de\be_{\MSbar,\mathrm{tad}}={}& 
-\frac{\Sigma^{A_0 G_0}_{\mathrm{tad}}}{M_{\PHa}^2}
\bigg|_{\UV},
\label{eq:dbtadTHDM}
\end{align}
where $\Sigma^{A_0 G_0}_{\mathrm{tad}}$ is the sum of all
(momentum-independent) explicit tadpole diagrams and tadpole counterterms to 
the self-energy $\Sigma^{A_0 G_0}(p^2)$.
Inserting all tadpole counterterms contributing to the 
$A_0 G_0$~mixing self-energy in the various tadpole schemes into 
Eq.~\refeq{eq:dbtadTHDM}, we obtain
\begin{align}
\de\be_{\MSbar,\mathrm{tad}}^\FJTS ={}& 
\biggl[c_{\beta-\alpha}\frac{\Delta v_h^\FJTS}{v} \biggl(\frac{\Mh^2}{M_{\PHa}^2}-1\biggr)
-s_{\beta-\alpha}\frac{\Delta v_H^\FJTS}{v}  \biggl(\frac{\MH^2}{M_{\PHa}^2}-1\biggr)
\biggr]\bigg|_{\UV},
\label{eq:dbetatadFJTSTHDM}
\\
\de\be_{\MSbar,\mathrm{tad}}^\PRTS ={}& 
\biggl(-c_{\beta-\alpha}\frac{\de t^\PRTS_h}{v M_{\PHa}^2}
+\frac{s_{\beta-\alpha} \de t^\PRTS_H}{v M_{\PHa}^2} 
\biggr)\bigg|_{\UV},
\label{eq:dbetatadPRTSTHDM}
\\
\de\be_{\MSbar,\mathrm{tad}}^\GIVS ={}&
\biggl(-c_{\beta-\alpha}\frac{\de t^\GIVS_{h}}{v M_{\PHa}^2}
+\frac{s_{\beta-\alpha} \de t^\GIVS_{H}}{v M_{\PHa}^2} 
\biggr)\bigg|_{\UV}
+\biggl(-c_{\beta-\alpha}\frac{\Delta v_h^\GIVS}{v} 
+s_{\beta-\alpha}\frac{\Delta v_H^\GIVS}{v}
\biggr)\bigg|_{\UV}
\label{eq:dbetatadGIVSTHDM}
\nonumber\\
={}& \de\be_{\MSbar,\mathrm{tad}}^\PRTS,
\end{align}
where Eq.~\refeq{eq:DvhHGIVS} was used in the last equation.
The agreement of $\de\be_{\MSbar,\mathrm{tad}}$ in the PRTS and GIVS,
which we have established in the class of $R_\xi$ gauges,
is particularly interesting. It means that the $\MSbar$ 
renormalizations of $\beta$ in the PRTS and GIVS 
in $R_\xi$ gauges are the same, at least at the one-loop level. 
Thus, all results obtained with 
$\MSbar$-renormalized $\beta$ a la PRTS in $R_\xi$ gauges can be 
reinterpreted as
results obtained in the GIVS, which is gauge independent.
It would be interesting to investigate in how far this 
statement holds beyond the one-loop level as well, but this task is
beyond the scope of this paper.
According to the arguments given in App.~D of \citere{Denner:2016etu},
where it is shown that $\de\be_{\MSbar}^\PRTS$ in fact depends on the gauge
in general,
the equivalence of $\be_{\MSbar}^\PRTS$ and $\be_{\MSbar}^\GIVS$ should, however,
not hold outside the class of $R_\xi$ gauges.

Analogously to the situation in the SESM discussed in \refse{se:SESMmixingangle},
and as discussed in 
\citeres{Denner:2016etu,Altenkamp:2017ldc,Denner:2017vms,Denner:2018opp},
changing the tadpole treatment in the $\MSbar$ renormalization of the mixing angles~$\alpha$ 
and $\beta$ changes the physical meaning of $\alpha$ and $\beta$, similar to a change in their
renormalization conditions. 
The corresponding finite shifts in the numerical
values of the renormalized parameters~$\alpha^{\TS}_{\MSbar}$ and $\beta^{\TS}_{\MSbar}$ 
when the tadpole scheme TS is changed from the PRTS to the FJTS
were calculated in \citeres{Denner:2016etu,Denner:2017vms} by considering NLO
corrections to specific vertices.
In \citere{Altenkamp:2017ldc}
arguments for a universal relation between 
($\alpha^\PRTS_{\MSbar},\beta^\PRTS_{\MSbar})$ 
and ($\alpha^\FJTS_{\MSbar},\beta^\FJTS_{\MSbar}$)
were given, based on the observation that
the FJTS is equivalent to a procedure in which no tadpole renormalization
is performed at all. 
In the following we proceed as in \refse{se:SESMmixingangle} for the SESM
and derive the general relation between the input parameters
of the THDM when changing the tadpole scheme, again basing the arguments on the
equivalence of the FJTS and performing no tadpole renormalization.
As in the SESM, the tadpole renormalization constants entering the relations
between the bare input parameters can be obtained via the substitutions
\refeq{eq:THDMdtPTRSgeneration}, which generate the PRTS tadpole counterterms in
the bare Lagrangian without tadpole terms as follows,
\begin{align}
\alpha_{0}^\FJTS &{} \;=\; 
\alpha_{0}\Big|_{\footnotesize\refeq{eq:THDMdtPTRSgeneration},\, p_0\to p_0^\PRTS},
\qquad
\beta_{0}^\FJTS {} \;=\; 
\beta_{0}\Big|_{\footnotesize\refeq{eq:THDMdtPTRSgeneration},\, p_0\to p_0^\PRTS},
\end{align}
which means that the substitution \refeq{eq:THDMdtPTRSgeneration} is applied
to $\alpha_0$ and $\beta_0$ expressed in terms of the bare parameters 
$\{p_0\}$ = $\{m_{11,0}^2$, $m_{22,0}^2$, $m_{12,0}^2$,
$\lambda_{1,0}$, \dots, $\lambda_{5,0}\}$
in the absence of tadpole terms.
Note that there are no
shifts in the OS-renormalized parameters for
$v$, $\Mh$, $\MH$, $\MA$, and $\MHpm$.
Owing to the last relation of Eq.~\refeq{eq:THDMdtPTRSgeneration},
there is no shift in the $\MSbar$-renormalized $\lambda_{5}$ either.
For $\alpha$ and $\beta$, however, Eq.~\refeq{eq:THDMdtPTRSgeneration}
implies the one-loop relations
\begin{align}
\alpha_{0}^\PRTS &{} \;=\; \alpha_{0}^\FJTS
+ \frac{e}{\MH^2-\Mh^2} \left( C_{hhH}\,\frac{\delta t_h^\FJTS}{\Mh^2} 
+ C_{hHH}\,\frac{\delta t_H^\FJTS}{\MH^2} \right),
\label{eq:THDMalphaconvPRTSFJTS}
\\
\be^\PRTS_0 &{} \;=\;  \be^\FJTS_0 +
\frac{1}{v} \left( -c_{\beta-\alpha}\frac{\delta t_h^\FJTS}{\Mh^2}
+s_{\beta-\alpha}\frac{\delta t_H^\FJTS}{\MH^2} \right)
\label{eq:THDMbetaconvPRTSFJTS},
\end{align}
the derivation of which is again somewhat cumbersome but straightforward.
The relations between the two renormalized mixing angles in the PRTS and FJTS
follows from the renormalization transformations
\begin{align}
\alpha_{0}^\TS = \alpha^\TS + \delta\alpha^\TS, \qquad
\beta_{0}^\TS = \beta^\TS + \delta\beta^\TS
\end{align}
with TS being the PRTS or FJTS.
The resulting shifts between the $\MSbar$-renormalized renormalized parameters in the two tadpole
schemes read
\begin{align}
\alpha^\PRTS_{\MSbar} - \alpha^\FJTS_{\MSbar} ={}&
\left(\alpha_0^\PRTS - \alpha_0^\FJTS \right)
-\left(\delta\alpha^\PRTS_{\MSbar} - \delta\alpha^\FJTS_{\MSbar} \right),
\\
\beta^\PRTS_{\MSbar} - \beta^\FJTS_{\MSbar} ={}&
\left(\beta_0^\PRTS - \beta_0^\FJTS \right)
-\left(\delta\beta^\PRTS_{\MSbar} - \delta\beta^\FJTS_{\MSbar} \right).
\end{align}
The differences between the bare mixing angles can be read from 
Eqs.~\refeq{eq:THDMalphaconvPRTSFJTS} and \refeq{eq:THDMbetaconvPRTSFJTS},
and the differences between the corresponding renormalization constants
$\delta\alpha^\TS_{\MSbar}$ and $\delta\beta^\TS_{\MSbar}$ again receive only
contributions from the corresponding tadpole contributions, which can be read
from Eqs.~\refeq{eq:dalphatadFJTSTHDM}, \refeq{eq:dalphatadPRTSTHDM},
\refeq{eq:dbetatadFJTSTHDM}, and \refeq{eq:dbetatadPRTSTHDM}.
Combining all those parts, we get
\begin{align}
\alpha_{\MSbar}^\PRTS - \alpha_{\MSbar}^\FJTS ={}&
- \frac{e}{\MH^2-\Mh^2} \left( C_{hhH}\,\frac{T^h}{\Mh^2} 
+ C_{hHH}\,\frac{T^H}{\MH^2} \right) \bigg|_{\mathrm{finite}},
\\
\be^\PRTS_{\MSbar}  - \be^\FJTS_{\MSbar} ={}&
\frac{1}{v} \left( c_{\beta-\alpha}\frac{T^h}{\Mh^2}
-s_{\beta-\alpha}\frac{T^H}{\MH^2} \right) \bigg|_{\mathrm{finite}}.
\label{eq:THDMbetaconvPRTSFJTS3}
\end{align}

As already explained in \refse{se:SESMmixingangle} for the SESM in more detail,
the conversion between the FJTS and the GIVS proceeds analogously, with
the GIVS constants $\delta t^\GIVS_{X,1}$ ($X=\eta_1,\eta_2,H,h$) playing the
roles of $\delta t^\PRTS_X$ in the PRTS.
The final results for the conversion are
\begin{align}
\alpha_{\MSbar}^\GIVS - \alpha_{\MSbar}^\FJTS ={}&
- \frac{e}{\MH^2-\Mh^2} \left( C_{hhH}\,\frac{T^h_{\mathrm{nl}}}{\Mh^2} 
+ C_{hHH}\,\frac{T^H_{\mathrm{nl}}}{\MH^2} \right) \bigg|_{\mathrm{finite}},
\\
\be^\GIVS_{\MSbar}  - \be^\FJTS_{\MSbar} ={}&
\frac{1}{v} \left( c_{\beta-\alpha}\frac{T^h_{\mathrm{nl}}}{\Mh^2}
-s_{\beta-\alpha}\frac{T^H_{\mathrm{nl}}}{\MH^2} \right) \bigg|_{\mathrm{finite}},
\label{eq:THDMbetaconvGIVSFJTS3}
\end{align}
which have the same form as the conversion between PRTS and FJTS with the only difference
that the tadpole contributions are calculated in the
non-linear Higgs representation. 
Owing to the gauge independence of the tadpole constants in the non-linear
representation, the relation between
($\alpha^\GIVS_{\MSbar},\beta^\GIVS_{\MSbar}$)
and ($\alpha^\FJTS_{\MSbar},\beta^\FJTS_{\MSbar}$)
is gauge independent.
Note that Eqs.~\refeq{eq:THDMbetaconvPRTSFJTS3} and \refeq{eq:THDMbetaconvGIVSFJTS3}
together with Eq.~\refeq{eq:DvhHGIVS} imply the notable relation
\begin{align}
\be^\GIVS_{\MSbar} = \be^\PRTS_{\MSbar},
\label{eq:betaTHDM_GIVS=PRTS}
\end{align}
i.e.\ the GIVS and PRTS $\MSbar$ renormalizations of $\beta$ coincide
in the class of $R_\xi$ gauges
(at least at the one-loop level).

The equality \refeq{eq:betaTHDM_GIVS=PRTS} can be used to put some 
phenomenological predictions based on $\MSbar$-renormalized angles $\beta$
in the PRTS on more solid theoretical ground. If the $R_\xi$~gauge is 
employed in those predictions, as most frequently done, the PRTS results
can be reinterpreted as GIVS results, which are gauge independent, at least as long
as no other sources of gauge dependences exist (as, e.g., from the renormalization of
$\alpha$ in the THDM). This arguments, for instance, holds for a scheme proposed and 
used in \citeres{Bredenstein:2006rh,Bredenstein:2006ha} for the THDM, as discussed in the 
next section in more detail.
Note that the argument is also applicable to many higher-order calculations
within the Minimal Supersymmetric extension of the SM (MSSM) which contains a THDM
of Type~II as Higgs sector. The corresponding parameter $\tan\beta$ is very often
renormalized in the $\MSbar$ scheme with PRTS tadpole treatment, and the $R_\xi$~gauge
is most often used in higher-order calculations,
see for example \citeres{Freitas:2002um, Frank:2006yh, Degrassi:2014pfa, Fritzsche:2013fta},
where this renormalization condition is applied in Higgs-mass-spectrum calculations 
or other higher-order calculations in the MSSM. 
In those applications
the $\MSbar$ PRTS $\tan \beta$ renormalization is found to be a convenient scheme 
leading to perturbatively well-behaved results.
The NLO results obtained in this way can, thus,
be interpreted as gauge-independent results obtained within the GIVS. 
We even expect that this statement carries over to predictions beyond the NLO level,
but the investigation of this conjecture is beyond the scope of this paper.

\subsubsection[NLO decay widths for $\Ph\to4f$ in the THDM]%
{\boldmath{NLO decay widths for $\Ph\to4f$ in the THDM}}
\label{se:THDMnumerics}

In this section we extend the discussion of NLO predictions for the
Higgs decay processes $\Ph\to\PW\PW/\PZ\PZ\to4f$ in the THDM started in 
\citeres{Altenkamp:2017ldc,Altenkamp:2017kxk,Denner:2018opp} with the
Monte Carlo program 
\Prophecy~3.0~\cite{Bredenstein:2006rh,Bredenstein:2006ha,Denner:2019fcr}.
To this end, we have implemented the GIVS tadpole treatment as new
option for $\MSbar$ renormalization schemes of the THDM in \Prophecy.

In detail, in \citeres{Altenkamp:2017ldc,Altenkamp:2017kxk} two types of
$\MSbar$ renormalization schemes for the mixing angles $\alpha$ and $\beta$ 
were considered, both with PRTS and FJTS tadpole treatments:
one scheme with $\MSbar$-renormalized angles $\alpha$, $\beta$,
and another scheme in which the coupling constant $\lambda_3$ was taken as independent
variable instead of the angle $\alpha$.
According to Eq.~\refeq{eq:betaTHDM_GIVS=PRTS} of the previous section,
there is no difference between the PRTS (in the $R_\xi$~gauge) and the GIVS tadpole treatment
in the renormalization of $\beta$, so that the $\MSbar$
scheme with input $(\lambda_3,\beta)$, which was called 
``$\MSbar(\lambda_3)$'' in \citeres{Altenkamp:2017ldc,Altenkamp:2017kxk},
is the same in the PRTS and GIVS tadpole variants.
This, in particular, means that the $\MSbar(\lambda_3)$ results of 
\citeres{Altenkamp:2017ldc,Altenkamp:2017kxk} are fully gauge independent when
interpreted as results obtained in the GIVS for the tadpoles.
Ref.~\cite{Denner:2019fcr} extended the discussion 
of \citeres{Altenkamp:2017ldc,Altenkamp:2017kxk}
by including
renormalization schemes based on OS or symmetry-inspired renormalization conditions
for $\alpha$ and $\beta$.

The numerical input for the scenarios called A1,%
\footnote{A1 is called Aa in \citeres{Altenkamp:2017ldc,Altenkamp:2017kxk}.}
A2, B1, B2
as well as the details of the calculational setup
can be found in \citeres{Altenkamp:2017ldc,Altenkamp:2017kxk,Denner:2018opp}.
The mass of the lighter Higgs boson is set to $\Mh=125\GeV$ in all 
considered scenarios, and the mixing angles are varied in their
empirically allowed range with central values in the vicinity of
$|c_{\beta-\alpha}|=|\cos(\beta-\alpha)|=0.1$.
The scenarios A1 and A2 are ``low-mass scenarios'' with
$\MH=300\GeV$ and $\MHa = \MHpm = 460\GeV$;
B1 and B2 are ``high-mass scenarios'' with
$\MH=600\GeV$ and $\MHa = \MHpm = 690\GeV$.
Results of the scheme conversion of the input values for the $\MSbar$ 
parameters $c_{\alpha\beta}=c_{\beta-\alpha}$
and $\beta$,
are given in \refapp{app:THDM}.
The following results for $\Ph\to4f$
are given for a Type~I THDM, but they hardly change
when going over to Types~II, ``lepton-specific'', or ``flipped'',
as shown in \citeres{Altenkamp:2017ldc,Altenkamp:2017kxk}.

Table~\ref{tab:THDM-H24f} shows NLO predictions for
the partial decay width of $\Ph\to4f$ for the considered scenarios
based on the various $\MSbar$ schemes using the OS-type scheme
OS12 as reference scheme in which the input parameters are defined.
\begin{table}
\centerline{\renewcommand{\arraystretch}{1.25} \small \tabcolsep 2pt
\begin{tabular}{|c|c||l|l|l|l|}
\hline
 & & \multicolumn{2}{c|}{A1} & \multicolumn{2}{c|}{A2} 
\\
Ren.\ scheme & tadpoles &
\multicolumn{1}{c|}{LO} & \multicolumn{1}{c|}{NLO}  & \multicolumn{1}{c|}{LO} & \multicolumn{1}{c|}{NLO}  
\\
\hline
\hline
OS12\,($\alpha,\beta$) & &
$0.89832(3)$ & $0.96194(7)_{+0.1\%}^{-0.1\%}$ &
$0.87110(3)$ & $0.92947(7)_{+0.1\%}^{-0.2\%}$
\\
\hline
\MSbar($\alpha,\beta$) & FJTS &
$0.89996(3)_{-7.4\%}^{+0.7\%}$ & $0.96283(7)_{-0.2\%}^{+0.8\%}$ &
$0.88508(3)_{-10.0\%}^{+2.2\%}$ & $0.93604(7)_{-11.0\%}^{+3.1\%}$ 
\\
& & {\footnotesize\;$\Delta_{\mathrm{OS12}}=+0.2\%$} & {\footnotesize\;$\Delta_{\mathrm{OS12}}=+0.1\%$}
  & {\footnotesize\;$\Delta_{\mathrm{OS12}}=+1.6\%$} & {\footnotesize$\;\Delta_{\mathrm{OS12}}=+0.7\%$}
\\
\hline
\MSbar($\alpha,\beta$) & PRTS &
$0.89035(3)_{+0.9\%}^{-2.8\%}$ & $0.96103(7)_{+0.4\%}^{+1.2\%}$ &
$0.86130(3)_{+2.3\%}^{-6.1\%}$ & $0.92784(7)_{+1.3\%}^{+1.3\%}$ 
\\
& & {\footnotesize\;$\Delta_{\mathrm{OS12}}=-0.9\%$} & {\footnotesize\;$\Delta_{\mathrm{OS12}}=-0.1\%$}
  & {\footnotesize\;$\Delta_{\mathrm{OS12}}=-1.1\%$} & {\footnotesize$\;\Delta_{\mathrm{OS12}}=-0.2\%$}
\\
\hline
\MSbar($\alpha,\beta$) & GIVS &
$0.89082(3)_{+0.9\%}^{-2.7\%}$ & $0.96106(7)_{+0.5\%}^{+1.2\%}$ &
$0.86249(3)_{+2.3\%}^{-5.8\%}$ & $0.92808(7)_{+1.3\%}^{+1.3\%}$ 
\\
& & {\footnotesize\;$\Delta_{\mathrm{OS12}}=-0.8\%$} & {\footnotesize\;$\Delta_{\mathrm{OS12}}=-0.1\%$}
  & {\footnotesize\;$\Delta_{\mathrm{OS12}}=-1.0\%$} & {\footnotesize$\;\Delta_{\mathrm{OS12}}=-0.1\%$}
\\
\hline
\MSbar($\lambda_3,\beta$) & FJTS &
$0.89246(3)_{+1.6\%}^{-15.1\%}$ & $0.96108(7)_{+1.9\%}^{+17.3\%}$ &
$0.85590(3)_{+5.5\%}^{-29.8\%}$ & $0.92723(7)_{+2.8\%}^{+18.3\%}$ 
\\
& & {\footnotesize\;$\Delta_{\mathrm{OS12}}=-0.7\%$} & {\footnotesize\;$\Delta_{\mathrm{OS12}}=-0.1\%$}
  & {\footnotesize\;$\Delta_{\mathrm{OS12}}=-1.7\%$} & {\footnotesize$\;\Delta_{\mathrm{OS12}}=-0.2\%$}
\\
\hline
\MSbar($\lambda_3,\beta$) & PRTS/GIVS &
$0.89156(3)_{+1.7\%}^{-8.4\%}$ & $0.96111(7)_{+2.1\%}^{+3.8\%}$ &
$0.85841(3)_{+5.0\%}^{-12.7\%}$ & $0.92729(7)_{+2.6\%}^{+4.6\%}$ 
\\
& & {\footnotesize\;$\Delta_{\mathrm{OS12}}=-0.8\%$} & {\footnotesize\;$\Delta_{\mathrm{OS12}}=-0.1\%$}
  & {\footnotesize\;$\Delta_{\mathrm{OS12}}=-1.5\%$} & {\footnotesize$\;\Delta_{\mathrm{OS12}}=-0.2\%$}
\\
\hline
\multicolumn{5}{c}{}\\
\hline
&& \multicolumn{2}{c|}{B1} & \multicolumn{2}{c|}{B2} 
\\
Ren.\ scheme & tadpoles &
\multicolumn{1}{c|}{LO} & \multicolumn{1}{c|}{NLO}  & \multicolumn{1}{c|}{LO} & \multicolumn{1}{c|}{NLO}  
\\
\hline
\hline
OS12\,($\alpha,\beta$) & &
$0.88698(3)$ & $0.94070(8)_{+0.2\%}^{-0.5\%}$ & $0.82573(3)$ & $0.87192(6)_{+0.3\%}^{-0.5\%}$ 
\\
\hline
\MSbar($\alpha,\beta$) & FJTS &
$0.90406(3)_{+0.4\%}^{+0.4\%}$ & $0.96069(7)_{-1.1\%}^{+1.5\%}$ & $0.90654(3)_{-2.5\%}^{-87.3\%}$ &  $0.96981(7)_{-0.0\%}^{>+100\%}$
\\
& & {\footnotesize\;$\Delta_{\mathrm{OS12}}=+1.9\%$} & {\footnotesize\;$\Delta_{\mathrm{OS12}}=+2.1\%$}
  & {\footnotesize\;$\Delta_{\mathrm{OS12}}=+9.8\%$} & {\footnotesize$\;\Delta_{\mathrm{OS12}}=+11.2\%$}
\\
\hline
\MSbar($\alpha,\beta$) & PRTS &
$0.88608(3)_{+0.4\%}^{-4.9\%}$ & $0.94049(8)_{+1.0\%}^{+1.1\%}$ & $0.87943(3)_{+2.6\%}^{-42.7\%}$ & $0.92421(7)_{+4.2\%}^{-39.2\%}$
\\
& & {\footnotesize\;$\Delta_{\mathrm{OS12}}=-0.1\%$} & {\footnotesize\;$\Delta_{\mathrm{OS12}}=-0.0\%$}
  & {\footnotesize\;$\Delta_{\mathrm{OS12}}=+6.5\%$} & {\footnotesize$\;\Delta_{\mathrm{OS12}}=+6.0\%$}
\\
\hline
\MSbar($\alpha,\beta$) & GIVS &
$0.88661(3)_{+0.4\%}^{-4.7\%}$ & $0.94082(8)_{+1.0\%}^{+1.0\%}$ & $0.87990(3)_{+2.6\%}^{-42.6\%}$ &  $0.92469(7)_{+4.1\%}^{-38.7\%}$
\\
& & {\footnotesize\;$\Delta_{\mathrm{OS12}}=-0.0\%$} & {\footnotesize\;$\Delta_{\mathrm{OS12}}=+0.0\%$}
  & {\footnotesize\;$\Delta_{\mathrm{OS12}}=+6.6\%$} & {\footnotesize$\;\Delta_{\mathrm{OS12}}=+6.1\%$}
\\
\hline
\MSbar($\lambda_3,\beta$) & FJTS &
$0.75976(3)_{+19.4\%}^{-66.9\%}$ &  $1.0889(3)_{-16.4\%}^{<-100\%}$ &
$0.90615(3)\pm?$ & $0.97290(7)\pm?$
\\
& & {\footnotesize\;$\Delta_{\mathrm{OS12}}=-14.3\%$} & {\footnotesize\;$\Delta_{\mathrm{OS12}}=+15.8\%$}
  & {\footnotesize\;$\Delta_{\mathrm{OS12}}=+9.7\%$} & {\footnotesize$\;\Delta_{\mathrm{OS12}}=+11.6\%$}
\\
\hline
\MSbar($\lambda_3,\beta$) & PRTS/GIVS &
$0.88417(3)_{+1.1\%}^{-7.7\%}$ & $0.93934(8)_{+1.4\%}^{+8.1\%}$ &
$0.87475(3)\pm?$ & $0.91958(7)\pm?$
\\
& & {\footnotesize\;$\Delta_{\mathrm{OS12}}=-0.3\%$} & {\footnotesize\;$\Delta_{\mathrm{OS12}}=-0.1\%$}
  & {\footnotesize\;$\Delta_{\mathrm{OS12}}=+5.9\%$} & {\footnotesize$\;\Delta_{\mathrm{OS12}}=+5.5\%$}
\\
\hline
\end{tabular}
}
\caption{LO and NLO decay widths $\Gamma^{\Ph\to4f}$[MeV]
of the light CP-even Higgs boson $\Ph$ of the \THDM\ for various 
scenarios in different renormalization schemes,
with the OS12 scheme as input scheme (and full conversion
of the input parameters into the other schemes).
The scale variation (given in percent) corresponds to the 
scales $\mu=\mu_0/2$ and $\mu=2\mu_0$ with central
scale $\mu_0=(M_{\PH_2}+M_{\PH_1}+M_{\Ha}+2M_{\PH^+})/5$.
{The quantity $\Delta_{\mathrm{OS12}}=\Gamma_{\MSbar}/\Gamma_{\mathrm{OS12}}-1$ shows the
relative
difference between the $\MSbar$ predictions and the OS12 value; its spread 
illustrates the scheme dependence of the prediction at LO and NLO.}}
\label{tab:THDM-H24f}
\end{table}
{In the reference scheme OS12, which was suggested in \citere{Denner:2018opp},
both $\alpha$ and $\beta$ are renormalized with OS conditions imposed on
appropriate ratios of Higgs-boson decay amplitudes into neutrinos. 
Although involving some process dependence, results obtained in the OS12 scheme
exhibit many desirable features such as gauge independence and excellent
perturbative stability even in exceptional parameter regions (mass degeneracies,
large Higgs masses, extreme mixing angles). The OS12 scheme is, thus, an
appropriate choice as reference scheme.}
The $\MSbar$ results with GIVS tadpole treatment extend Tab.~7 of
\citere{Denner:2018opp}, where results for further renormalization schemes
can be found.
Since the input values of the mixing angles $\alpha$, $\beta$, 
and the parameter $\lambda_{5}$
are properly converted from the OS12 scheme to the other $\MSbar$ schemes,
ideal predictions would not show any residual dependence on the renormalization 
scale $\mu$ and on the choice of renormalization scheme.
Recall that the small $\mu$ dependence of the OS12 predictions only originates
from the dependence of the genuine NLO correction on $\lambda_{5}$, while
$\alpha$ and $\beta$ are fixed. 
The comprehensive comparison of predictions based on several other schemes
carried out in \citere{Denner:2018opp},
moreover, shows that the OS12 predictions are perturbatively
perfectly stable with very small corrections.
The comparison of $\MSbar$ numbers to the OS12 numbers, thus, 
nicely quantify the scheme dependence of the LO and NLO results.

Compared to the discussion of the renormalization scale and scheme dependence
of the corresponding results in the SESM in \refse{se:SESMnumerics}, the
situation for the THDM results is much more diverse, and the results 
for the low-mass and high-mass scenarios A1, A2 and B1, B2, respectively,
differ considerably.
In the low-mass scenarios A1, A2 the $\MSbar$ predictions based on 
PRTS or GIVS show a nice reduction of their scale and scheme uncertainty in the
transition from LO to NLO, as already discussed in 
\citeres{Altenkamp:2017ldc,Altenkamp:2017kxk,Denner:2018opp} for the
PRTS variant. As also pointed out there, the FJTS variants already show
issues with perturbative stability in A1, A2;
nevertheless the FJTS $\MSbar$ predictions look still consistent with
a reasonable, albeit significant, scale uncertainty reflecting the
overall theoretical uncertainty realistically.

In the high-mass scenario B1, the $\MSbar(\alpha,\beta)$ schemes
behave properly for all tadpole schemes.
For B1, the $\MSbar(\lambda_3,\beta)$ {scheme} produces still reasonable results
for the PRTS/GIVS tadpole variants, though without significant reduction of
the scale uncertainty. However, the scheme $\MSbar(\lambda_3,\beta)$ FJTS totally fails,
as can already be conjectured from the corrections observed in the
parameter conversion shown in \refapp{app:THDM}.
The situation for high-mass scenario B2 is even more extreme.
The failure of the FJTS $\MSbar$ schemes can already be anticipated from the
corrections to the parameter conversions, see again \refapp{app:THDM}.
The PRTS and GIVS $\MSbar$ variants behave better, but do not show
uncertainty reductions in the transition from LO to NLO.
The question marks in \refta{tab:THDM-H24f} indicate that the
scale variation could not be evaluated consistently, since the
conversion effects pushed the parameter $s_\alpha$ out of its
bounds $|s_\alpha|\le1$ for the extreme scales.

In summary, we confirm the expectation that the new (gauge-independent)
GIVS tadpole variant
produces results in the $\MSbar$-type schemes that are very close to
the corresponding (gauge-dependent) PRTS results;
for the $\MSbar(\lambda_3,\beta)$ scheme the variants are even identical.
The superiority of the PRTS variant over the FJTS variant in view of
perturbative stability, which was discussed in
\citeres{Altenkamp:2017ldc,Altenkamp:2017kxk,Denner:2018opp} in
greater detail, thus, nicely carries over to the GIVS,
which rescues gauge independence in addition.

It should, however, also be mentioned that none of the 
$\MSbar$ schemes with any tadpole treatment is able to produce
fully satisfactory results for the four-body decays of the
heavier CP-even neutral Higgs boson~H of the THDM.
Predictions for the decays $\PH\to4f$ typically receive larger corrections
than for $\Ph\to4f$, because the leading-order decay widths of the former are reduced
by small values of $c_{\beta-\alpha}^2$ w.r.t.\ the latter which are proportional
to $s_{\beta-\alpha}^2\lsim1$. The potentially larger corrections to $\PH\to4f$
render the corresponding NLO predictions more prone to perturbative instabilities.
The GIVS results for the NLO decay widths 
are again similar to the PRTS results,
which are included in the detailed study of NLO predictions for
$\PH\to\PW\PW/\PZ\PZ\to4f$ based on various types of renormalization 
schemes presented in \citere{Denner:2018opp}. 
For $\PH\to4f$, predictions, thus, should be based on the OS-type
or symmetry-inspired schemes proposed in \citere{Denner:2018opp}.

\section{Conclusions}
\label{se:Conclusions}

Applying $\MSbar$ renormalization conditions to mass parameters
or mixing angles leads to subtle issues in the precise definition
of vev parameters
if EW corrections are included in predictions.
Technically, these issues concern the treatment of tadpoles in the
course of renormalization.
In the past, mostly two tadpole schemes have been in use, called
PRTS and FJTS in this paper, which, however, both have benefits and drawbacks.
The PRTS is based on the idea to renormalize parameters in such a way
that a field is expanded about the 
minimum of the effective Higgs potential.
For Higgs fields developing vevs this means that no potentially large corrections
to the vevs enter predictions, but gauge-dependent tadpole loop contributions
enter the relations between bare parameters of the theory (at least in the 
usually adopted Higgs representations), leading to gauge-dependent parametrizations of
observables.
The alternative scheme, called FJTS, avoids gauge dependences by just
redistributing tadpole contributions in predictions 
via field shifts without altering parameter
relations, but on cost of an expansion about a field configuration that corresponds
to the vacuum only in 
lowest order. By experience this procedure
is prone to artificially large corrections which jeopardize the stability of
perturbation theory.

To improve on this unsatisfactory situation, in a preceding paper~\cite{Dittmaier:2022maf}
we have proposed a hybrid scheme for treating tadpole contributions,
dubbed {\it Gauge-Invariant Vacuum expectation value 
Scheme (GIVS)},
which unifies the benefits and avoids the shortcomings of the PRTS and FJTS.
Perturbative stability is achieved by an expansion about the true
vacuum field configuration, and gauge dependences are avoided by employing
non-linear Higgs re\-presentations,
in which CP-even Higgs fields and their
vevs are gauge-invariant quantities.
As an application in the SM, we 
had shown in \citere{Dittmaier:2022maf} that the conversion of
OS-renormalized to $\MSbar$-renormalized masses involves only very moderate
EW corrections, in contrast to the situation in the FJTS.
In this paper, we have applied the GIVS to the $\MSbar$ renormalization
of Higgs mixing angles 
in two scalar extensions of the SM:
a Higgs singlet extension (called SESM),
and the Two-Higgs-Doublet model (THDM), the renormalization of which has been
extensively discussed in the literature in recent years.

To apply the GIVS to the SESM and THDM, we have first formulated
the Higgs sectors in 
a non-linear way. While the SESM requires only minor
modifications beyond the SM formulation, the generalization to the THDM
is quite non-trivial and interesting in its own right.
We have described the transition from linearly to non-linearly realized Higgs doublets
in detail and have presented the (non-polynomial) kinetic Lagrangians of the THDM Higgs
field in a way that renders the generation of all Feynman
rules (which involve 
arbitrarily many Goldstone legs) quite simple.
While kinetic Lagrangians become more complicated in non-linear Higgs representations,
the Higgs potentials become simpler, because all would-be Goldstone fields
drop out there by construction.

To illustrate the perturbative stability of the results based on
$\MSbar$ renormalization with the GIVS tadpole treatment,
we have discussed the NLO predictions for the important four-body decays of
CP-even neutral Higgs bosons, $\Ph/\PH\to\PW\PW/\PZ\PZ\to4\,$fermions
based on $\MSbar$-renormalized Higgs mixing angles.
In all cases, we have found that the (gauge-independent) GIVS results are very close
to the corresponding (gauge-dependent) PRTS results, thus, sharing the
superiority over the FJTS results w.r.t.\ perturbative stability.
The new GIVS variants of the $\MSbar$ schemes will be made available 
in a forthcoming public version of the Monte Carlo program \Prophecy\
for the considered four-body Higgs decays.

Finally, in the THDM we have made the observation that the 
GIVS variant for an $\MSbar$-renormalized parameter $\tan\beta=v_2/v_1$
produces exactly the same one-loop renormalization constants for
$\tan\beta$ as the PRTS variant 
in the class of $R_\xi$ gauges, i.e.\ that effectively the $\MSbar$
PRTS renormalization for $\tan\beta$ does not suffer from gauge dependences
after reinterpretation as GIVS results. This observation is particularly
interesting in the context of EW renormalization of supersymmetric
models, in which $\tan\beta$ is often $\MSbar$-renormalized with
PRTS-like tadpole prescriptions. We conjecture that this
interesting fact generalizes to higher orders as well and, thus, 
can put higher-order predictions in supersymmetric models obtained with the PRTS
in $R_\xi$ gauges on a gauge-invariant basis.

\subsection*{Acknowledgements}

\begin{sloppypar}
Ansgar Denner is gratefully acknowledged
for helpful discussions and for comments on the manuscript.
Moreover, the authors acknowledge support by the state of 
Baden-W\"urttemberg through bwHPC, by the DFG through 
grant no INST 39/963-1 FUGG, and by the BMBF under contract 05H21VFCA1.
H.R.'s research is funded by  the
Deutsche Forschungsgemeinschaft (DFG, German Research
Foundation)---project no.\ 442089526; 442089660.
\end{sloppypar}


\appendix
\section*{Appendix}

\section{Parameter conversion tables}
\label{se:conversiontables}

The full set of input parameters of the various SESM and THDM scenarios,
as used in \refses{se:SESMnumerics} and \ref{se:THDMnumerics}, can be found
in \citeres{Altenkamp:2018bcs} and \cite{Altenkamp:2017ldc,Altenkamp:2017kxk}, respectively.
Here we list results for the conversion of those
non-standard input parameters that are subject to different 
renormalization schemes for the SESM and THDM.
The input parameters are defined in
the on-shell input schemes OS and OS12%
\footnote{Although NLO results in the OS and OS12 schemes are completely
independent of the tadpole treatment, there is some slight dependence on the
tadpole scheme (beyond NLO accuracy) used in these schemes if input parameter
conversions between schemes are done in a self-consistent way
i.e.\ not truncated at the NLO level 
(see \citere{Denner:2019vbn} for more details). 
For completeness, we mention that we apply the PRTS
variant in the OS and OS12 schemes.}
in the SESM and THDM,
respectively, and subsequently converted into the various 
$\MSbar$ renormalization schemes. 
Details on the conversion procedure can also
be found in Sect.~3.5 of \citere{Denner:2018opp}.

\subsection{SESM scenarios}
\label{app:SESM}

Table~\ref{tab:conversion-SESM-OS} shows the results for the (full) conversion
of parameters from the OS renormalization scheme used
as input scheme to $\MSbar$ schemes for the
considered benchmark scenarios. 
\begin{table}
\centerline{\renewcommand{\arraystretch}{1.25} \small \tabcolsep 5pt
\begin{tabular}{|c|c||c|c|c|c|c|c|c|c|c|c|c|c|c|c|c|c|}
\hline
&& \multicolumn{2}{c|}{BHM200$^+$} & \multicolumn{2}{c|}{BHM200$^-$} 
& \multicolumn{2}{c|}{BHM400} & \multicolumn{2}{c|}{BHM600} 
\\
Ren.\ scheme  & tadpoles
  & \multicolumn{1}{c|}{\sa} & \multicolumn{1}{c|}{$\lambda_{12}$}  
  & \multicolumn{1}{c|}{\sa} & \multicolumn{1}{c|}{$\lambda_{12}$}  
  & \multicolumn{1}{c|}{\sa} & \multicolumn{1}{c|}{$\lambda_{12}$}  
  & \multicolumn{1}{c|}{\sa} & \multicolumn{1}{c|}{$\lambda_{12}$}  
\\
\hline
\hline
  OS &         & $0.29$&  $0.07$ & $-0.29$ &$-0.07$ & $0.26$ & $0.17$ & $0.22$ & $0.23$\\
\hline
\hline
  \MSbar& FJTS  & $0.321$& $0.077$& $-0.316$& $-0.076$ & $0.264$ & $0.173$ & $0.212$ & $0.222$\\
\hline
  \MSbar& PRTS  & $0.302$& $0.073$& $-0.304$& $-0.073$ & $0.267$ & $0.175$ & $0.226$ & $0.236$\\
\hline
  \MSbar& GIVS  & $0.301$ & $0.073$ & $-0.302$& $-0.073$ & $0.266$ & $0.174$ & $0.225$ & $0.236$\\
\hline
\end{tabular}
}
  \caption{Conversion of parameters from the OS scheme as
    input scheme to the different $\MSbar$ renormalization schemes at the central scale
    $\mu_0=\Mh$ performed in the PRTS, using the bare parameters for
  $\tan\alpha$ and $\lambda_{12}$ in the matching.}
\label{tab:conversion-SESM-OS}
\end{table}
This table completes Tab.~14 of \citere{Denner:2018opp}, where the input
conversion to other schemes can be found.
We recall that the included renormalization schemes employ $\lambda_1$ (not $\lambda_{12}$)
as independent running $\MSbar$-renormalized parameter, although the scenarios are fixed
by the values of $\lambda_{12}$.

The moderate corrections in the input parameter conversions reflect
the perturbative stability of the $\MSbar$ schemes with all the considered
tadpole treatments in 
all considered scenarios.
Note also that the corrections are almost identical for the PRTS and GIVS
variants, as expected from the small differences in the tadpole terms of
the renormalization constant $\delta\alpha$ of the mixing angle $\alpha$.

\subsection{THDM scenarios}
\label{app:THDM}

Table~\ref{tab:conversion-THDM-OS12} shows the corresponding conversion in
the THDM from the OS12 renormalization scheme used
as input scheme to the different $\MSbar$
renormalization schemes.
\begin{table}
\centerline{\renewcommand{\arraystretch}{1.25} \small \tabcolsep 5pt
\begin{tabular}{|c|c||c|c|c|c|c|c|c|c|c|c|c|c|c|c|c|c|}
\hline
&& \multicolumn{2}{c|}{A1} & \multicolumn{2}{c|}{A2} 
& \multicolumn{2}{c|}{B1} & \multicolumn{2}{c|}{B2} 
\\
Ren.\ scheme & tadpoles
  & \multicolumn{1}{c|}{\cab} & \multicolumn{1}{c|}{\tb}  
  & \multicolumn{1}{c|}{\cab} & \multicolumn{1}{c|}{\tb}  
  & \multicolumn{1}{c|}{\cab} & \multicolumn{1}{c|}{\tb}  
  & \multicolumn{1}{c|}{\cab} & \multicolumn{1}{c|}{\tb}  
\\
\hline
\hline
  OS12\,($\alpha,\beta$) &  & $0.1$ & $2.0$ & $0.2$ & $2.0$ & $0.15$ & $4.5$  & $0.3$ & $3.0$ \\
\hline
\hline
  \MSbar($\alpha,\beta$)& FJTS  & $0.090$ & $1.93$ & $0.157$ & $1.91$ & $0.061$ & $3.82$ & $-0.031$ & $3.70$ \\
\hline
  \MSbar($\alpha,\beta$)& PRTS  & $0.137$ & $1.90$ & $0.225$ & $1.91$ & $0.153$ & $4.40$  & $0.176$ & $2.83$\\
\hline
  \MSbar($\alpha,\beta$)& GIVS  & $0.135$ & $1.90$ & $0.222$ & $1.91$ & $0.151$ & $4.40$ & $0.174$ & $2.83$ \\ 
\hline
  \MSbar($\lambda_3,\beta$)& FJTS  & $0.128$ & $1.92$ & $0.238$ & $1.88$ & $0.403$ & $2.04$ & $-0.037$ & $3.62$ \\
\hline
  \MSbar($\lambda_3,\beta$)& PRTS/GIVS  & $0.132$ & $1.90$ & $0.232$ & $1.91$ & $0.160$ & $4.40$ & $0.190$ & $2.83$ \\
\hline
\end{tabular}
}
  \caption{Conversion of parameters from the OS12 scheme as
    input scheme to the different $\MSbar$ renormalization schemes at the central scale 
    $\mu_0=(\Mh+\MH+\MHa+2\MHpm)/5$
    performed in the PRTS, using the bare parameters for $\tan\alpha$ and $\tan\beta$ 
    in the matching.}
\label{tab:conversion-THDM-OS12}
\end{table}
This table completes Tab.~15 of \citere{Denner:2018opp}, where the input
conversion to other schemes can be found.
As compared to the SESM, the conversion effects for the considered THDM scenarios 
are more pronounced.
The conversion to
\MSbarFJTS{} becomes perturbatively unstable (and fails
completely) for scenarios B1 and B2.
On the other hand, the PRTS and GIVS tadpole treatments lead to moderate corrections,
reflecting perturbative stability. As expected, the results of the PRTS and GIVS variants
almost agree for the \MSbar($\alpha,\beta$) scheme; 
in the \MSbar($\lambda_3,\beta$) scheme these 
two tadpole variants produce
identical results as explained in \refse{se:GIVschemeTHDM}.

\bibliographystyle{JHEPmod}

\begin{thebibliography}{10}

\bibitem{LHCHiggsCrossSectionWorkingGroup:2011wcg}
{\bf LHC Higgs Cross Section Working Group} Collaboration, S.~Dittmaier et~al.,
  {\it {Handbook of LHC Higgs Cross Sections: 1. Inclusive Observables}},
  \href{http://arxiv.org/abs/1101.0593}{{\tt arXiv:1101.0593}}.

\bibitem{Dittmaier:2012vm}
{\bf LHC Higgs Cross Section Working Group} Collaboration, S.~Dittmaier et~al.,
  {\it {Handbook of LHC Higgs Cross Sections: 2. Differential Distributions}},
  \href{http://arxiv.org/abs/1201.3084}{{\tt arXiv:1201.3084}}.

\bibitem{LHCHiggsCrossSectionWorkingGroup:2013rie}
{\bf LHC Higgs Cross Section Working Group} Collaboration, J.~R. Andersen
  et~al., {\it {Handbook of LHC Higgs Cross Sections: 3. Higgs Properties}},
  \href{http://arxiv.org/abs/1307.1347}{{\tt arXiv:1307.1347}}.

\bibitem{LHCHiggsCrossSectionWorkingGroup:2016ypw}
{\bf LHC Higgs Cross Section Working Group} Collaboration, D.~de~Florian
  et~al., {\it {Handbook of LHC Higgs Cross Sections: 4. Deciphering the Nature
  of the Higgs Sector}},  \href{http://arxiv.org/abs/1610.07922}{{\tt
  arXiv:1610.07922}}.

\bibitem{Amoroso:2020lgh}
S.~Amoroso et~al., {\it {Les Houches 2019: Physics at TeV Colliders: Standard
  Model Working Group Report}},  in {\em {11th Les Houches Workshop on Physics
  at TeV Colliders}: {PhysTeV Les Houches}}, 3, 2020.
\newblock \href{http://arxiv.org/abs/2003.01700}{{\tt arXiv:2003.01700}}.

\bibitem{Heinrich:2020ybq}
G.~Heinrich, {\it {Collider Physics at the Precision Frontier}},  {\em Phys.
  Rept.} {\bf 922} (2021) 1--69, [\href{http://arxiv.org/abs/2009.00516}{{\tt
  arXiv:2009.00516}}].

\bibitem{Denner:2019vbn}
A.~Denner and S.~Dittmaier, {\it {Electroweak Radiative Corrections for
  Collider Physics}},  {\em Phys. Rept.} {\bf 864} (2020) 1--163,
  [\href{http://arxiv.org/abs/1912.06823}{{\tt arXiv:1912.06823}}].

\bibitem{Krause:2016oke}
M.~Krause, R.~Lorenz, M.~M{\"u}hlleitner, R.~Santos, and H.~Ziesche, {\it
  {Gauge-independent Renormalization of the 2-Higgs-Doublet Model}},  {\em
  JHEP} {\bf 09} (2016) 143, [\href{http://arxiv.org/abs/1605.04853}{{\tt
  arXiv:1605.04853}}].

\bibitem{Denner:2016etu}
A.~Denner, L.~Jenniches, J.-N. Lang, and C.~Sturm, {\it {Gauge-independent
  $\overline{MS}$ renormalization in the 2HDM}},  {\em JHEP} {\bf 09} (2016)
  115, [\href{http://arxiv.org/abs/1607.07352}{{\tt arXiv:1607.07352}}].

\bibitem{Altenkamp:2017ldc}
L.~Altenkamp, S.~Dittmaier, and H.~Rzehak, {\it {Renormalization schemes for
  the Two-Higgs-Doublet Model and applications to $h\to WW/ZZ\to4\,$fermions}},
   {\em JHEP} {\bf 09} (2017) 134, [\href{http://arxiv.org/abs/1704.02645}{{\tt
  arXiv:1704.02645}}].

\bibitem{Altenkamp:2018bcs}
L.~Altenkamp, M.~Boggia, and S.~Dittmaier, {\it {Precision calculations for $h
  \to WW/ZZ \to 4$ fermions in a Singlet Extension of the Standard Model with
  Prophecy4f}},  \href{http://arxiv.org/abs/1801.07291}{{\tt
  arXiv:1801.07291}}.

\bibitem{Denner:2018opp}
A.~Denner, S.~Dittmaier, and J.-N. Lang, {\it {Renormalization of mixing
  angles}},  {\em JHEP} {\bf 11} (2018) 104,
  [\href{http://arxiv.org/abs/1808.03466}{{\tt arXiv:1808.03466}}].

\bibitem{Bohm:1986rj}
M.~B{\"o}hm, H.~Spiesberger, and W.~Hollik, {\it {On the One-Loop
  Renormalization of the Electroweak Standard Model and Its Application to
  Leptonic Processes}},  {\em Fortsch. Phys.} {\bf 34} (1986) 687--751.

\bibitem{Denner:1991kt}
A.~Denner, {\it {Techniques for calculation of electroweak radiative
  corrections at the one-loop level and results for W physics at LEP-200}},
  {\em Fortsch. Phys.} {\bf 41} (1993) 307--420,
  [\href{http://arxiv.org/abs/0709.1075}{{\tt arXiv:0709.1075}}].

\bibitem{Fleischer:1980ub}
J.~Fleischer and F.~Jegerlehner, {\it {Radiative corrections to Higgs decays in
  the extended Weinberg-Salam Model}},  {\em Phys. Rev.} {\bf D23} (1981)
  2001--2026.

\bibitem{Actis:2006ra}
S.~Actis, A.~Ferroglia, M.~Passera, and G.~Passarino, {\it {Two-Loop
  Renormalization in the Standard Model. Part I: Prolegomena}},  {\em Nucl.
  Phys. B} {\bf 777} (2007) 1--34,
  [\href{http://arxiv.org/abs/hep-ph/0612122}{{\tt hep-ph/0612122}}].

\bibitem{Jegerlehner:2012kn}
F.~Jegerlehner, M.~Y. Kalmykov, and B.~A. Kniehl, {\it {On the difference
  between the pole and the $\overline {MS}$ masses of the top quark at the
  electroweak scale}},  {\em Phys. Lett. B} {\bf 722} (2013) 123--129,
  [\href{http://arxiv.org/abs/1212.4319}{{\tt arXiv:1212.4319}}].

\bibitem{Kniehl:2015nwa}
B.~A. Kniehl, A.~F. Pikelner, and O.~L. Veretin, {\it {Two-loop electroweak
  threshold corrections in the Standard Model}},  {\em Nucl. Phys. B} {\bf 896}
  (2015) 19--51, [\href{http://arxiv.org/abs/1503.02138}{{\tt
  arXiv:1503.02138}}].

\bibitem{Kataev:2022dua}
A.~L. Kataev and V.~S. Molokoedov, {\it {Notes on interplay of the QCD and EW
  perturbative corrections to the pole-running top-quark mass ratio}},
  \href{http://arxiv.org/abs/2201.12073}{{\tt arXiv:2201.12073}}.

\bibitem{Dudenas:2020ggt}
V.~D\={u}d\.{e}nas and M.~L\"oschner, {\it {Vacuum expectation value
  renormalization in the Standard Model and beyond}},  {\em Phys. Rev. D} {\bf
  103} (2021) 076010, [\href{http://arxiv.org/abs/2010.15076}{{\tt
  arXiv:2010.15076}}].

\bibitem{Dittmaier:2022maf}
S.~Dittmaier and H.~Rzehak, {\it {Electroweak renormalization based on
  gauge-invariant vacuum expectation values of non-linear Higgs
  representations. Part I. Standard Model}},  {\em JHEP} {\bf 05} (2022) 125,
  [\href{http://arxiv.org/abs/2203.07236}{{\tt arXiv:2203.07236}}].

\bibitem{Lee:1972yfa}
B.~W. Lee and J.~Zinn-Justin, {\it {Spontaneously Broken Gauge Symmetries Part
  3: Equivalence}},  {\em Phys. Rev. D} {\bf 5} (1972) 3155--3160.

\bibitem{Grosse-Knetter:1992tbp}
C.~Grosse-Knetter and R.~{K\"ogerler}, {\it {Unitary gauge, Stuckelberg
  formalism and gauge invariant models for effective lagrangians}},  {\em Phys.
  Rev. D} {\bf 48} (1993) 2865--2876,
  [\href{http://arxiv.org/abs/hep-ph/9212268}{{\tt hep-ph/9212268}}].

\bibitem{Schabinger:2005ei}
R.~M. Schabinger and J.~D. Wells, {\it {A Minimal spontaneously broken hidden
  sector and its impact on Higgs boson physics at the large hadron collider}},
  {\em Phys. Rev.} {\bf D72} (2005) 093007,
  [\href{http://arxiv.org/abs/hep-ph/0509209}{{\tt hep-ph/0509209}}].

\bibitem{Patt:2006fw}
B.~Patt and F.~Wilczek, {\it {Higgs-field portal into hidden sectors}},
  \href{http://arxiv.org/abs/hep-ph/0605188}{{\tt hep-ph/0605188}}.

\bibitem{Bowen:2007ia}
M.~Bowen, Y.~Cui, and J.~D. Wells, {\it {Narrow trans-TeV Higgs bosons and H
  $\to$ hh decays: Two LHC search paths for a hidden sector Higgs boson}},
  {\em JHEP} {\bf 03} (2007) 036,
  [\href{http://arxiv.org/abs/hep-ph/0701035}{{\tt hep-ph/0701035}}].

\bibitem{Gunion:2002zf}
J.~F. Gunion and H.~E. Haber, {\it {The CP conserving two Higgs doublet model:
  The Approach to the decoupling limit}},  {\em Phys. Rev. D} {\bf 67} (2003)
  075019, [\href{http://arxiv.org/abs/hep-ph/0207010}{{\tt hep-ph/0207010}}].

\bibitem{Branco:2011iw}
G.~C. Branco, et~al., {\it {Theory and phenomenology of two-Higgs-doublet
  models}},  {\em Phys. Rept.} {\bf 516} (2012) 1--102,
  [\href{http://arxiv.org/abs/1106.0034}{{\tt arXiv:1106.0034}}].

\bibitem{Altenkamp:2017kxk}
L.~Altenkamp, S.~Dittmaier, and H.~Rzehak, {\it {Precision calculations for $h
  \to WW/ZZ \to 4$ fermions in the Two-Higgs-Doublet Model with Prophecy4f}},
  {\em JHEP} {\bf 03} (2018) 110, [\href{http://arxiv.org/abs/1710.07598}{{\tt
  arXiv:1710.07598}}].

\bibitem{Bredenstein:2006rh}
A.~Bredenstein, A.~Denner, S.~Dittmaier, and M.~M. Weber, {\it {Precise
  predictions for the Higgs-boson decay $H\to WW/ZZ\to4\,$ leptons}},  {\em
  Phys. Rev.} {\bf D74} (2006) 013004,
  [\href{http://arxiv.org/abs/hep-ph/0604011}{{\tt hep-ph/0604011}}].

\bibitem{Bredenstein:2006ha}
A.~Bredenstein, A.~Denner, S.~Dittmaier, and M.~M. Weber, {\it {Radiative
  corrections to the semileptonic and hadronic Higgs-boson decays $H\to W W / Z
  Z\to4\,$ fermions}},  {\em JHEP} {\bf 02} (2007) 080,
  [\href{http://arxiv.org/abs/hep-ph/0611234}{{\tt hep-ph/0611234}}].

\bibitem{Denner:2019fcr}
A.~Denner, S.~Dittmaier, and A.~M\"uck, {\it {PROPHECY4F 3.0: A Monte Carlo
  program for Higgs-boson decays into four-fermion final states in and beyond
  the Standard Model}},  {\em Comput. Phys. Commun.} {\bf 254} (2020) 107336,
  [\href{http://arxiv.org/abs/1912.02010}{{\tt arXiv:1912.02010}}].

\bibitem{Dittmaier:1995cr}
S.~Dittmaier and C.~Grosse-Knetter, {\it {Deriving nondecoupling effects of
  heavy fields from the path integral: A Heavy Higgs field in an SU(2) gauge
  theory}},  {\em Phys. Rev. D} {\bf 52} (1995) 7276--7293,
  [\href{http://arxiv.org/abs/hep-ph/9501285}{{\tt hep-ph/9501285}}].

\bibitem{Dittmaier:1995ee}
S.~Dittmaier and C.~Grosse-Knetter, {\it {Integrating out the standard Higgs
  field in the path integral}},  {\em Nucl. Phys. B} {\bf 459} (1996) 497--536,
  [\href{http://arxiv.org/abs/hep-ph/9505266}{{\tt hep-ph/9505266}}].

\bibitem{Kanemura:2015fra}
S.~Kanemura, M.~Kikuchi, and K.~Yagyu, {\it {Radiative corrections to the Higgs
  boson couplings in the model with an additional real singlet scalar field}},
  {\em Nucl. Phys.} {\bf B907} (2016) 286--322,
  [\href{http://arxiv.org/abs/1511.06211}{{\tt arXiv:1511.06211}}].

\bibitem{Bojarski:2015kra}
F.~Bojarski, G.~Chalons, D.~Lopez-Val, and T.~Robens, {\it {Heavy to light
  Higgs boson decays at NLO in the Singlet Extension of the Standard Model}},
  {\em JHEP} {\bf 02} (2016) 147, [\href{http://arxiv.org/abs/1511.08120}{{\tt
  arXiv:1511.08120}}].

\bibitem{Denner:2017vms}
A.~Denner, J.-N. Lang, and S.~Uccirati, {\it {NLO electroweak corrections in
  extended Higgs Sectors with RECOLA2}},  {\em JHEP} {\bf 07} (2017) 087,
  [\href{http://arxiv.org/abs/1705.06053}{{\tt arXiv:1705.06053}}].

\bibitem{Kluberg-Stern:1974iel}
H.~Kluberg-Stern and J.~B. Zuber, {\it {Ward Identities and Some Clues to the
  Renormalization of Gauge Invariant Operators}},  {\em Phys. Rev. D} {\bf 12}
  (1975) 467--481.

\bibitem{Nielsen:1975fs}
N.~K. Nielsen, {\it {On the Gauge Dependence of Spontaneous Symmetry Breaking
  in Gauge Theories}},  {\em Nucl. Phys. B} {\bf 101} (1975) 173--188.

\bibitem{Gambino:1999ai}
P.~Gambino and P.~A. Grassi, {\it {The Nielsen identities of the SM and the
  definition of mass}},  {\em Phys. Rev. D} {\bf 62} (2000) 076002,
  [\href{http://arxiv.org/abs/hep-ph/9907254}{{\tt hep-ph/9907254}}].

\bibitem{Kanemura:2004mg}
S.~Kanemura, Y.~Okada, E.~Senaha, and C.~P. Yuan, {\it {Higgs coupling
  constants as a probe of new physics}},  {\em Phys. Rev. D} {\bf 70} (2004)
  115002, [\href{http://arxiv.org/abs/hep-ph/0408364}{{\tt hep-ph/0408364}}].

\bibitem{Lopez-Val:2009xtx}
D.~Lopez-Val and J.~Sola, {\it {Neutral Higgs-pair production at Linear
  Colliders within the general 2HDM: Quantum effects and triple Higgs boson
  self-interactions}},  {\em Phys. Rev. D} {\bf 81} (2010) 033003,
  [\href{http://arxiv.org/abs/0908.2898}{{\tt arXiv:0908.2898}}].

\bibitem{Kanemura:2014dja}
S.~Kanemura, M.~Kikuchi, and K.~Yagyu, {\it {Radiative corrections to the
  Yukawa coupling constants in two Higgs doublet models}},  {\em Phys. Lett. B}
  {\bf 731} (2014) 27--35, [\href{http://arxiv.org/abs/1401.0515}{{\tt
  arXiv:1401.0515}}].

\bibitem{Krause:2018wmo}
M.~Krause, M.~M\"uhlleitner, and M.~Spira, {\it {2HDECAY \textemdash{}A program
  for the calculation of electroweak one-loop corrections to Higgs decays in
  the Two-Higgs-Doublet Model including state-of-the-art QCD corrections}},
  {\em Comput. Phys. Commun.} {\bf 246} (2020) 106852,
  [\href{http://arxiv.org/abs/1810.00768}{{\tt arXiv:1810.00768}}].

\bibitem{Krause:2019qwe}
M.~Krause and M.~M\"uhlleitner, {\it {Impact of Electroweak Corrections on
  Neutral Higgs Boson Decays in Extended Higgs Sectors}},  {\em JHEP} {\bf 04}
  (2020) 083, [\href{http://arxiv.org/abs/1912.03948}{{\tt arXiv:1912.03948}}].

\bibitem{Dittmaier:2021fls}
S.~Dittmaier, S.~Schuhmacher, and M.~Stahlhofen, {\it {Integrating out heavy
  fields in the path integral using the background-field method: general
  formalism}},  {\em Eur. Phys. J. C} {\bf 81} (2021) 826,
  [\href{http://arxiv.org/abs/2102.12020}{{\tt arXiv:2102.12020}}].

\bibitem{Freitas:2002um}
A.~Freitas and D.~{St\"ockinger}, {\it {Gauge dependence and renormalization of
  tan beta in the MSSM}},  {\em Phys. Rev. D} {\bf 66} (2002) 095014,
  [\href{http://arxiv.org/abs/hep-ph/0205281}{{\tt hep-ph/0205281}}].

\bibitem{Frank:2006yh}
M.~Frank, et~al., {\it {The Higgs Boson Masses and Mixings of the Complex MSSM
  in the Feynman-Diagrammatic Approach}},  {\em JHEP} {\bf 02} (2007) 047,
  [\href{http://arxiv.org/abs/hep-ph/0611326}{{\tt hep-ph/0611326}}].

\bibitem{Degrassi:2014pfa}
G.~Degrassi, S.~Di~Vita, and P.~Slavich, {\it {Two-loop QCD corrections to the
  MSSM Higgs masses beyond the effective-potential approximation}},  {\em Eur.
  Phys. J. C} {\bf 75} (2015) 61, [\href{http://arxiv.org/abs/1410.3432}{{\tt
  arXiv:1410.3432}}].

\bibitem{Fritzsche:2013fta}
T.~Fritzsche, et~al., {\it {The Implementation of the Renormalized Complex MSSM
  in FeynArts and FormCalc}},  {\em Comput. Phys. Commun.} {\bf 185} (2014)
  1529--1545, [\href{http://arxiv.org/abs/1309.1692}{{\tt arXiv:1309.1692}}].

\end{thebibliography}
\providecommand{\href}[2]{#2}\begingroup\raggedright\endgroup

\end{document}